\documentclass[aps,prl,twocolumn,10pt,superscriptaddress]{revtex4-2}

\usepackage{mathrsfs}
\usepackage{bbm}
\usepackage{amsmath}
\usepackage{amssymb}
\usepackage{amsthm}
\usepackage{graphicx}
\usepackage{MnSymbol}
\usepackage{mathtools}
\usepackage{physics}
\usepackage{dsfont}

\makeatletter
\usepackage{hyperref}
\usepackage{color}
\definecolor{supcol}{RGB}{10,50,180}
\definecolor{eqcol}{RGB}{220,10,100}
\hypersetup{
	colorlinks,
	citecolor=supcol,
	linkcolor=eqcol,
	urlcolor=supcol
}

\allowdisplaybreaks

\newcommand{\E}{\mathbb{E}}
\newcommand{\Lg}{\mathcal{L}}
\newcommand{\Rc}{\mathcal{R}}
\newcommand{\Rm}{\mathsf{R}}
\newcommand{\Ac}{\mathcal{A}}
\newcommand{\Am}{\mathsf{A}}
\newcommand{\Fh}{\mathcal{F}}

\newcommand{\Cm}{\mathsf{C}}
\newcommand{\bj}{\vb*{j}}
\newcommand{\bu}{\vb*{u}}
\newcommand{\vf}{\vb*{f}}
\newcommand{\bg}{\vb*{g}}
\newcommand{\bh}{\vb*{h}}
\newcommand{\bx}{\vb*{x}}
\newcommand{\bv}{\vb*{v}}
\newcommand{\bz}{\vb*{z}}
\newcommand{\bw}{\vb*{w}}
\newcommand{\bW}{\vb*{W}}

\newcommand{\bnu}{\vb*{\nu}}
\newcommand{\D}{\mathsf{D}}
\newcommand{\T}{\mathsf{T}}
\newcommand{\jr}{\bar{\jmath}}
\newcommand{\st}{{\rm ss}}
\newcommand{\vbar}{\bar{v}}
\newcommand{\cond}{\,|\,}
\newcommand{\inp}[2]{\langle\!\langle #1 , #2 \rangle\!\rangle}

\newcommand{\sectionprl}[1]{{\em #1}\/---}

\DeclareMathOperator{\Var}{Var}
\DeclareMathOperator{\Cov}{Cov}

\begin{document}

\title{Exact Fluctuation-Response Relations for Underdamped Langevin Dynamics}

\author{Tan Van Vu}
\email{tan.vu@yukawa.kyoto-u.ac.jp}
\affiliation{Center for Gravitational Physics and Quantum Information, Yukawa Institute for Theoretical Physics, Kyoto University, Kitashirakawa Oiwakecho, Sakyo-ku, Kyoto 606-8502, Japan}

\author{Van Tuan Vo}
\affiliation{Quantum AI \& Cyber Security Research Institute, FPT Corporation, 10 Pham Van Bach, Cau Giay, Ha Noi 100000, Vietnam}

\author{Ruicheng Bao}
\affiliation{Department of Physics, Graduate School of Science, The University of Tokyo, Hongo, Bunkyo-ku, Tokyo 113-0033, Japan}

\author{Keiji Saito}
\affiliation{Department of Physics, Kyoto University, Kyoto 606-8502, Japan}

\date{\today}

\begin{abstract}
Thermodynamic uncertainty relations connect current fluctuations to dissipation and, in certain settings, can emerge from underlying fluctuation-response relations. However, in underdamped dynamics, conventional mean-current uncertainty relations can fail, and the underlying connection between fluctuations, response, and dissipation remains elusive. Here, we uncover this connection by deriving an exact finite-time fluctuation-response equality for underdamped Langevin dynamics, which is valid for arbitrary time-dependent driving and general additive observables. The equality yields sharp response bounds and a variational characterization of the dynamically generated variance. Choosing the perturbation along the irreversible probability flow, we obtain an experimentally accessible friction-response thermodynamic uncertainty relation that is saturable at any finite observation time. It provides both an exact variational principle for entropy production and a weak-noise coherence-dissipation bound for underdamped limit cycles. We further show that the conventional uncertainty factor of velocity-resolved currents can decay exponentially with dissipation, even in driven free diffusion, whereas the friction-response factor retains its universal lower bound. These results establish response, rather than the mean current itself, as the quantity directly linking fluctuations and dissipation in underdamped dynamics.
\end{abstract}

\maketitle

\sectionprl{Introduction}Thermodynamic uncertainty relations (TURs) express a fundamental cost-precision trade-off in nonequilibrium systems: suppressing the relative fluctuations of a sustained current requires dissipation \cite{Barato.2015.PRL,Gingrich.2016.PRL,Horowitz.2020.NP}. Extensions to finite times, arbitrary initial states, and driven systems have greatly broadened their scope \cite{Dechant.2018.JSM,Hasegawa.2019.PRL,Timpanaro.2019.PRL,Liu.2020.PRL,Koyuk.2020.PRL,Dieball.2023.PRL}. Beyond constraining fluctuations, TURs also provide an operational tool for thermodynamic inference because measurable current statistics can be used to bound entropy production \cite{Li.2019.NC,Manikandan.2020.PRL,Vu.2020.PRE,Otsubo.2020.PRE}, which is often difficult to access directly. However, for inference, validity alone is not sufficient; a useful bound must also be tight and experimentally accessible \cite{Seifert.2019.ARCMP}.

A complementary and broader perspective emerges from fluctuation-response relations \cite{Kubo.1991,Agarwal.1972,Marconi.2008.PR,Harada.2005.PRL,Baiesi.2009.PRL,Prost.2009.PRL,Seifert.2010.EPL,Owen.2020.PRX,Dechant.2020.PNAS,Ptaszynski.2024.PRL,Bao.2024.arxiv,Zheng.2025.PRE,Vu.2025.PRXQ,Liu.2025.CP}, which connect fluctuations to the susceptibility of an observable under perturbations of the dynamics. Recent developments have promoted this connection from inequalities to exact nonequilibrium identities. For Markov jump processes, current covariances can be expressed as quadratic forms of local responses, with finite-time extensions to general time-integrated observables \cite{Aslyamov.2025.PRL,Kwon.2025.PRL,Ptaszynski.2026.PRE,Ptaszynski.2026.PRE2,Aslyamov.2026.arxiv}. Related exact results have been obtained for overdamped Langevin dynamics in the stationary long-time limit \cite{Chun.2026.arxiv} and in frequency-resolved settings \cite{Dechant.2026.PRL,Zheng.2026.arxiv,Kwon.2026.arxiv}. These results suggest that TURs are manifestations of a more fundamental fluctuation-response structure: when a perturbation is aligned with the irreversible probability flow, its response becomes related to the mean current, while its quadratic cost becomes the entropy production.

Underdamped dynamics obstruct this structure in two ways. First, thermal noise acts directly only on the velocity degrees of freedom, whereas experimentally relevant transport currents often accumulate through position-space motion. Consequently, fluctuations and currents generally probe different sectors of phase space, and the response of a current need not be determined by its mean. This mismatch is reflected in the violation \cite{Fischer.2020.PRE,Pietzonka.2022.PRL} or modification \cite{Vu.2019.PRE.UnderdampedTUR,Lee.2021.PRE.TUR,Kwon.2022.NJP,Fu.2022.PRE,Dechant.2022.arxiv} of conventional TURs in underdamped systems. Second, physically relevant perturbations of underdamped dynamics can modify not only the drift but also the diffusion, whereas standard likelihood-ratio approaches naturally describe drift perturbations at fixed noise strength. A paradigmatic thermodynamic example is a variation of the friction coefficient at fixed temperature: through the fluctuation-dissipation relation, it simultaneously modifies the dissipative drift and noise amplitude. In the continuous-time limit, altering the diffusion changes the quadratic variation of almost every trajectory, rendering the corresponding path measures mutually singular \cite{Baiesi.2014.EPJB}. These features raise a fundamental question: can fluctuations, responses, and dissipation be organized within a single exact framework for underdamped dynamics?

Here, we answer this question affirmatively. First, using the Doob martingale associated with a general finite-time additive observable, we derive a fluctuation-response identity that is valid for arbitrary initial ensembles and time-dependent driving, without imposing time-reversal parity [Eq.~\eqref{eq:fre}]. The identity decomposes the observable variance into a contribution inherited from the initial ensemble and a dynamically generated contribution expressed as a quadratic form of the local response kernel. Casting the response directly in terms of injected probability currents places drift and diffusion perturbations on an equal footing and yields sharp response bounds together with a variational representation of the dynamically generated variance. The same framework also extends directly to Fourier-filtered observables. Second, choosing the irreversible probability flow as the perturbation leads to a friction-response TUR governed by the friction susceptibility rather than by the observable mean [Eq.~\eqref{eq:fric.TUR}]. Moreover, by explicitly constructing an observable that realizes the equality condition, we show that the bound is saturable at arbitrary finite times, yielding an exact variational principle for the total entropy production. Applying this TUR to the asymptotic phase yields a weak-noise coherence-dissipation bound for underdamped limit cycles, governed by the friction sensitivity of their oscillation frequency [Eq.~\eqref{eq:coherence.bound}]. Third, we demonstrate a striking violation of the conventional TUR in driven free diffusion: for suitable velocity-resolved antisymmetric currents, the uncertainty factor decays \emph{exponentially} with thermodynamic cost, whereas the corresponding friction-response factor remains bounded below by unity. Together, these results establish a finite-time framework that unifies fluctuation-response relations, precision bounds, and thermodynamic inference in underdamped systems.

\sectionprl{Martingale representation and exact variance identity}Consider the $d$-dimensional underdamped Langevin dynamics,
\begin{align}
	\dd{\bx_t}&=\bv_t\dd{t},\notag\\
	\dd{\bv_t}&=[\vf_t(\bx_t)-\gamma\bv_t]\dd{t}+\sqrt{2\D}\dd{\bW_t},
\label{eq:langevin}
\end{align}
where $\vf_t$ is the force acting on the system, $\gamma$ is the friction coefficient, $\D$ is a symmetric positive-definite diffusion matrix, and $\bW_t$ is a $d$-dimensional Wiener process. Here, we use units in which the particle mass and Boltzmann constant are unity. Writing the phase-space state as $\bz=(\bx,\bv)$, let $p_t(\bz)$ denote the probability density. The evolution $\partial_t p_t=\Lg_t(p_t)$ is generated by the Fokker-Planck operator $\Lg_t$, whose backward (adjoint) generator is given by
\begin{equation}
	\Lg_t^\dagger\equiv\bv\cdot\nabla_{\bx}+(\vf_t-\gamma\bv)\cdot\nabla_{\bv}+\nabla_{\bv}\cdot\D\nabla_{\bv}.
\label{eq:generator}
\end{equation}
We consider a general additive phase-space observable accumulated over the interval $[0,\tau]$,
\begin{equation}
	J_\tau=\int_0^\tau[\phi_t(\bz_t)\dd{t}+\bw_t(\bz_t)\circ\dd{\bv_t}],
\label{eq:und.general.observable}
\end{equation}
where $\phi_t$ is an arbitrary scalar function, $\bw_t$ is an arbitrary $d$-dimensional vector field, and $\circ$ denotes the Stratonovich product. This class includes time-symmetric observables such as integrated kinetic energy and occupation times, time-antisymmetric transport currents, and observables of mixed parity. Position-space currents $\int_0^\tau\vb*{\psi}_t(\bz_t)\circ\dd{\bx_t}$ are included through $\phi_t=\vb*{\psi}_t\cdot\bv$. Therefore, the symbol $J_\tau$ does not imply any definite time-reversal parity.

To characterize the fluctuations of $J_\tau$, we introduce the conditional expectation of its remaining contribution after time $t$,
\begin{equation}
	G_t(\bz)\equiv\E\qty[\int_t^\tau[\phi_s(\bz_s)\dd{s}+\bw_s(\bz_s)\circ\dd{\bv_s}]\,\big|\,\bz_t=\bz].
\label{eq:value}
\end{equation}
Thus, $G_t(\bz)$ provides the expected future contribution to $J_\tau$ from a trajectory currently at $\bz$. It obeys the backward equation $\partial_tG_t+\Lg_t^\dagger G_t=-\Phi_t$ with the terminal condition $G_\tau=0$, where $\Phi_t\equiv\phi_t+\bw_t\cdot(\vf_t-\gamma\bv)+\D:\nabla_{\bv}\bw_t$ \cite{fnt1}. Let $\Fh_t$ denote the information generated by the trajectory up to time $t$. The Doob martingale associated with $J_\tau$ is then
\begin{equation}
	M_t\equiv J_t+G_t(\bz_t)=\E[J_\tau\cond\Fh_t].
\label{eq:martingale}
\end{equation}
Martingale methods have played an important role in stochastic thermodynamics \cite{Roldan.2023.AP}, particularly in establishing the universal properties of entropy production fluctuations and stopping-time statistics \cite{Pigolotti.2017.PRL,Neri.2017.PRX}. Here, we exploit the Doob martingale of the observable itself to connect its fluctuations directly to the local response. Applying It\^o's formula to Eq.~\eqref{eq:martingale} and using the backward equation gives
\begin{equation}
	\dd{M_t}=\sqrt{2\D}[\bw_t(\bz_t)+\nabla_{\bv}G_t(\bz_t)]\cdot\dd{\bW_t}.
\label{eq:dM}
\end{equation}
Equation \eqref{eq:dM} is the basic martingale representation underlying our results. 

The martingale representation \eqref{eq:dM} immediately yields an exact expression for the fluctuations of $J_\tau$. Defining the martingale amplitude $\bu_t\equiv\bw_t+\nabla_{\bv}G_t$ and integrating Eq.~\eqref{eq:dM} over $[0,\tau]$, we obtain, by the It\^o isometry,
\begin{equation}
	\Var(J_\tau)=\Var_0(J_\tau) +2\int_0^\tau\lVert\bu_t\rVert_{\D,p_t}^2\dd{t},
\label{eq:exactvar}
\end{equation}
where $\Var_0(J_\tau)\equiv\Var(\E[J_\tau\cond\bz_0])$ and $\lVert\bu\rVert_{\D,p}^2\equiv\int\bu^{\top}\D\bu\, p\dd{\bz}$. Equation \eqref{eq:exactvar} decomposes the total variance into two nonnegative contributions: $\Var_0(J_\tau)$ captures the variability of the conditional mean inherited from the random initial state, whereas the integral is the variance generated dynamically by subsequent noise. An analogous initial-variability term appears in recent finite-time fluctuation-response relations for nonautonomous Markov jump processes \cite{Aslyamov.2026.arxiv}. For a fixed initial state, $\Var_0(J_\tau)=0$; for stationary long-time measurements, it is subextensive in $\tau$. Importantly, Eq.~\eqref{eq:exactvar} is exact at finite times and requires neither stationarity nor Gaussian fluctuations.

A corresponding stationary representation follows naturally in the long-time limit. For time-independent dynamics and observable fields $\phi$ and $\bw$, let $\Psi(\bz)$ solve the Poisson equation $\Lg^\dagger\Psi=-(\Phi-\jr_{\st})$ \cite{Cattiaux.2012}, where $\jr_{\st}\equiv\ev{\Phi}_{\st}$ and $\ev{\cdot}_{\st}$ denotes the steady-state average. Applying It\^o's formula to $\Psi(\bz_t)$ gives the decomposition $J_\tau-\jr_{\st}\tau=\int_0^\tau\sqrt{2\D}(\bw+\nabla_{\bv}\Psi)\cdot\dd{\bW_t}+\Psi(\bz_0)-\Psi(\bz_\tau)$. The boundary term contributes only subextensively to the variance, while the martingale term controls the fluctuations at long times. Hence, we obtain
\begin{equation}
	D_J\equiv\lim_{\tau\to\infty}\frac{\Var(J_\tau)}{2\tau} = \ev{(\bw+\nabla_{\bv}\Psi)^{\top}\D(\bw+\nabla_{\bv}\Psi)}_{\st}.
\label{eq:longvar}
\end{equation}
Equation~\eqref{eq:longvar} expresses the long-time current diffusivity as a local quadratic form of $\bw+\nabla_{\bv}\Psi$ in phase space, providing an alternative to the usual correlation representation. Together, Eqs.~\eqref{eq:exactvar} and \eqref{eq:longvar} provide the finite- and long-time fluctuation structures from which the response relations follow.

\sectionprl{Exact fluctuation-response relations}We now connect the martingale representation to linear response. Writing the Fokker-Planck equation in continuity form, $\partial_t p_t=-\nabla_{\bx}\cdot\bj_t^{\bx}-\nabla_{\bv}\cdot\bj_t^{\bv}$, we consider an infinitesimal source $\delta\bj_{t,\mathrm{src}}^{\bv}$ injected into the velocity-space probability current. Such perturbations may arise from changes in the drift, the diffusion, or both. The induced variation of the probability density obeys the corresponding linearized continuity equation with source $-\nabla_{\bv}\cdot\delta\bj_{t,\mathrm{src}}^{\bv}$. Combining this equation with the backward representation of $G_t$ yields the variation of the observable average \cite{fnt2},
\begin{equation}
 	\delta\E[J_\tau]=\int_0^\tau\int\bu_t(\bz)\cdot\delta\bj_{t,\mathrm{src}}^{\bv}(\bz)\dd{\bz}\dd{t}.
\label{eq:response}
\end{equation}
Thus, the martingale amplitude $\bu_t=\bw_t+\nabla_{\bv}G_t$ is simultaneously the local response kernel, $\bu_t=\delta\E[J_\tau]/\delta\bj_{t,\mathrm{src}}^{\bv}$. Combining this response representation with Eq.~\eqref{eq:exactvar} gives the exact finite-time fluctuation-response equality
\begin{equation}
	\Var(J_\tau)-\Var_0(J_\tau)=2\int_0^\tau\left\lVert\frac{\delta\E[J_\tau]}{\delta\bj_{t,\mathrm{src}}^{\bv}}\right\rVert_{\D,p_t}^2\dd{t}.
\label{eq:fre}
\end{equation}
Equation~\eqref{eq:fre} identifies the dynamically generated variance exactly with the squared local susceptibility measured in the diffusion metric.

This equality immediately yields response bounds for specific perturbation directions in the space of velocity-space probability currents. Consider an injected current of the form $\delta\bj_{t,\mathrm{src}}^{\bv}=\epsilon\bg_t p_t$, where $\epsilon$ is a small scalar parameter, and $\bg_t(\bz)\in\mathbb{R}^d$ is an arbitrary field. For example, such a current is generated by the drift perturbation $\vf_t\to\vf_t+\epsilon\bg_t$. We define the corresponding linear response and quadratic cost by
\begin{equation}
	\Rc[\bg]\equiv\left.\partial_\epsilon\E_\epsilon[J_\tau]\right|_{\epsilon=0},\quad\Ac[\bg]\equiv\int_0^\tau\lVert\bg_t\rVert_{\D^{-1},p_t}^2\dd{t},
\label{eq:RA}
\end{equation}
where $\E_\epsilon$ denotes the expectation under the perturbed dynamics. Using Eq.~\eqref{eq:response}, the linear response becomes $\Rc[\bg]=\int_0^\tau\ev{\bu_t\cdot\bg_t}_{t}\dd{t}$, where $\ev{\cdot}_t$ denotes the ensemble average at time $t$. Applying the Cauchy-Schwarz inequality then yields
\begin{equation}
	\Var(J_\tau)-\Var_0(J_\tau)\ge\frac{2\Rc[\bg]^2}{\Ac[\bg]}.
\label{eq:fri}
\end{equation}
Therefore, each admissible perturbation provides a lower bound on the dynamically generated variance. Moreover, this bound is tight: optimizing over perturbation fields gives the equivalent dual representation
\begin{equation}
	\frac{\Var(J_\tau)-\Var_0(J_\tau)}{2}=\max_{\bg_t}\qty(2\Rc[\bg]-\Ac[\bg]),
\label{eq:dual}
\end{equation}
with the maximum attained at $\bg_t=\D\bu_t$. Equations \eqref{eq:fre}, \eqref{eq:fri}, and \eqref{eq:dual} establish an exact finite-time fluctuation-response structure for general additive observables in underdamped dynamics and constitute our first main result. The framework also generalizes to multiple observables and perturbations, yielding matrix fluctuation-response inequalities and tighter dissipation bounds through observable cross-correlations. It further extends directly to Fourier-filtered observables, yielding finite-time frequency-resolved relations that preserve the temporal structure of the response and reduce, in the stationary long-time limit, to the known finite-frequency bounds (see End Matter).

The same martingale construction applies naturally to other Markov processes, with the fluctuation-response relations carrying over under the corresponding process-dependent metrics. Detailed derivations are provided in the Supplemental Material (SM) \cite{Supp.PhysRev}. These extensions recover recently obtained finite-time covariance relations for nonautonomous Markov jump processes and long-time results for overdamped Langevin dynamics \cite{Aslyamov.2026.arxiv,Chun.2026.arxiv}.

\sectionprl{Friction-response TUR, thermodynamic inference, and coherent oscillation cost}We next extract a thermodynamic consequence of the fluctuation-response inequality \eqref{eq:fri}. Consider dynamics satisfying the fluctuation-dissipation relation $\D=\gamma\T$, where $\T$ is a temperature matrix independent of $\gamma$. The total entropy production is
\begin{equation}
	\Sigma_\tau=\int_0^\tau\lVert\bnu_t^{\mathrm{irr}}\rVert_{\D^{-1},p_t}^2\dd{t},
\label{eq:entropy}
\end{equation}
where $\bnu_t^{\mathrm{irr}}\equiv -\gamma\bv-\D\nabla_{\bv}\ln p_t$ is the local irreversible velocity. Choosing $\bg_t=\bnu_t^{\mathrm{irr}}$ in Eq.~\eqref{eq:fri} makes the perturbation cost equal to the entropy production, $\Ac[\bnu^{\mathrm{irr}}]=\Sigma_\tau$. This perturbation has a direct physical interpretation. Under the rescaling $\gamma\to\gamma(1+\epsilon)$ at fixed $\T$, the diffusion matrix changes simultaneously as $\D\to\D(1+\epsilon)$, preserving the fluctuation-dissipation relation. The resulting source contribution to the velocity-space probability current is $\delta\bj_{t,\mathrm{src}}^{\bv}=\epsilon\bnu_t^{\mathrm{irr}}p_t$. Provided that the initial density $p_0$ and the observable fields $\phi_t$ and $\bw_t$ have no explicit $\gamma$ dependence \cite{fnt3}, the corresponding response is therefore
\begin{equation}
	\Rc[\bnu^{\mathrm{irr}}]=\gamma\partial_\gamma\E[J_\tau].
\label{eq:gammaresponse}
\end{equation}
Substitution into Eq.~\eqref{eq:fri} yields the friction-response TUR, 
\begin{equation}
	\frac{\Var(J_\tau)-\Var_0(J_\tau)}{(\gamma\partial_\gamma\E[J_\tau])^2}\geq\frac{2}{\Sigma_\tau}.
\label{eq:fric.TUR}
\end{equation}
Unlike conventional TURs, Eq.~\eqref{eq:fric.TUR} involves the friction susceptibility rather than the observable mean and holds for any additive observable of the form \eqref{eq:und.general.observable}, irrespective of time-reversal parity and under arbitrary time-dependent driving. This relation also holds for overdamped dynamics and, for time-antisymmetric currents, reduces to a strengthened form of the conventional TUR \cite{Koyuk.2020.PRL}.

Equation \eqref{eq:fric.TUR} is closely related to perturbation-based TURs previously derived from the Cram\'er-Rao inequality \cite{Hasegawa.2019.PRE,Vu.2019.PRE.UnderdampedTUR,Lee.2021.PRE.TUR,Kwon.2022.NJP}. The exact fluctuation-response identity underlying our results, however, reveals additional structure. The earlier bounds involve the full variance \cite{Lee.2021.PRE.TUR,Kwon.2022.NJP}, whereas Eq.~\eqref{eq:fre} separates it into a contribution inherited from the initial state and a dynamically generated contribution, with only the latter entering Eq.~\eqref{eq:fric.TUR}. The variance in our bound is therefore reduced by precisely $\Var_0(J_\tau)$, a sharpening that can be particularly significant at finite times when preparation uncertainty contributes appreciably to the total fluctuations \cite{Supp.PhysRev}. More importantly, the explicit response kernel makes the saturation condition transparent. For any finite observation time, optimization over the generalized observable class gives
\begin{equation}
	\Sigma_\tau=\max_{\phi_t,\bw_t}\frac{2(\gamma\partial_\gamma\E[J_\tau])^2}{\Var(J_\tau)-\Var_0(J_\tau)}.
\label{eq:EP.var}
\end{equation}
Equations \eqref{eq:fric.TUR} and \eqref{eq:EP.var} together constitute our second main result: the friction-response TUR is valid at arbitrary finite times and is saturable within the generalized observable class, thereby providing an exact variational principle for the total entropy production. The maximizing observable can be constructed analytically from the response kernel (see End Matter) and, in practice, approximated using a finite basis of observables \cite{Vu.2020.PRE} or data-driven optimization \cite{Otsubo.2020.PRE}.

The quantities entering this inference scheme are experimentally accessible whenever the friction coefficient can be varied while maintaining the fluctuation-dissipation relation. The friction susceptibility can be obtained by comparing otherwise identical measurements at slightly different friction coefficients, while the variance terms follow from trajectory statistics. Several platforms naturally allow the required perturbation, with friction and thermal noise varying together according to $\D=\gamma\T$. For an optically levitated nanoparticle in a dilute gas, the gas damping is controlled by the background pressure through Epstein drag and can be tuned over a wide range while the chamber temperature is held fixed \cite{Li.2011.NP,Rondin.2017.NN}. Electrical circuits provide an analogous realization. In a series RLC circuit, charge and current correspond to position and velocity, respectively, while $L$, $R$, and $1/C$ play the roles of mass, friction, and stiffness, respectively \cite{Freitas.2020.PRX}. At fixed temperature, changing the passive resistance modifies both the damping and the Johnson-Nyquist noise in accordance with the fluctuation-dissipation relation \cite{Johnson.1928.PR,Nyquist.1928.PR}. Thermal electrical fluctuations of this type have been used directly in stochastic-thermodynamic experiments \cite{Garnier.2005.PRE,Ciliberto.2013.PRL}.

The friction-response TUR also constrains the thermodynamic cost of coherent oscillations \cite{Oberreiter.2022.PRE}. Consider an autonomous underdamped system with a stable deterministic limit cycle, and introduce weak thermal noise by scaling $\T\mapsto\epsilon\T$, where $\epsilon\ll1$. Applying Eq.~\eqref{eq:fric.TUR} to the increment of the reference asymptotic phase yields a weak-noise coherence-dissipation bound \cite{Supp.PhysRev}
\begin{equation}
	\Sigma_{\rm cyc}^{\epsilon}\geq 4\pi^2\mathcal{N}_\epsilon(\gamma\partial_\gamma\ln\omega_\gamma)^2.
\label{eq:coherence.bound}
\end{equation}
Here, $\Sigma_{\rm cyc}^{\epsilon}$ is the entropy produced per oscillation, $\mathcal N_\epsilon$ is the number of oscillations completed before the amplitude of the oscillatory correlation falls to $e^{-1}$ of its initial value, and $\omega_\gamma$ is the deterministic oscillation frequency. A driven rotor exactly saturates this bound at every positive noise strength \cite{Supp.PhysRev}. This result generalizes the existing coherence-dissipation bound for overdamped limit cycles \cite{Santolin.2025.PRL,Nagayama.2025.arxiv,Kolchinsky.2026.PRE}, which is recovered in the overdamped limit, where $\gamma\partial_\gamma\ln\omega_\gamma\to-1$. At finite inertia, the factor $|\gamma\partial_\gamma\ln\omega_\gamma|$ can be arbitrarily large, yielding a correspondingly stronger coherence-dissipation bound \cite{Supp.PhysRev}.

\sectionprl{Exponential violation of the conventional TUR}We now use the exact long-time representation \eqref{eq:longvar} to show that velocity resolution can produce arbitrarily strong violations of the conventional TUR, even in the simplest linear underdamped dynamics. Consider a unit-mass particle diffusing on a ring under a constant force $f_t=f>0$, where the positive direction is chosen without loss of generality. Its steady-state velocity distribution is Gaussian with mean $\vbar\equiv f/\gamma$ and variance $T$. In terms of the dimensionless driving strength $\kappa\equiv\vbar^2/T$, the steady-state entropy production rate is $\sigma_{\st}=\gamma\kappa$. We consider the time-antisymmetric velocity-resolved current
\begin{equation}
    J_\tau=\int_0^\tau \psi(v_t)\circ\dd{x_t},
    \qquad
    \psi(v)=\frac{1}{v}\erf\qty(\frac{\alpha v}{\sqrt{T}}),
\label{eq:weightedcurrent}
\end{equation}
where $\alpha>0$. The apparent singularity at $v=0$ is removable: $\psi(v)$ is smooth, bounded, and even under velocity reversal, so that $J_\tau$ is a well-defined antisymmetric current.

The current statistics can be evaluated analytically \cite{Supp.PhysRev}. With $\lambda\equiv2\alpha^2/(1+2\alpha^2)\in(0,1)$, the steady-state mean is $\jr_{\st}
    =\erf(\sqrt{\lambda\kappa/2})$, and therefore approaches unity exponentially fast as $\kappa$ increases. The diffusivity obtained from Eq.~\eqref{eq:longvar} admits the exact representation
\begin{equation}
	D_J=\frac{2}{\pi\gamma}\int_0^\lambda\frac{\ln(\lambda/r)}{\sqrt{1-r^2}}\exp\qty(-\frac{\lambda\kappa}{1+r})\dd{r}.
\label{eq:varfree}
\end{equation}
As shown in the SM \cite{Supp.PhysRev}, for any fixed $0<\lambda<1$, it has the large-drive asymptotic form $D_J\sim\gamma^{-1}\ell(\lambda)\kappa^{-2}e^{-a\kappa}$ for $\kappa\gg 1$, where $\ell(\lambda)\equiv 2(1+\lambda)^4/(\pi\lambda^3\sqrt{1-\lambda^2})$ and $a\equiv\lambda/(1+\lambda)$. The conventional long-time TUR requires the uncertainty factor $\mathcal{Q}\equiv D_J\sigma_{\st}/\jr_{\st}^2$ to satisfy $\mathcal{Q}\ge1$. Instead,
\begin{equation}
	\mathcal{Q}\sim \ell(\lambda)\kappa^{-1}e^{-a\kappa}\longrightarrow 0
\label{eq:exp.Q}
\end{equation}
as $\kappa\to\infty$. Thus, the conventional uncertainty factor is exponentially, rather than algebraically, suppressed with dissipation. This result lies outside the free-diffusion conjecture of Ref.~\cite{Fischer.2020.PRE}, which considers currents with integrands $\theta(x)v^n$ of a fixed odd velocity order $n$, whereas Eq.~\eqref{eq:weightedcurrent} contains contributions from arbitrarily high odd velocity orders. Exponential enhancements of precision have also been reported in quantum many-body settings \cite{Meier.2025.NP}, where quantum coherence plays a central role \cite{Vu.2025.PRXQ}. Here, by contrast, the underlying classical dynamics is linear and Gaussian; the exponential enhancement arises entirely from the nonlinear velocity-resolved readout. A simpler example exhibiting algebraic suppression is given in the SM \cite{Supp.PhysRev}.

\begin{figure}[t]
\centering
\includegraphics[width=1.0\linewidth]{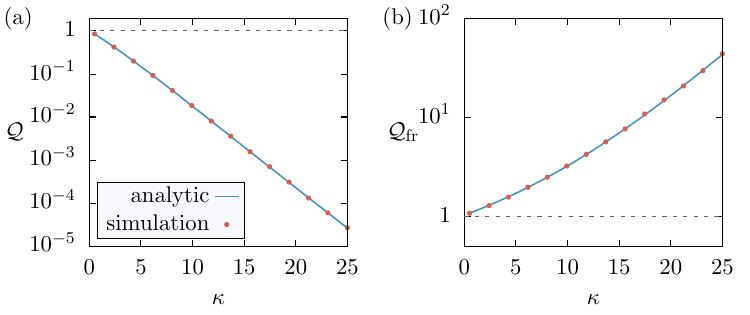}
\protect\caption{Comparison of the conventional and friction-response TURs. (a) Conventional and (b) friction-response uncertainty factors as functions of the driving strength $\kappa$. Circles denote Langevin simulations at $\tau=400$, while solid lines show the analytical $\tau\to\infty$ results. Parameters are $\gamma=1$, $T=1$, $\alpha=1$, and $f=\sqrt{\kappa}$.}
\label{fig:1D}
\end{figure}

The friction-response TUR, by contrast, remains fully valid. At fixed $f$, $T$, and $\alpha$, the current response reads $\gamma\partial_\gamma \jr_{\st}=-\sqrt{2\lambda\kappa/\pi}\exp(-\lambda\kappa/2)$, and hence
\begin{equation}
	\mathcal{Q}_{\mathrm{fr}}\equiv\frac{D_J\sigma_{\st}}{(\gamma\partial_\gamma \jr_{\st})^2}=1+\sum_{n=1}^{\infty}\frac{\lambda^n}{2^n n!}\frac{H_n(\eta)^2}{(n+1)^2}\ge 1,
\label{eq:exp.Qfr}
\end{equation}
where $\eta=\sqrt{\lambda\kappa/2}$ and $H_n$ denotes the physicists' Hermite polynomial of degree $n$. Figure~\ref{fig:1D} confirms both analytical predictions. The contrast has a simple physical interpretation. At strong driving, $\jr_{\st}\simeq 1$: the mean current remains finite while its fluctuations are exponentially suppressed. Yet the same current also becomes exponentially insensitive to friction. The mean current therefore no longer reflects the fluctuation scale, whereas the friction susceptibility does. This example demonstrates explicitly why response, rather than the mean current, provides the appropriate thermodynamic quantity for constraining precision in underdamped dynamics.

\sectionprl{Conclusion}We have established an exact finite-time fluctuation-response framework for underdamped Langevin dynamics that separates initial-state fluctuations from dynamically generated ones and yields sharp response bounds for general additive observables. Choosing the irreversible probability flow as the perturbation gives a friction-response TUR governed by the friction susceptibility rather than the observable mean; the bound is saturable at arbitrary finite times, yielding an exact variational principle for the total entropy production and a coherence-dissipation bound for underdamped limit cycles. For driven free diffusion, we have further shown that a nonlinear velocity-resolved current can violate the conventional TUR arbitrarily strongly, with the uncertainty factor decaying exponentially with dissipation even in linear Gaussian dynamics.

\begin{acknowledgments}
\sectionprl{Acknowledgments}We thank Hyun-Myung Chun for a fruitful discussion. We also acknowledge the use of ChatGPT Sol 5.6 and Claude Optus 5.5 for assistance with manuscript preparation and analytical checks. TVV was supported by JSPS KAKENHI Grants No.~JP23K13032, No.~JP26K00019, No.~JP26K00022, and No.~JP26H02015. VTV was supported by QACI, FPT Corporation. RB was supported by JSPS KAKENHI Grant No.~JP25KJ0766. KS was supported by JSPS KAKENHI Grants No.~JP26H02015 and No.~JP26H00388.
\end{acknowledgments}

\onecolumngrid
\begin{center}
    \textbf{End Matter}
\end{center}
\twocolumngrid

\sectionprl{Appendix A: Multivariate generalization of fluctuation-response relations}The fluctuation-response structure extends naturally to multiple observables and perturbations. Consider $m$ additive observables $\{J_\tau^\alpha\}_{\alpha=1}^m$, with martingale amplitudes $\bu_t^\alpha=\bw_t^\alpha+\nabla_{\bv}G_t^\alpha$. Since the corresponding martingales are driven by the same Wiener process, the polarized It\^o isometry gives the exact covariance identity
\begin{equation}
	\Cm^{\alpha\beta}-\Cm_0^{\alpha\beta}=2\int_0^\tau\inp{\bu_t^\alpha}{\bu_t^\beta}_{\D,p_t}\dd{t},
\label{eq:C}
\end{equation}
where
\begin{align}
	\Cm^{\alpha\beta}&\equiv\Cov(J_\tau^\alpha,J_\tau^\beta),\notag\\
	\Cm_0^{\alpha\beta}&\equiv\Cov(\E[J_\tau^\alpha\cond\bz_0],\E[J_\tau^\beta\cond\bz_0]),
\end{align}
and $\inp{\vb*{a}}{\vb*{b}}_{\D,p}\equiv\int\vb*{a}^{\top}\D\vb*{b} p\dd{\bz}$. Thus, Eq.~\eqref{eq:exactvar} is the diagonal part of an exact covariance relation.

We next consider $n$ independently controlled perturbation directions in the space of injected velocity-space currents, $\delta\bj_{t, \mathrm{src}}^{\bv}=\sum_{i=1}^n\epsilon_i\bg_t^ip_t$, where $\{\epsilon_i\}_{i=1}^n$ are small independent perturbation parameters. Define the response matrix and the quadratic cost matrix by
\begin{align}
	\Rm_{\alpha i}&\equiv\partial_{\epsilon_i}\E_{\vb*{\epsilon}}[J_\tau^\alpha]|_{\vb*{\epsilon}=0},\notag\\
	\Am_{ij}&\equiv\int_0^\tau\inp{\bg_t^i}{\bg_t^j}_{\D^{-1},p_t}\dd{t}.
\label{eq:matrixRA}
\end{align}
Writing $\bh_t^i\equiv\D^{-1}\bg_t^i$, the response relation gives
\begin{equation}
	\Rm_{\alpha i}=\int_0^\tau\inp{\bu_t^\alpha}{\bh_t^i}_{\D,p_t}\dd{t}.
\end{equation}
Hence, under the time-integrated inner product $\int_0^\tau\inp{\cdot}{\cdot}_{\D,p_t}\dd{t}$, the fields $\{\bu^\alpha\}$ and $\{\bh^i\}$ have the positive semidefinite Gram matrix
\begin{equation}
	\begin{pmatrix}
		[\Cm-\Cm_0]/2 & \Rm\\
		\Rm^{\top} & \Am
	\end{pmatrix}
	\succeq 0.
\label{eq:matrixGram}
\end{equation}
Taking its Schur complement yields the matrix fluctuation-response inequality
\begin{equation}
	\Cm-\Cm_0 \succeq 2\Rm\Am^{-1}\Rm^{\top},
\label{eq:mat.fri}
\end{equation}
for $\Am\succ0$. Similarly, when $\Cm-\Cm_0\succ0$, we have
\begin{equation}
	\Am\succeq2\Rm^{\top}(\Cm-\Cm_0)^{-1}\Rm.
\label{eq:mat.fri.dual}
\end{equation}
These relations generalize Eq.~\eqref{eq:fri} to simultaneous measurements of multiple observables and responses.

As an important special case, choose a single perturbation $\bg_t=\bnu_t^{\mathrm{irr}}$ (i.e., $n=1$). Then $\Am=\Sigma_\tau$ and the response vector has components $r_\alpha\equiv\Rm_{\alpha 1}=\gamma\partial_\gamma\E[J_\tau^\alpha]$. Equation~\eqref{eq:mat.fri.dual} gives the multivariate friction-response TUR
\begin{equation}
	\Sigma_\tau\ge 2\vb*{r}^{\top}(\Cm-\Cm_0)^{-1}\vb*{r},
\label{eq:multiTUR}
\end{equation}
where $\vb*{r}=(r_1,\ldots,r_m)^{\top}$. Thus, correlations among measured observables can be exploited systematically to tighten dissipation inference \cite{Dechant.2019.JPA}. In fact, Eq.~\eqref{eq:multiTUR} is the optimal friction-response bound obtainable from all linear combinations of the chosen observables \cite{Vu.2020.PRE}.

\sectionprl{Appendix B: Frequency-resolved fluctuation-response relations}The finite-time framework admits a direct frequency-domain representation without any additional stochastic construction: a Fourier-filtered observable remains within the class \eqref{eq:und.general.observable}, with explicitly time-dependent weights. Detailed derivations are given in the SM \cite{Supp.PhysRev}. For a real frequency $\omega$, define
\begin{equation}
    J_\tau[\omega]\equiv\int_0^\tau e^{-i\omega t}\dd{J_t},
\label{eq:em.freq.obs}
\end{equation}
which is obtained from Eq.~\eqref{eq:und.general.observable} by the replacements $\phi_t\to e^{-i\omega t}\phi_t$ and $\bw_t\to e^{-i\omega t}\bw_t$. For a complex random variable $X$, we use $\Var(X)\equiv\E[|X-\E[X]|^2]$. The martingale construction then gives the exact finite-time identity
\begin{equation}
    \Var(J_\tau[\omega])-\Var_0(J_\tau[\omega])=2\int_0^\tau\lVert\bu_t[\omega]\rVert_{\D,p_t}^2\dd{t},
\label{eq:em.freq.identity}
\end{equation}
where $\bu_t[\omega]\equiv e^{-i\omega t}\bw_t+\nabla_{\bv}G_t[\omega]$, and $G_t[\omega]$ solves $\partial_tG_t[\omega]+\Lg_t^\dagger G_t[\omega]=-e^{-i\omega t}\Phi_t$ with $G_\tau[\omega]=0$. For complex fields, the norm is understood in the Hermitian sense, $\lVert\bu\rVert_{\D,p}^2\equiv\int\bu^\dagger\D\bu\,p\dd{\bz}$. Equation~\eqref{eq:em.freq.identity} is valid for arbitrary initial ensembles and arbitrary time-dependent driving.

For an injected current $\delta\bj_{t,\mathrm{src}}^{\bv}=\epsilon\bg_t p_t$, the response of $J_\tau[\omega]$ is $\Rc_\omega[\bg]=\int_0^\tau r_t^\omega[\bg]\dd{t}$, where $r_t^\omega[\bg]\equiv\ev{\bu_t[\omega]\cdot\bg_t}_t$. Since $\Phi_t$ and $\bw_t$ are real, $\bu_t[\omega]^*=\bu_t[-\omega]$. Applying the Cauchy-Schwarz inequality first locally in time and then over the full observation interval yields the following hierarchy of fluctuation-response inequalities:
\begin{align}
    \Var(J_\tau[\omega])-\Var_0(J_\tau[\omega])&\ge2\int_0^\tau\frac{\abs{r_t^\omega[\bg]}^2}{a_t[\bg]}\dd{t}\notag\\
    &\ge\frac{2\abs{\Rc_\omega[\bg]}^2}{\Ac[\bg]},
\label{eq:em.freq.hierarchy}
\end{align}
where $a_t[\bg]\equiv\ev{\bg_t^\top\D^{-1}\bg_t}_t$ and $\Ac[\bg]=\int_0^\tau a_t[\bg]\dd{t}$. The first inequality is generally more informative: temporal integration can strongly suppress an oscillatory response through phase cancellation, whereas the time-resolved form retains its local amplitude.

We next specialize to autonomous dynamics in the stationary state, with time-independent observable fields $\phi$ and $\bw$. The finite-window spectrum $S_\tau(\omega)\equiv\Var(J_\tau[\omega])/\tau$ converges under standard mixing conditions to the stationary power spectrum
\begin{equation}
    S_{\dot J}(\omega)\equiv\int_{-\infty}^{\infty}e^{-i\omega t}C_{\dot J}(t)\dd{t},
\label{eq:em.freq.fejer}
\end{equation}
where $C_{\dot J}(t)\equiv\Cov(\dot J_t,\dot J_0)$. Let $\Psi_\omega$ be the centered solution of $(\Lg^\dagger-i\omega)\Psi_\omega=-(\Phi-\jr_{\st})$, with $\ev{\Psi_\omega}_{\st}=0$, and $\bu_\omega\equiv\bw+\nabla_{\bv}\Psi_\omega$ be the stationary kernel. Applying It\^o's formula to $e^{-i\omega t}\Psi_\omega(\bz_t)$ decomposes the centered observable $J_\tau[\omega]-\E[J_\tau[\omega]]$ into a martingale whose variance grows as $2\tau\lVert\bu_\omega\rVert_{\D,p_{\st}}^2$ and subextensive boundary terms. Hence,
\begin{equation}
    S_{\dot J}(\omega)=2\lVert\bu_\omega\rVert_{\D,p_{\st}}^2=2\ev{\bu_\omega^\dagger\D\bu_\omega}_{\st}.
\label{eq:em.freq.spectrum}
\end{equation}
At $\omega=0$, $\Psi_0$ reduces to the stationary Poisson solution and $S_{\dot J}(0)=2D_J$, recovering Eq.~\eqref{eq:longvar}. Moreover, since $\Psi_\omega=O(|\omega|^{-1})$ for large $|\omega|$ under regularity conditions, the direct velocity-current term leaves a nonzero high-frequency plateau, $S_{\dot J}(\omega)\to2\ev{\bw^\top\D\bw}_{\st}$, which is absent for pure rate observables with $\bw=0$.

The stationary response is governed by the same kernel. For a time-independent perturbation direction $\bg$, define
\begin{equation}
    \chi_{\bg}(\omega)\equiv\ev{\bu_\omega\cdot\bg}_{\st},
    \qquad
    a_{\st}[\bg]\equiv\ev{\bg^\top\D^{-1}\bg}_{\st} .
\end{equation}
Here, $\chi_{\bg}(\omega)$ is the Fourier transform of the corresponding causal response kernel. Away from the terminal boundary, $r_t^\omega[\bg]\to e^{-i\omega t}\chi_{\bg}(\omega)$ as $\tau-t\to\infty$, while $a_t[\bg]=a_{\st}[\bg]$. Consequently, the first inequality in Eq.~\eqref{eq:em.freq.hierarchy} yields in the long-time limit
\begin{equation}
    2\abs{\chi_{\bg}(\omega)}^2\le S_{\dot J}(\omega)a_{\st}[\bg],
\label{eq:em.freq.FRI}
\end{equation}
recovering the stationary finite-frequency response inequality of Ref.~\cite{Dechant.2026.PRL}. The same result follows directly by applying the Cauchy-Schwarz inequality to the exact spectral identity \eqref{eq:em.freq.spectrum}. Thus, the known stationary bound emerges as the long-time limit of the finite-time relation \eqref{eq:em.freq.hierarchy}. Finally, choosing the perturbation along the irreversible velocity flow, $\bg=\bnu^{\rm irr}$, gives $a_{\st}[\bg]=\sigma_{\st}$ and hence
\begin{equation}
    \sigma_{\st}\ge\frac{2\abs{\chi_{\bnu^{\rm irr}}(\omega)}^2}{S_{\dot J}(\omega)}.
\label{eq:end.freq.entropic}
\end{equation}
Here, $\sigma_{\st}$ denotes the entropy production rate.
At $\omega=0$, $\chi_{\bnu^{\rm irr}}(0)=\gamma\partial_\gamma\jr_{\st}$, so Eq.~\eqref{eq:end.freq.entropic} reduces to the stationary long-time friction-response TUR \eqref{eq:fric.TUR}.

\sectionprl{Appendix C: Finite-time saturation and exact entropy-production inference}We now establish that the friction-response TUR can be saturated at an arbitrary finite observation time. Equality in the Cauchy-Schwarz step leading to Eq.~\eqref{eq:fric.TUR} requires the response kernel to be proportional to the diffusion-weighted field,
\begin{equation}
    \bu_t=c\,\D^{-1}\bnu_t^{\mathrm{irr}},
\label{eq:und.finite.saturation.condition}
\end{equation}
where $c$ is a constant. For sufficiently regular observables, a nonzero velocity-current weight is required for exact finite-time saturation when the terminal irreversible velocity field is nonzero. Indeed, $G_\tau=0$ implies $\bu_\tau=\bw_\tau$. For a pure rate observable, $\bu_\tau=0$, which is incompatible with the nontrivial saturation condition when $\bnu_\tau^{\mathrm{irr}}\neq 0$, provided that the relevant fields are continuous up to the terminal time. The generalized observable class of Eq.~\eqref{eq:und.general.observable} allows saturation in this case as well. Since an overall rescaling of the observable leaves the corresponding uncertainty factor unchanged, we set $c=1$ without loss of generality. We next construct explicitly an observable satisfying this condition for any finite $\tau$.

Assuming that $p_t(\bz)$ is smooth and strictly positive, define
\begin{equation}
    \Lambda_t(\bz)\equiv-\ln\qty[p_t(\bz)\exp\qty(\frac{1}{2}\bv^{\top}\T^{-1}\bv)].
\label{eq:und.psi.t}
\end{equation}
Using $\D=\gamma\T$, its velocity gradient satisfies $\D\nabla_{\bv}\Lambda_t=-\gamma\bv-\D\nabla_{\bv}\ln p_t=\bnu_t^{\mathrm{irr}}$. Thus, $\nabla_{\bv}\Lambda_t$ is precisely the field required by the saturation condition. We now show that this field can be realized by an observable of the form \eqref{eq:und.general.observable}. Moreover, the velocity-current weight need not have any explicit time dependence. Choose
\begin{equation}
    \bw^{\mathrm{opt}}(\bz)\equiv\nabla_{\bv}\Lambda_\tau(\bz),~ G_t^{\mathrm{opt}}(\bz)\equiv\Lambda_t(\bz)-\Lambda_\tau(\bz).
\label{eq:und.chi.G.opt}
\end{equation}
The terminal condition $G_\tau^{\mathrm{opt}}=0$ is automatically satisfied, while the corresponding response kernel becomes
\begin{equation}
    \bu_t^{\mathrm{opt}}=\bw^{\mathrm{opt}}+\nabla_{\bv}G_t^{\mathrm{opt}}=\nabla_{\bv}\Lambda_t=\D^{-1}\bnu_t^{\mathrm{irr}}.
\label{eq:und.u.opt.finite}
\end{equation}
Hence, the finite-time saturation condition \eqref{eq:und.finite.saturation.condition} is satisfied throughout the entire observation interval. It remains to construct the state-dependent part of the observable so that $G_t^{\mathrm{opt}}$ is indeed its conditional future value. Define the effective It\^o rate by $\Phi_t^{\mathrm{opt}}\equiv-\partial_tG_t^{\mathrm{opt}}-\Lg_t^\dagger G_t^{\mathrm{opt}}$. Since the effective rate associated with Eq.~\eqref{eq:und.general.observable} is $\Phi_t=\phi_t+\bw_t\cdot(\vf_t-\gamma\bv)+\D:\nabla_{\bv}\bw_t$, the required state-dependent rate is
\begin{equation}
    \phi_t^{\mathrm{opt}}\equiv\Phi_t^{\mathrm{opt}}-\bw^{\mathrm{opt}}\cdot(\vf_t-\gamma\bv)-\D:\nabla_{\bv}\bw^{\mathrm{opt}}.
\label{eq:und.phi.opt.finite}
\end{equation}
Equations~\eqref{eq:und.chi.G.opt} and \eqref{eq:und.phi.opt.finite} therefore define an explicit observable whose response kernel satisfies $\bu_t^{\mathrm{opt}}=\D^{-1}\bnu_t^{\mathrm{irr}}$. All coefficients entering this observable are constructed from the reference process and are held fixed when the friction derivative is evaluated.

For this choice, the variance and response terms can be evaluated as
\begin{align}
    \Var(J_\tau^{\mathrm{opt}})-\Var_0(J_\tau^{\mathrm{opt}})&=2\Sigma_\tau,\\
    \gamma\partial_\gamma\E[J_\tau^{\mathrm{opt}}]&=\Sigma_\tau.
\end{align}
The friction-response TUR is therefore saturated exactly for arbitrary finite $\tau$. The construction above also shows that the maximum remains attainable when the velocity-current weight is restricted to a time-independent field $\bw(\bz)$, provided that the state-dependent rate $\phi_t$ is allowed to depend explicitly on time.


\begin{thebibliography}{67}
\bibitem [{fnt({\natexlab{a}})}]{fnt1}%
  \BibitemOpen
  \href@noop {} {}Here, we use $\mathsf{X}:\mathsf{Y}\equiv\tr(\mathsf{X}^{\top}\mathsf{Y})$ for any matrices $\mathsf{X}$ and $\mathsf{Y}$.\BibitemShut {Stop}%
\bibitem [{fnt({\natexlab{b}})}]{fnt2}%
  \BibitemOpen
  \href@noop {} {}Both the initial density $p_0$ and the observable fields $\phi_t$ and $\bw_t$ are fixed under the perturbation.\BibitemShut {Stop}%
\bibitem [{fnt({\natexlab{c}})}]{fnt3}%
  \BibitemOpen
  \href@noop {} {}If $p_0$ depends explicitly on $\gamma$, an additional initial-preparation term appears in the response. For time-independent dynamics and observable fields, under suitable mixing and integrability conditions, this term is $O(1)$ and does not contribute to the response per unit time in the long-time limit.\BibitemShut {Stop}%
\end{thebibliography}
\end{document}


\title{Supplemental Material for\\ ``Exact Fluctuation-Response Relations for Underdamped Langevin Dynamics''}

\author{Tan Van Vu}
\email{tan.vu@yukawa.kyoto-u.ac.jp}
\affiliation{Center for Gravitational Physics and Quantum Information, Yukawa Institute for Theoretical Physics, Kyoto University, Kitashirakawa Oiwakecho, Sakyo-ku, Kyoto 606-8502, Japan}

\author{Van Tuan Vo}
\affiliation{Quantum AI \& Cyber Security Research Institute, FPT Corporation, 10 Pham Van Bach, Cau Giay, Ha Noi 100000, Vietnam}

\author{Ruicheng Bao}
\affiliation{Department of Physics, Graduate School of Science, The University of Tokyo, Hongo, Bunkyo-ku, Tokyo 113-0033, Japan}

\author{Keiji Saito}
\affiliation{Department of Physics, Kyoto University, Kyoto 606-8502, Japan}

\begin{abstract}
This Supplemental Material provides detailed derivations of the martingale representations, exact variance-response identities, and friction-response thermodynamic uncertainty relation (TUR) for underdamped dynamics. We examine initial-state fluctuations, short-time behavior, finite-time saturation, and entropy-production inference. We develop multivariate and frequency-resolved extensions and derive a weak-noise coherence-dissipation bound with an exactly saturating rotor example. Driven-diffusion models demonstrate arbitrarily strong violations of the conventional TUR while respecting the friction-response bound. We also present corresponding constructions for overdamped Langevin dynamics and Markov jump processes, including improved finite-time TURs and their saturation conditions. Equation and figure numbers are prefixed with S [e.g., Eq.~(S1) or Fig.~S1]. The numbers without this prefix [e.g., Eq.~(1) or Fig.~1] refer to the items in the main text.
\end{abstract}

\pacs{}
\maketitle

\tableofcontents

\vspace{1cm}

For convenience, we use the following notation throughout this Supplemental Material:
\begin{align}
	\E[J(\Gamma)]&=\int J(\Gamma) P(\Gamma)\dd{\Gamma},\\
	\ev{a}_t &= \int a(\bz)p_t(\bz)\dd{\bz},\\
	\inprod{\vb*{a}}{\vb*{b}} &\equiv \int \vb*{a}(\bz)^{\top} \vb*{b}(\bz)\dd{\bz}.
\end{align}
Here, $\E[J(\Gamma)]$ denotes the ensemble average of an arbitrary trajectory-dependent quantity $J$, $\ev{a}_t$ denotes the ensemble average of a state-dependent function $a$ at time $t$, and $\inprod{\vb*{a}}{\vb*{b}}$ denotes the standard inner product.

The forward generator $\Lg$ and the corresponding backward (adjoint) generator $\Lg^\dagger$ satisfy
\begin{equation}
	\inprod{\Lg^\dagger a}{b} = \inprod{a}{\Lg b}.
\end{equation}

\section{Underdamped Langevin dynamics}
\label{sec:exactmart}
Consider a $d$-dimensional underdamped system governed by the Langevin equation
\begin{equation}
	\dd{\bx_t}=\bv_t\dd{t},~\dd{\bv_t}=[-\gamma\bv_t+\bF_t(\bx_t)]\dd{t}+\sqrt{2\D}\dd{\bW_t}.
\end{equation}
Here, $\gamma$ is the friction coefficient, $\bF_t$ is the force exerted on the system, $\D$ is a symmetric positive-definite diffusion matrix, and $\bW_t$ is a $d$-dimensional Wiener process.
Let $p_t(\bx,\bv)$ denote the probability density of the system being in the state $\bz=(\bx,\bv)$ at time $t$.
The time evolution of $p_t$ is governed by the Fokker-Planck equation
\begin{align}
	\dot p_t=\Lg_t(p_t)&\equiv -\bv\cdot\nabla_{\bx}p_t - \nabla_{\bv}\cdot[(\bF_t-\gamma\bv)p_t - \D\nabla_{\bv}p_t]\\
	&=-\nabla_{\bx}\cdot \bj_t^{\bx} - \nabla_{\bv}\cdot \bj_t^{\bv},
\end{align}
where $\bj_{t}^{\bx}=\bv p_t$ and $\bj_{t}^{\bv}=(\bF_t-\gamma\bv)p_t - \D\nabla_{\bv}p_t$ are the probability currents in position and velocity space, respectively.
The backward generator corresponding to the forward generator $\Lg_t$ is
\begin{equation}
	\Lg_t^\dagger =\bv\cdot\nabla_{\bx} +(\bF_t-\gamma\bv)\cdot\nabla_{\bv} + \nabla_{\bv}\cdot(\D\nabla_{\bv}).
\end{equation}
In stochastic thermodynamics, the total entropy production is given by
\begin{equation}
	\Sigma_\tau=\int_0^\tau \ev{(\vb*{\nu}_t^{\mathrm{irr}})^{\top}\D^{-1}\vb*{\nu}_t^{\mathrm{irr}}}_t\dd{t},
\end{equation}
where $\vb*{\nu}_t^{\mathrm{irr}}\equiv \bj_t^{\mathrm{irr}}/p_t$ and $\bj_t^{\mathrm{irr}}\equiv -\gamma\bv p_t-\D\nabla_{\bv}p_t$.

We consider the time-integrated observable 
\begin{equation}
	J_\tau=\int_0^\tau[\phi_t(\bz_t)\dd{t}+\bw_t(\bz_t)\circ\dd{\bv_t}],
\label{eq:und.general.observable}
\end{equation}
where $\circ$ denotes the Stratonovich product, $\phi_t$ is an arbitrary scalar function, and $\bw_t$ is an arbitrary $d$-dimensional vector field.

Converting the Stratonovich integral in Eq.~\eqref{eq:und.general.observable} to It\^o form gives
\begin{equation}
    \dd{J_t}=\Phi_t(\bz_t)\dd{t}+\sqrt{2\D}\bw_t(\bz_t)\cdot\dd{\bW_t},
\label{eq:und.J.ito}
\end{equation}
where the effective It\^o rate is
\begin{equation}
    \Phi_t\equiv\phi_t+\bw_t\cdot(\bF_t-\gamma\bv)+\D:\nabla_{\bv}\bw_t.
\label{eq:und.effective.rate}
\end{equation}
Here, $\cdot$ denotes the It\^o product and $\msf{X}:\msf{Y}\equiv\tr(\msf{X}^{\top}\msf{Y})$ for any matrices $\msf{X}$ and $\msf{Y}$. Equation~\eqref{eq:und.effective.rate} follows from
\begin{equation}
    \bw_t\circ\dd{\bv_t}=\bw_t\cdot\dd{\bv_t}+\D:\nabla_{\bv}\bw_t\dd{t}.
\end{equation}

\subsection{Doob martingale of the observable}
Let $\mca{F}_t$ denote the filtration generated by the history $\bz_{[0,t]}$ of the process up to time $t$.
We define the stochastic process
\begin{align}
	M_t\equiv\E\qty[J_\tau\cond \mca{F}_t]&=J_t+G_t(\bz_t),\\
	G_t(\bz)&\equiv\E\Big[\int_t^\tau(\phi_s(\bz_s)\dd{s}+\bw_s(\bz_s)\circ\dd{\bv_s})\,\Big|\, \bz_t=\bz\Big].\label{eq:Y}
\end{align}
Thus, $G_t(\bz_t)$ is the expected remaining observable conditioned on the state at time $t$.
By construction, $\{M_t\}_{0\le t\le \tau}$ is a martingale with endpoints $M_0=\E[J_\tau\cond \bz_0]$ and $M_\tau=J_\tau$.
Indeed, for $0\le s\le t\le\tau$, the martingale property follows from
\begin{align}
	\E[M_t\cond\mca{F}_s] &= \E\qty[J_t + G_t(\bz_t)\cond\mca{F}_s]\notag\\
	&=J_s + \E\Big[\int_s^\tau(\phi_{s'}(\bz_{s'})\dd{s'}+\bw_{s'}(\bz_{s'})\circ\dd{\bv_{s'}})\,\Big|\, \bz_s\Big]\notag\\
	&=J_s + G_s(\bz_s)\notag\\
	&=M_s.
\end{align}
Furthermore, $G_t$ satisfies the terminal-value backward Kolmogorov equation
\begin{align}
	\partial_t G_t+\Lg_t^\dagger G_t=-\Phi_t, \label{eq:backward}
\end{align}
with terminal condition $G_\tau(\bz)=0$.
A detailed derivation follows.

Define $\Exv[\cdot]\equiv\E[\cdot\cond\bz_t=\bz]$. For a small increment $h>0$, we split the remaining observable at time $t+h$ to obtain
\begin{equation}
	\Exv\Big[\int_t^\tau(\phi_s(\bz_s)\dd{s}+\bw_s(\bz_s)\circ\dd{\bv_s})\Big]=\Exv\Big[\int_t^{t+h}(\phi_s(\bz_s)\dd{s}+\bw_s(\bz_s)\circ\dd{\bv_s})\Big]+\Exv\Big[\int_{t+h}^\tau(\phi_s(\bz_s)\dd{s}+\bw_s(\bz_s)\circ\dd{\bv_s})\Big].
\end{equation}
Using the tower property of conditional expectation together with the Markov property, we rewrite the second term on the right-hand side as
\begin{equation}
	\Exv\qty[\int_{t+h}^\tau(\phi_s(\bz_s)\dd{s}+\bw_s(\bz_s)\circ\dd{\bv_s})]=\Exv\qty[\E\Big[\int_{t+h}^\tau(\phi_s(\bz_s)\dd{s}+\bw_s(\bz_s)\circ\dd{\bv_s})\,\Big|\,\bz_{t+h}\Big]].
\end{equation}
By definition, the inner conditional expectation is
\begin{equation}
	\E\qty[\int_{t+h}^\tau(\phi_s(\bz_s)\dd{s}+\bw_s(\bz_s)\circ\dd{\bv_s})\,\Big|\,\bz_{t+h}]=G_{t+h}(\bz_{t+h}).
\end{equation}
Combining these expressions yields the dynamic programming relation
\begin{equation}
	G_t(\bz)=\underbrace{\Exv\qty[\int_t^{t+h}(\phi_s(\bz_s)\dd{s}+\bw_s(\bz_s)\circ\dd{\bv_s})]}_{\textstyle (\mathrm I)} + \underbrace{\Exv\qty[G_{t+h}(\bz_{t+h})]}_{\textstyle (\mathrm{II})}.
\label{eq:recursion}
\end{equation}
Assuming sufficient regularity, the first term has the short-time expansion
\begin{equation}
	(\mathrm I)=\Exv\qty[\int_t^{t+h}\qty(\Phi_s(\bz_s)\dd{s}+\sqrt{2\D}\bw_s(\bz_s)\cdot\dd{\bW_s})]=\Phi_t(\bz)h+O(h^2).\label{eq:I}
\end{equation}
The short-time expansion generated by the forward operator $\Lg_t$ gives
\begin{align}
	(\mathrm{II})&=\int G_{t+h}(\bz')\qty[1+h\Lg_t]\delta(\bz'-\bz)\dd{\bz'}+O(h^2)\notag\\
	&=\int \qty[1+h\Lg_t^\dagger]G_{t+h}(\bz')\delta(\bz'-\bz)\dd{\bz'}+O(h^2)\notag\\
	&=G_t(\bz) + h\partial_t G_t(\bz) + h\Lg_t^\dagger G_t(\bz) + O(h^2).
\end{align}
Substituting these expansions into Eq.~\eqref{eq:recursion} yields
\begin{equation}
	G_t=\Phi_t h+G_t+h\qty[\partial_t G_t+\Lg_t^\dagger G_t]+O(h^2).
\end{equation}
Dividing by $h$ and taking the limit $h\to0$ gives
\begin{equation}
	0=\Phi_t+\partial_t G_t+\Lg_t^\dagger G_t\quad\Rightarrow\quad\partial_t G_t+\Lg_t^\dagger G_t=-\Phi_t,
\end{equation}
together with the terminal condition $G_\tau(\bz)=0$.

\subsection{Quadratic expression for the observable variance}
Applying It\^o's formula to $M_t=J_t+G_t(\bz_t)$, together with $\dd{J_t}=\Phi_t(\bz_t)\dd{t}+\sqrt{2\D}\bw_t(\bz_t)\cdot\dd{\bW_t}$ and $\dd{\bv_t}=[\bF_t(\bx_t)-\gamma\bv_t]\dd{t}+\sqrt{2\D}\dd{\bW_t}$, gives
\begin{align}
	\dd{M_t}&=\qty[\Phi_t+\partial_t G_t+\Lg_t^\dagger G_t]\dd{t} + \sqrt{2\D}(\bw_t+\nabla_{\bv}G_t)\cdot\dd{\bW_t}\notag\\
	&=\sqrt{2\D}[\bw_t(\bz_t)+\nabla_{\bv}G_t(\bz_t)]\cdot\dd{\bW_t},\label{eq:dM}
\end{align}
where the second equality follows from Eq.~\eqref{eq:backward}.
We now provide a detailed derivation of Eq.~\eqref{eq:dM}.

Applying the multivariate It\^o formula to $G_t(\bz_t)$, where both $\bx_t$ and $\bv_t$ are It\^o processes, gives
\begin{equation}
	\dd{G_t}=\partial_t G_t\dd{t} + \dd{\bx_t}\cdot \nabla_{\bx}G_t + \dd{\bv_t}\cdot \nabla_{\bv}G_t + \frac{1}{2}\dd{\bx_t}^{\top} \nabla_{\bx}\nabla_{\bx}G_t \dd{\bx_t} + \dd{\bx_t}^{\top} \nabla_{\bx}\nabla_{\bv}G_t \dd{\bv_t} + \frac{1}{2}\dd{\bv_t}^{\top} \nabla_{\bv}\nabla_{\bv}G_t \dd{\bv_t}.\label{eq:fullito}
\end{equation}
The relevant It\^o multiplication rules are
\begin{equation}
(\dd{t})^2=0,\qquad (\dd{t})(\dd{\bW_t})=0,\qquad (\dd{\bW_t}^{\top}\msf{A}\dd{\bW_t})=\tr(\msf{A})\dd{t}.
\label{eq:table}
\end{equation}
Because the position increment has no noise component, $\dd{\bx_t}=\bv_t\dd{t}$, every second-order term containing $\dd{\bx_t}$ pairs $\dd{t}$ with either another $\dd{t}$ or $\dd{\bW_t}$ and therefore vanishes by Eq.~\eqref{eq:table}:
\begin{align}
	\dd{\bx_t}^{\top} \nabla_{\bx}\nabla_{\bx}G_t \dd{\bx_t}&={\bv_t}^{\top} \nabla_{\bx}\nabla_{\bx}G_t {\bv_t}(\dd{t})^2=0,\label{eq:dx2}\\
	\dd{\bx_t}^{\top} \nabla_{\bx}\nabla_{\bv}G_t \dd{\bv_t}&=({\bv_t}^{\top} \dd{t}) \nabla_{\bx}\nabla_{\bv}G_t ([\bF_t-\gamma\bv_t]\dd{t}+\sqrt{2\D}\dd{\bW_t})\notag\\
	&=\bv_t^{\top} \nabla_{\bx}\nabla_{\bv}G_t(\bF_t-\gamma\bv_t){(\dd{t})^2} + \bv_t^{\top} \nabla_{\bx}\nabla_{\bv}G_t \sqrt{2\D}{\dd{t}\dd{\bW_t}}=0.\label{eq:dxdv}
\end{align}
Only the velocity increment contains a Wiener component and thus gives a nonvanishing second-order term:
\begin{equation}
	\dd{\bv_t}^{\top} \nabla_{\bv}\nabla_{\bv}G_t \dd{\bv_t}= \dd{\bW_t}^{\top} \sqrt{2\D}\nabla_{\bv}\nabla_{\bv}G_t \sqrt{2\D}\dd{\bW_t} = 2 \nabla_{\bv}\cdot(\D\nabla_{\bv})G_t\dd{t}.\label{eq:dv2}
\end{equation}
Substituting the surviving terms into Eq.~\eqref{eq:fullito} yields
\begin{equation}
	\dd{G_t}=\partial_t G_t\dd{t}+\bv_t\cdot\nabla_{\bx} G_t\dd{t}+(\nabla_{\bv}G_t)\cdot \qty([\bF_t-\gamma\bv_t]\dd{t}+\sqrt{2\D}\dd{\bW_t}) + \nabla_{\bv}\cdot(\D\nabla_{\bv})G_t\dd{t}.
\end{equation}
Separating the finite-variation part from the local-martingale part, we obtain
\begin{equation}
	\dd{G_t}=\Big[\partial_t G_t+\underbrace{\bv_t\cdot\nabla_{\bx} G_t+(\bF_t-\gamma\bv_t)\cdot\nabla_{\bv}G_t + \nabla_{\bv}\cdot(\D\nabla_{\bv})G_t}_{\textstyle =\Lg_t^\dagger G_t}\Big]\dd{t} + (\sqrt{2\D}\nabla_{\bv}G_t)\cdot\dd{\bW_t}.\label{eq:dG}
\end{equation}
Adding the observable increment $\dd{J_t}=\Phi_t(\bz_t)\dd{t}+\sqrt{2\D}\bw_t(\bz_t)\cdot\dd{\bW_t}$ gives
\begin{equation}
	\dd{M_t}=(\Phi_t+\partial_t G_t+\Lg_t^\dagger G_t)\dd{t}+\sqrt{2\D}(\bw_t+\nabla_{\bv}G_t)\cdot\dd{\bW_t}=\sqrt{2\D}(\bw_t+\nabla_{\bv}G_t)\cdot\dd{\bW_t},
\end{equation}
where the final equality follows from the backward equation $\partial_tG_t+\Lg_t^\dagger G_t=-\Phi_t$.

\subsubsection{Finite-time expression}
Integrating $\dd{M_t}=\sqrt{2\D}(\bw_t+\nabla_{\bv}G_t)\cdot\dd{\bW_t}$ over $[0,\tau]$ and using the endpoint values of $M_t$ yields
\begin{equation}
	J_\tau = \E[J_\tau\cond\bz_0] + \int_0^\tau\sqrt{2\D}\qty[\bw_t(\bz_t)+\nabla_{\bv}G_t(\bz_t)]\cdot\dd{\bW_t}.
\end{equation}
Subtracting $\E[J_\tau]$ from both sides, taking the expectation of the square, and applying the It\^o isometry yields the exact finite-time variance
\begin{equation}
\tcboxmath{
	\Var(J_\tau)=\Var_0(J_\tau) + 2\int_0^\tau\ev{\vb*{u}_t^{\top} \D\vb*{u}_t}_t\dd{t},
}
\label{eq:varexact}
\end{equation}
where $\Var_0(J_\tau)\equiv\Var(\E[J_\tau\cond\bz_0])$ and $\vb*{u}_t(\bz)\equiv \bw_t(\bz)+\nabla_{\bv}G_t(\bz)$. The cross term vanishes because the stochastic integral has zero conditional mean given $\bz_0$.

\subsubsection{Long-time expression}
For time-independent dynamics and observable fields $\phi$ and $\bw$, consider the scaled long-time variance in the steady state,
\begin{equation}
	D_J\equiv\lim_{\tau\to\infty}\frac{\mathrm{Var}(J_\tau)}{2\tau}.
\end{equation}
Let $\Psi(\bz)$ solve the stationary Poisson equation
\begin{equation}
	\Lg^\dagger \Psi=-(\Phi-\jr_{\st}),\qquad \jr_{\st}\equiv\ev{\Phi}_{\st}=\ev{\phi}_{\st} + \inprod{\bw}{\bj_{\st}^{\bv}}.
\label{eq:poisson}
\end{equation}
Here, $\ev{\cdot}_{\st}=\int(\cdot)p_{\st}(\bz)\dd{\bz}$ denotes the steady-state average.
Define $Y_t\equiv J_t-\jr_{\st}t+\Psi(\bz_t)$. It\^o's formula gives
\begin{equation}
	\dd{Y_t}=(\Phi-\jr_{\st}+\Lg^\dagger\Psi)\dd{t}+\sqrt{2\D}\qty(\bw+\nabla_{\bv}\Psi)\cdot\dd{\bW_t}=\sqrt{2\D}\qty(\bw+\nabla_{\bv}\Psi)\cdot\dd{\bW_t},
\end{equation}
where the drift term vanishes by Eq.~\eqref{eq:poisson}. Integrating from $t=0$ to $t=\tau$ therefore gives
\begin{equation}
	J_\tau-\jr_{\st}\tau=\int_0^\tau\sqrt{2\D}\qty[\bw+\nabla_{\bv} \Psi](\bz_t)\cdot\dd{\bW_t}+\Psi(\bz_0)-\Psi(\bz_\tau).
\end{equation}
The boundary term has bounded variance, whereas the variance of the martingale term grows linearly in time by the It\^o isometry. Consequently, the scaled long-time variance is
\begin{equation}
\tcboxmath{
	D_J=\ev{(\bw+\nabla_{\bv}\Psi)^{\top} \D (\bw+\nabla_{\bv}\Psi)}_{\st}.
}
\label{eq:long.var}
\end{equation}
To justify this limit, define the boundary term $B_\tau\equiv \Psi(\bz_0)-\Psi(\bz_\tau)$.
In the steady state, the observable variance can be written as
\begin{equation}
	\Var(J_\tau)=2\tau \ev{(\bw+\nabla_{\bv}\Psi)^{\top} \D (\bw+\nabla_{\bv}\Psi)}_{\st} + \E[B_\tau^2] + 2\E\qty[B_\tau\int_0^\tau\sqrt{2\D}(\bw+\nabla_{\bv} \Psi)\cdot\dd{\bW_t}].
\end{equation}
Using $(a-b)^2\le 2(a^2+b^2)$ and stationarity, the boundary contribution is bounded uniformly in $\tau$:
\begin{align}
	\E[B_\tau^2]&\le 4\ev{\Psi^2}_{\st}.
\end{align}
The Cauchy-Schwarz inequality and the It\^o isometry bound the cross term as
\begin{align}
	\qty|\E\qty[B_\tau\int_0^\tau\sqrt{2\D}(\bw+\nabla_{\bv} \Psi)\cdot\dd{\bW_t}]|&\le \sqrt{\E[B_\tau^2]}\sqrt{\E\qty[\qty(\int_0^\tau\sqrt{2\D}(\bw+\nabla_{\bv} \Psi)\cdot\dd{\bW_t})^2]}\notag\\
	&\le\sqrt{4\ev{\Psi^2}_{\st} }\sqrt{2\tau\ev{(\bw+\nabla_{\bv}\Psi)^{\top} \D(\bw+\nabla_{\bv}\Psi)}_{\st}}.
\end{align}
Thus, the boundary and cross terms are $O(1)$ and $O(\sqrt{\tau})$, respectively, while the martingale variance is $O(\tau)$.
Dividing by $2\tau$ and taking the limit $\tau\to\infty$ yields Eq.~\eqref{eq:long.var}.

\subsection{Fluctuation-response relations}
We formulate the response directly in terms of an arbitrary infinitesimal source $\delta\bj_{t,\mathrm{src}}^{\bv}$ injected into the velocity-space probability current. Such a source may arise from a perturbation of the drift, the diffusion, or both. Proposition~\ref{prop:und.perturb} gives the corresponding first variation of the observable average as
\begin{equation}
	\delta\E[J_\tau]=\int_0^\tau \inprod{\vb*{u}_t}{\delta\bj_{t,\mathrm{src}}^{\bv}}\dd{t},
\end{equation}
where $\vb*{u}_t(\bz)=\bw_t(\bz)+\nabla_{\bv}G_t(\bz)$.
It follows from the definition of the functional derivative that
\begin{equation}
	\frac{\delta\E[J_\tau]}{\delta\bj_{t,\mathrm{src}}^{\bv}}=\vb*{u}_t.
\end{equation}
Combining this identity with Eq.~\eqref{eq:varexact} yields the exact fluctuation-response relation
\begin{equation}
\tcboxmath{
	\Var(J_\tau)-\Var_0(J_\tau) = 2\int_0^\tau\ev{\qty(\frac{\delta\E[J_\tau]}{\delta\bj_{t,\mathrm{src}}^{\bv}})^{\top}\D\qty(\frac{\delta\E[J_\tau]}{\delta\bj_{t,\mathrm{src}}^{\bv}})}_t\dd{t}.
}
\end{equation}

We next consider a specific perturbation direction in the space of injected velocity-space currents, $\delta\bj_{t,\mathrm{src}}^{\bv}=\epsilon\bg_t p_t$, where $\epsilon$ is a small scalar perturbation parameter and $\bg_t(\bz)\in\mathbb{R}^d$ is an arbitrary field. For example, this source current can be generated by the drift perturbation $\bF_t\to\bF_t+\epsilon\bg_t$. We define the corresponding linear response and quadratic cost by
\begin{equation}
	\Rc[\bg]\equiv\left.\partial_\epsilon\E_\epsilon[J_\tau]\right|_{\epsilon=0},
    \qquad
    \Ac[\bg]\equiv\int_0^\tau\ev{\bg_t^{\top}\D^{-1}\bg_t}_t\dd{t},
\end{equation}
where $\E_\epsilon$ denotes the expectation under the perturbed dynamics.
Using the response relation above gives
\begin{equation}
    \Rc[\bg]=\int_0^\tau\inprod{\bu_t}{\bg_t p_t}\dd{t}=\int_0^\tau\ev{\bu_t\cdot\bg_t}_t\dd{t}.
\end{equation}
The Cauchy-Schwarz inequality then gives the fluctuation-response inequality
\begin{equation}
\tcboxmath{
	\Rc[\bg]^2\le \frac{\Ac[\bg]}{2}\qty[\Var(J_\tau) - \Var_0(J_\tau)].
}
\label{eq:fri}
\end{equation}

We now derive the dual variational representation associated with Eq.~\eqref{eq:fri}. Using
\begin{equation}
	\Rc[\bg]=\int_0^\tau\ev{\bu_t^{\top}\bg_t}_t\dd{t},
	\qquad
	\Ac[\bg]=\int_0^\tau\ev{\bg_t^{\top}\D^{-1}\bg_t}_t\dd{t},
\end{equation}
and completing the square, we obtain
\begin{align}
	2\Rc[\bg]-\Ac[\bg]
	&=\int_0^\tau\ev{2\bu_t^{\top}\bg_t
	-\bg_t^{\top}\D^{-1}\bg_t}_t\dd{t}\notag\\
	&=\int_0^\tau\ev{\bu_t^{\top}\D\bu_t}_t\dd{t}-\int_0^\tau\ev{(\bg_t-\D\bu_t)^{\top}
	\D^{-1}(\bg_t-\D\bu_t)}_t\dd{t}\notag\\
	&\le\int_0^\tau\ev{\bu_t^{\top}\D\bu_t}_t\dd{t}\notag\\
	&=\frac{\Var(J_\tau)-\Var_0(J_\tau)}{2}.
\end{align}
The upper bound is attained for $\bg_t=\D\bu_t$. Therefore, the excess variance admits the dual variational representation
\begin{equation}
\tcboxmath{
	\frac{\Var(J_\tau)-\Var_0(J_\tau)}{2}
	=\max_{\bg_t}\qty(2\Rc[\bg]-\Ac[\bg]).
}
\end{equation}

\begin{proposition}\label{prop:und.perturb}
Consider an arbitrary admissible infinitesimal injection $\delta\bj_{t,\mathrm{src}}^{\bv}(\bz)$ of velocity-space probability current, while the initial density and the observable fields $\phi_t$ and $\bw_t$ are held fixed. Then the corresponding variation of the observable average is
\begin{equation}
    \delta\E[J_\tau]=\int_0^\tau\inprod{\bu_t}{\delta\bj_{t,\mathrm{src}}^{\bv}}\dd{t},
\label{eq:underresp}
\end{equation}
where $\bu_t=\bw_t+\nabla_{\bv}G_t$.
\end{proposition}

\begin{proof}
The induced variation $\delta p_t$ of the probability density satisfies
\begin{equation}
    \partial_t\delta p_t=\Lg_t(\delta p_t)-\nabla_{\bv}\cdot\delta\bj_{t,\mathrm{src}}^{\bv},
    \qquad
    \delta p_0=0.
\label{eq:und.forced}
\end{equation}
The mean of the additive observable can be expressed in terms of the velocity-space probability current as
\begin{equation}
    \E[J_\tau]=\int_0^\tau\qty[\inprod{\phi_t}{p_t}+\inprod{\bw_t}{\bj_t^{\bv}}]\dd{t}.
\label{eq:mean.current.rep}
\end{equation}
Under an arbitrary infinitesimal current injection $\delta\bj_{t,\mathrm{src}}^{\bv}$, the total variation of the velocity-space current is
\begin{equation}
    \delta\bj_{t}^{\bv}=\Jc_t^{\bv}(\delta p_t)+\delta\bj_{t,\mathrm{src}}^{\bv},
\label{eq:und.total.current.variation}
\end{equation}
where $\Jc_t^{\bv}(p)\equiv(\bF_t-\gamma\bv)p-\D\nabla_{\bv}p$ denotes the probability-current operator in the velocity sector.
Taking the variation of Eq.~\eqref{eq:mean.current.rep} therefore gives
\begin{align}
    \delta\E[J_\tau]&=\int_0^\tau\qty[\inprod{\phi_t}{\delta p_t}+\inprod{\bw_t}{\Jc_t^{\bv}(\delta p_t)}+\inprod{\bw_t}{\delta\bj_{t,\mathrm{src}}^{\bv}}]\dd{t}\notag\\
    &=\int_0^\tau\qty[\inprod{\Phi_t}{\delta p_t}+\inprod{\bw_t}{\delta\bj_{t,\mathrm{src}}^{\bv}}]\dd{t},
\label{eq:response.current.rep}
\end{align}
where integration by parts in velocity space and $\Phi_t=\phi_t+\bw_t\cdot(\bF_t-\gamma\bv)+\D:\nabla_{\bv}\bw_t$ have been used.

Let $\mca{U}(t,s)$ and $\mca{U}^\dagger(t,s)$ denote the forward propagator generated by $\Lg_t$ and its adjoint propagator, respectively, given by
\begin{equation}
	\mca{U}(t,s)=\mca{T}\exp(\int_s^t\Lg_r\dd{r}),\quad \mca{U}^\dagger(t,s)=\overline{\mca{T}}\exp(\int_s^t\Lg_r^\dagger\dd{r}).
\end{equation}
Here, $\mca{T}$ and $\overline{\mca{T}}$ denote chronological and anti-chronological time ordering, respectively. The backward equation $\partial_t G_t+\Lg_t^\dagger G_t=-\Phi_t$, together with the terminal condition $G_\tau=0$, gives 
\begin{equation}
	G_s=\int_s^\tau\mca{U}^\dagger(t,s)\Phi_t\dd{t}.
\label{eq:und.G.prop}
\end{equation}
The variation of the probability density can also be written directly as
\begin{equation}
    \delta p_t=-\int_0^t\mca{U}(t,s)\nabla_{\bv}\cdot\delta\bj_{s,\mathrm{src}}^{\bv}\dd{s}.
\label{eq:und.delta.p}
\end{equation}
Using Eqs.~\eqref{eq:und.G.prop} and \eqref{eq:und.delta.p}, the first term becomes
\begin{align}
    \int_0^\tau\inprod{\Phi_t}{\delta p_t}\dd{t}&=-\int_0^\tau\int_0^t\inprod{\Phi_t}{\mca{U}(t,s)\nabla_{\bv}\cdot\delta\bj_{s,\mathrm{src}}^{\bv}}\dd{s}\dd{t}\notag\\
    &=-\int_0^\tau\int_s^\tau\inprod{\mca{U}^\dagger(t,s)\Phi_t}{\nabla_{\bv}\cdot\delta\bj_{s,\mathrm{src}}^{\bv}}\dd{t}\dd{s}\notag\\
    &=-\int_0^\tau\inprod{G_s}{\nabla_{\bv}\cdot\delta\bj_{s,\mathrm{src}}^{\bv}}\dd{s}\notag\\
    &=\int_0^\tau\inprod{\nabla_{\bv}G_t}{\delta\bj_{t,\mathrm{src}}^{\bv}}\dd{t}.
\end{align}
Substitution into Eq.~\eqref{eq:response.current.rep} gives
\begin{align}
    \delta\E[J_\tau]&=\int_0^\tau\inprod{\bw_t+\nabla_{\bv}G_t}{\delta\bj_{t,\mathrm{src}}^{\bv}}\dd{t}\notag\\
    &=\int_0^\tau\inprod{\bu_t}{\delta\bj_{t,\mathrm{src}}^{\bv}}\dd{t},
\end{align}
which proves the result.
\end{proof}

\subsection{Operationally accessible friction-response TUR}
Consider dynamics satisfying the fluctuation-dissipation relation $\D=\gamma\T$.
For the velocity-space drift perturbation, choose $\bg_t(\bz)=\vb*{\nu}_t^{\mathrm{irr}}(\bz)$. The corresponding cost is
\begin{equation}
	\Ac[\bg]=\int_0^\tau\ev{(\vb*{\nu}_t^{\mathrm{irr}})^{\top}\D^{-1}\vb*{\nu}_t^{\mathrm{irr}}}_t\dd{t}=\Sigma_\tau.
\end{equation}
If the initial probability density $p_0$, the state-dependent rate $\phi_t$, and the weight $\bw_t$ have no explicit $\gamma$ dependence, then
\begin{equation}
	\Rc[\bg]=\gamma\partial_\gamma\E[J_\tau].
\end{equation}
To see this, consider the perturbation $\gamma\to\gamma(1+\epsilon)$ while keeping $\T$ fixed, so that $\D\to\D(1+\epsilon)$ and the fluctuation-dissipation relation remains satisfied. The source current is then
\begin{equation}
    \delta\bj_{t,\mathrm{src}}^{\bv}=\epsilon(-\gamma\bv p_t-\D\nabla_{\bv}p_t)=\epsilon\bj_t^{\mathrm{irr}}=\epsilon\bnu_t^{\mathrm{irr}}p_t.
\label{eq:und.friction.source}
\end{equation}
Equation~\eqref{eq:underresp} therefore gives
\begin{equation}
    \gamma\partial_\gamma\E[J_\tau]=\int_0^\tau\inprod{\bu_t}{\bnu_t^{\mathrm{irr}}p_t}\dd{t}=\Rc[\bg].
\label{eq:und.gamma.response.general}
\end{equation}
Consequently, the fluctuation-response inequality yields the following operationally accessible TUR for underdamped dynamics:
\begin{equation}
\tcboxmath{
	\frac{\Var(J_\tau)- \Var_0(J_\tau)}{(\gamma\partial_\gamma\E[J_\tau])^2}\ge\frac{2}{\Sigma_\tau}.
}
\label{eq:fric.TUR}
\end{equation}

\subsubsection{Finite-time saturation and exact entropy-production inference}
We discuss the saturation condition of the friction-response TUR at finite times. The equality condition in the Cauchy-Schwarz step is
\begin{equation}
    \bu_t=c\,\D^{-1}\bnu_t^{\mathrm{irr}}
\label{eq:und.finite.saturation.condition}
\end{equation}
with a constant $c$. We set $c=1$ without loss of generality. We now show constructively that this condition can be realized at an arbitrary finite observation time $\tau$.

Assuming the probability density is smooth and strictly positive, define
\begin{equation}
    \Lambda_t(\bz)\equiv-\ln\qty[p_t(\bz)\exp\qty(\frac{1}{2}\bv^{\top}\T^{-1}\bv)].
\label{eq:und.psi.t}
\end{equation}
Then,
\begin{equation}
    \D\nabla_{\bv}\Lambda_t=-\gamma\bv-\D\nabla_{\bv}\ln p_t=\bnu_t^{\mathrm{irr}}.
\label{eq:und.psi.gradient}
\end{equation}
Even if the velocity-current weight is required to have no explicit time dependence, choose
\begin{equation}
    \bw^{\mathrm{opt}}(\bz)\equiv\nabla_{\bv}\Lambda_\tau(\bz),
    \qquad
    G_t^{\mathrm{opt}}(\bz)\equiv\Lambda_t(\bz)-\Lambda_\tau(\bz).
\label{eq:und.chi.G.opt}
\end{equation}
The terminal condition $G_\tau^{\mathrm{opt}}=0$ is immediately satisfied, and
\begin{equation}
    \bu_t^{\mathrm{opt}}=\bw^{\mathrm{opt}}+\nabla_{\bv}G_t^{\mathrm{opt}}=\nabla_{\bv}\Lambda_t=\D^{-1}\bnu_t^{\mathrm{irr}}.
\label{eq:und.u.opt.finite}
\end{equation}
To make $G_t^{\mathrm{opt}}$ the conditional future value of an observable of the form \eqref{eq:und.general.observable}, choose its effective It\^o rate as
\begin{equation}
    \Phi_t^{\mathrm{opt}}
    \equiv-\partial_tG_t^{\mathrm{opt}}-\Lg_t^\dagger G_t^{\mathrm{opt}},
\end{equation}
and hence the state-dependent rate is given by
\begin{equation}
    \phi_t^{\mathrm{opt}}\equiv\Phi_t^{\mathrm{opt}}-\bw^{\mathrm{opt}}\cdot(\bF_t-\gamma\bv)-\D:\nabla_{\bv}\bw^{\mathrm{opt}}.
\label{eq:und.phi.opt.finite}
\end{equation}
All coefficients in Eqs.~\eqref{eq:und.chi.G.opt} and \eqref{eq:und.phi.opt.finite} are defined using the reference process and are held fixed when the friction derivative is evaluated.

For this observable, Eqs.~\eqref{eq:varexact}, \eqref{eq:und.gamma.response.general}, and \eqref{eq:und.u.opt.finite} give
\begin{align}
    \Var(J_\tau^{\mathrm{opt}})-\Var_0(J_\tau^{\mathrm{opt}})&=2\Sigma_\tau,\\
    \gamma\partial_\gamma\E[J_\tau^{\mathrm{opt}}]&=\Sigma_\tau.
\end{align}
Consequently, the friction-response TUR \eqref{eq:fric.TUR} is exactly saturated at arbitrary finite times, and
\begin{equation}
\tcboxmath{
    \Sigma_\tau=\max_{\phi_t,\bw_t}\frac{2(\gamma\partial_\gamma\E[J_\tau])^2}{\Var(J_\tau)-\Var_0(J_\tau)}.
}
\label{eq:und.finite.variational}
\end{equation}
The construction above shows that the maximum remains attainable even when $\bw_t$ is restricted to a time-independent field $\bw(\bz)$, provided the state-dependent rate $\phi_t$ may depend on time. 

\subsubsection{Short-time behavior for the restricted class $\bw_t=0$}

We now examine the short-time behavior of the friction-response TUR for the restricted class of observables with $\bw_t=0$, for which there is no direct velocity-space Stratonovich current. In this case, the initial-state contribution $\Var_0(J_\tau)$ plays a particularly important role. We show that using the full variance generally leads to a divergent short-time uncertainty factor as $\tau\to 0$, whereas subtracting $\Var_0(J_\tau)$ isolates the dynamically generated fluctuations and yields a finite short-time limit.

For clarity, we present the derivation for time-independent dynamics and an additive observable $J_\tau=\int_0^\tau \phi(\bz_t)\dd{t}$. We assume throughout that the initial density $p_0$ and the observable rate $\phi$ have no explicit dependence on $\gamma$. The same short-time asymptotic analysis applies to smoothly time-dependent dynamics, with the relevant quantities evaluated at the initial time.

The conditional future observable introduced in the main text can be written as
\begin{equation}
	G_t(\bz)=\E\qty[\int_t^\tau\phi(\bz_s)\dd{s}\,\middle|\,\bz_t=\bz]=\int_0^{\tau-t}e^{s\Lg^\dagger}\phi(\bz)\dd{s}.
\label{eq:short.G}
\end{equation}
We use the two exact identities derived above,
\begin{align}
	\Var(J_\tau)-\Var_0(J_\tau)&=2\int_0^\tau\ev{(\nabla_{\bv}G_t)^{\top}\D\nabla_{\bv}G_t}_t\dd{t},\label{eq:short.varidentity}\\
	\gamma\partial_\gamma\E[J_\tau]&=\int_0^\tau\ev{\bnu_t^{\mathrm{irr}}\cdot\nabla_{\bv}G_t}_t\dd{t}.\label{eq:short.responseidentity}
\end{align}
We compare the uncertainty factor associated with our friction-response TUR,
\begin{equation}
	\Qc_{\mathrm{sub}}(\tau)\equiv\frac{[\Var(J_\tau)-\Var_0(J_\tau)]\Sigma_\tau}{2(\gamma\partial_\gamma\E[J_\tau])^2},
\label{eq:short.Qsubdef}
\end{equation}
with the corresponding expression obtained when the full variance is used \cite{Kwon.2022.NJP},
\begin{equation}
	\Qc_{\mathrm{full}}(\tau)\equiv\frac{\Var(J_\tau)\Sigma_\tau}{2(\gamma\partial_\gamma\E[J_\tau])^2}.
\label{eq:short.Qfulldef}
\end{equation}
Both satisfy $\Qc\ge 1$ whenever the corresponding uncertainty relation is applicable. Their short-time behaviors, however, are generally very different.

Expanding Eq.~\eqref{eq:short.G} in powers of $\tau-t$ gives
\begin{equation}
	G_t=(\tau-t)\phi+\frac{(\tau-t)^2}{2}\Lg^\dagger\phi+\frac{(\tau-t)^3}{3!}(\Lg^\dagger)^2\phi+\cdots.
\label{eq:short.Gexp}
\end{equation}
The fluctuation and friction response are both controlled by $\nabla_{\bv}G_t$. We therefore define $k\ge 1$ as the smallest integer such that
\begin{equation}
	\nabla_{\bv}(\Lg^\dagger)^{k-1}\phi\not\equiv0,
\end{equation}
and introduce
\begin{equation}
	\bchi(\bz)\equiv\nabla_{\bv}(\Lg^\dagger)^{k-1}\phi(\bz).
\end{equation}
Equation \eqref{eq:short.Gexp} then implies
\begin{equation}
	\nabla_{\bv}G_t(\bz)=\frac{(\tau-t)^k}{k!}\bchi(\bz)+O[(\tau-t)^{k+1}].
\label{eq:short.kernel}
\end{equation}
Thus, $k-1$ is the number of applications of the backward generator required before the observable becomes sensitive to the noisy velocity sector. Two cases are particularly relevant. If $\phi$ depends explicitly on velocity, then generically $k=1$ and $\bchi=\nabla_{\bv}\phi$. This includes velocity- or momentum-resolved transport observables. By contrast, if $\phi=\phi(\bx)$ is independent of velocity, then $\nabla_{\bv}\phi=0$ and
\begin{equation}
	\Lg^\dagger\phi=\bv\cdot\nabla_{\bx}\phi,
\end{equation}
so that $k=2$ and $\bchi=\nabla_{\bx}\phi$, provided $\nabla_{\bx}\phi\neq 0$. Position-dependent observables, including the rates entering work-like and occupation-type observables, therefore generically belong to the $k=2$ class.

We now derive the leading short-time behavior of the four quantities entering Eqs.~\eqref{eq:short.Qsubdef} and \eqref{eq:short.Qfulldef}. Throughout, $\ev{\cdot}_0$ denotes an average over the initial distribution $p_0$. First, using Eq.~\eqref{eq:short.Gexp} at $t=0$,
\begin{equation}
	\E[J_\tau\cond\bz_0]=G_0(\bz_0)=\tau\phi(\bz_0)+O(\tau^2).
\end{equation}
It follows that
\begin{equation}
	\Var_0(J_\tau)=\tau^2\Var_0(\phi)+O(\tau^3),
\label{eq:short.Var0}
\end{equation}
where $\Var_0(\phi)\equiv\ev{\phi^2}_0-\ev{\phi}_0^2$.
Importantly, this contribution is always $O(\tau^2)$, independently of $k$. It is determined directly by the variation of the observable rate across the initial ensemble and does not originate from noise accumulated during the observation interval.

Next, substituting Eq.~\eqref{eq:short.kernel} into the exact variance
identity \eqref{eq:short.varidentity} yields
\begin{align}
	\Var(J_\tau)-\Var_0(J_\tau)&=\frac{2}{(k!)^2}\int_0^\tau(\tau-t)^{2k}\ev{\bchi^{\top}\D\bchi}_0\dd{t}+O(\tau^{2k+2})\notag\\
	&=\frac{2\tau^{2k+1}}{(k!)^2(2k+1)}\ev{\bchi^{\top}\D\bchi}_0+O(\tau^{2k+2}).
\label{eq:short.Vardyn}
\end{align}
Here, we used $p_t=p_0+O(t)$ and
\begin{equation}
	\int_0^\tau(\tau-t)^{2k}\dd{t}=\frac{\tau^{2k+1}}{2k+1}.
\end{equation}
The dynamically generated variance is therefore $O(\tau^{2k+1})$, which is of higher order than the $O(\tau^2)$ preparation variance for every $k\ge 1$. Similarly, Eq.~\eqref{eq:short.responseidentity} gives
\begin{align}
	\gamma\partial_\gamma\E[J_\tau]&=\frac{1}{k!}\int_0^\tau(\tau-t)^k\ev{\bnu_0^{\mathrm{irr}}\cdot\bchi}_0\dd{t}+O(\tau^{k+2})\notag\\
	&=\frac{\tau^{k+1}}{(k+1)!}\ev{\bnu_0^{\mathrm{irr}}\cdot\bchi}_0+O(\tau^{k+2}).
\label{eq:short.R}
\end{align}
Finally, continuity of the instantaneous entropy production rate implies
\begin{equation}
	\Sigma_\tau=\tau\sigma_0+O(\tau^2),\qquad\sigma_0\equiv\ev{(\bnu_0^{\mathrm{irr}})^{\top}\D^{-1}\bnu_0^{\mathrm{irr}}}_0,
\label{eq:short.Sigma}
\end{equation}
where a non-vanishing initial entropy-production rate is assumed.
Equations~\eqref{eq:short.Var0}--\eqref{eq:short.Sigma} summarize the short-time scaling:
\begin{equation}
	\Var_0=O(\tau^2),\qquad
	\Var-\Var_0=O(\tau^{2k+1}),\qquad
	\gamma\partial_\gamma\E[J_\tau]=O(\tau^{k+1}),\qquad
	\Sigma_\tau=O(\tau).
\label{eq:short.orders}
\end{equation}
The first two relations reveal the essential asymmetry. The $O(\tau^2)$ fluctuations inherited from the preparation generally dominate the genuinely dynamical fluctuations at short times.

We first consider Eq.~\eqref{eq:short.Qsubdef}. Assuming $\ev{\bnu_0^{\mathrm{irr}}\cdot\bchi}_0\neq 0$, substitution of Eqs.~\eqref{eq:short.Vardyn}--\eqref{eq:short.Sigma} gives
\begin{align}
	\Qc_{\mathrm{sub}}(\tau)&=\frac{\dfrac{2\tau^{2k+1}}{(k!)^2(2k+1)}\ev{\bchi^{\top}\D\bchi}_0\tau\sigma_0}{2\dfrac{\tau^{2k+2}}{[(k+1)!]^2}\ev{\bnu_0^{\mathrm{irr}}\cdot\bchi}_0^2}+O(\tau)\notag\\
	&=\frac{(k+1)^2}{2k+1}\frac{\ev{\bchi^{\top}\D\bchi}_0\sigma_0}{\ev{\bnu_0^{\mathrm{irr}}\cdot\bchi}_0^2}+O(\tau).
\end{align}
Consequently,
\begin{equation}
\tcboxmath{
	\lim_{\tau\to0}\Qc_{\mathrm{sub}}(\tau)=\frac{(k+1)^2}{2k+1}\frac{\ev{\bchi^{\top}\D\bchi}_0\sigma_0}{\ev{\bnu_0^{\mathrm{irr}}\cdot\bchi}_0^2}.
}
\label{eq:short.Qsub}
\end{equation}
All powers of $\tau$ cancel exactly. The remaining ratio is bounded below by unity by the Cauchy-Schwarz inequality. Indeed,
\begin{align}
	\ev{\bnu_0^{\mathrm{irr}}\cdot\bchi}_0^2&=\ev{(\D^{-1}\bnu_0^{\mathrm{irr}})^{\top}\D\bchi}_0^2\notag\\
	&\le\ev{(\bnu_0^{\mathrm{irr}})^{\top}\D^{-1}\bnu_0^{\mathrm{irr}}}_0\ev{\bchi^{\top}\D\bchi}_0\notag\\
	&=\sigma_0\ev{\bchi^{\top}\D\bchi}_0 .
\end{align}
Therefore,
\begin{equation}
	\lim_{\tau\to0}\Qc_{\mathrm{sub}}(\tau)\ge\frac{(k+1)^2}{2k+1}>1.
\label{eq:short.Qsubbound}
\end{equation}
Equality is attained when $\bchi\propto\D^{-1}\bnu_0^{\mathrm{irr}}$. In particular,
\begin{equation}
	\Qc_{\mathrm{sub}}(0^+)\ge
	\begin{cases}
	4/3, & k=1,\\[2pt]
	9/5, & k=2.
	\end{cases}
\label{eq:short.values}
\end{equation}
Thus, although the generalized observable class admits exact saturation at arbitrary finite times, the restricted class $\bw_t=0$ is generically not saturable in the short-time limit. Even under optimal alignment, its limiting uncertainty factor is strictly larger than unity.

We next consider the response-based uncertainty relation constructed from the full variance. Suppose $\Var_0(\phi)>0$. Since the $O(\tau^2)$ preparation term dominates Eq.~\eqref{eq:short.Vardyn},
\begin{equation}
	\Var(J_\tau)=\tau^2\Var_0(\phi)+O(\tau^3).
\end{equation}
Using Eqs.~\eqref{eq:short.R} and \eqref{eq:short.Sigma}, we obtain
\begin{align}
	\Qc_{\mathrm{full}}(\tau)&\simeq\frac{\tau^2\Var_0(\phi)\tau\sigma_0}{2\tau^{2k+2}\ev{\bnu_0^{\mathrm{irr}}\cdot\bchi}_0^2/[(k+1)!]^2}\notag\\
	&=\frac{[(k+1)!]^2}{2}\frac{\Var_0(\phi)\sigma_0}{\ev{\bnu_0^{\mathrm{irr}}\cdot\bchi}_0^2}\tau^{-(2k-1)}.
\label{eq:short.Qfull}
\end{align}
Hence,
\begin{equation}
\tcboxmath{
	\Qc_{\mathrm{full}}(\tau)\propto\tau^{-(2k-1)}\longrightarrow\infty .
}
\label{eq:short.divergence}
\end{equation}
The divergence therefore originates entirely from the $O(\tau^2)$ variance inherited from the initial ensemble.

By comparison, inference based on the full variance becomes asymptotically uninformative. From Eq.~\eqref{eq:short.Qfull},
\begin{equation}
    \frac{2(\gamma\partial_\gamma\E[J_\tau])^2}{\tau\Var(J_\tau)}=O(\!\tau^{2k-1})\longrightarrow0.
\end{equation}
If the initial preparation is deterministic, $\Var_0(J_\tau)=0$, and the distinction between the full and subtracted variance forms disappears.

For convenience, the generic short-time behaviors are summarized as
\begin{center}
\begin{tabular}{c|c|c}
\hline\hline
Friction-response TUR & ~full variance~ \cite{Kwon.2022.NJP} & ~subtracted variance (this work)~\\
\hline
general $k$ & $\Qc_{\mathrm{full}}\sim\tau^{-(2k-1)}$ & $\Qc_{\mathrm{sub}}=O(1)$\\
$k=1$ (velocity-resolved) & $\sim\tau^{-1}$ & $\Qc_{\mathrm{sub}}(0^+)\ge4/3$\\
$k=2$ (position/work-like) & $\sim\tau^{-3}$ & $\Qc_{\mathrm{sub}}(0^+)\ge9/5$\\
\hline\hline
\end{tabular}
\end{center}
The variance subtraction therefore does more than quantitatively tighten the friction-response TUR. At short times, it removes the leading preparation-induced fluctuations that are unrelated to the dynamical response, converting a generically divergent uncertainty factor into a finite one.

\subsection{Multivariate generalization}
Consider $m$ time-integrated phase-space observables
\begin{equation}
	J^\alpha_\tau=\int_0^\tau[\phi_t^\alpha(\bz_t)\dd{t}+\bw_t^\alpha(\bz_t)\circ\dd{\bv_t}],\qquad \alpha=1,\dots,m,
\end{equation}
where $\phi_t^\alpha$ and $\bw_t^\alpha$ are arbitrary and need not have a definite parity under time reversal. For each observable, define the conditional future observable and its velocity-gradient kernel by
\begin{align}
	G_t^\alpha(\bz)&\equiv\E\Big[\int_t^\tau(\phi_s^\alpha(\bz_s)\dd{s}+\bw_s^\alpha(\bz_s)\circ\dd{\bv_s})\,\Big|\,\bz_t=\bz\Big],\\
	\partial_tG_t^\alpha+\Lg_t^\dagger G_t^\alpha&=-\Phi_t^\alpha,\qquad G^\alpha_\tau=0,\label{eq:mul.backward}\\
	\bu_t^\alpha &\equiv \bw_t^\alpha+\nabla_{\bv}G_t^\alpha.
\end{align}
Then $M_t^\alpha=J_t^\alpha+G_t^\alpha(\bz_t)=\E[J^\alpha_\tau\cond\mca{F}_t]$ is a martingale. Applying It\^o's formula and using Eq.~\eqref{eq:mul.backward} eliminates the drift term and gives
\begin{equation}
	\dd{M_t^\alpha}=\qty[\sqrt{2\D}\bu_t^\alpha(\bz_t)]\cdot\dd{\bW_t}.
\label{eq:mart}
\end{equation}
All $m$ martingales are driven by the same Wiener process $\bW_t$.
For later use, define the weighted inner product
\begin{equation}
	\inp{\vb*{a}}{\vb*{b}}_{\D,p}\equiv\int \vb*{a}(\bz)^{\top}\D\vb*{b}(\bz)p(\bz)\dd{\bz}.
\end{equation}

\begin{proposition}\label{prop:cov}
Let $\Cm^{\alpha\beta}\equiv\Cov\qty(J^\alpha_\tau,J^\beta_\tau)$ denote the observable covariance, and define its initial-state contribution by $\Cm_{0}^{\alpha\beta}\equiv\Cov\qty(\E[J^\alpha_\tau\cond\bz_0],\E[J^\beta_\tau\cond\bz_0])$. Then
\begin{equation}
	\Cm^{\alpha\beta} - \Cm^{\alpha\beta}_0 = 2\int_0^\tau\inp{\bu_t^\alpha}{\bu_t^\beta}_{\D,p_t}\dd{t}.
\label{eq:C}
\end{equation}
\end{proposition}

\begin{proof}
Integrating Eq.~\eqref{eq:mart} gives
\begin{equation}
	J^\alpha_\tau=\E[J^\alpha_\tau\cond\bz_0]+\int_0^\tau\qty[\sqrt{2\D}\bu_t^\alpha(\bz_t)]\cdot\dd{\bW_t}.
\end{equation}
The stochastic integral has zero conditional mean given $\mca{F}_0$. Hence, the cross terms between the initial-state contribution and the martingale increment vanish, and
\begin{equation}
	\Cm^{\alpha\beta}=\Cm_0^{\alpha\beta}+\E\qty[(M^\alpha_\tau-M^\alpha_0)(M^\beta_\tau-M^\beta_0)].
\end{equation}
The polarized It\^o isometry for two integrals driven by the same Wiener process reads
\begin{equation}
	\E\Big[\Big(\int_0^\tau\vb*{a}_t\cdot\dd{\bW_t}\Big)\Big(\int_0^\tau\vb*{b}_t\cdot\dd{\bW_t}\Big)\Big]=\E\int_0^\tau\vb*{a}_t\cdot\vb*{b}_t\dd{t}.
\end{equation}
Applying this identity with $\vb*{a}_t=\sqrt{2\D}\bu_t^\alpha(\bz_t)$ and $\vb*{b}_t=\sqrt{2\D}\bu_t^\beta(\bz_t)$ yields Eq.~\eqref{eq:C}.
\end{proof}

Consider $n$ independently controlled perturbation directions in the space of injected velocity-space currents, $\delta\bj_{t,\mathrm{src}}^{\bv}=\sum_{i=1}^n \epsilon_i\bg_t^i p_t$, where $\{\epsilon_i\}_{i=1}^n$ are small independent perturbation parameters and the fields $\{\bg_t^i(\bz)\}_{i=1}^n$ may depend on both the phase-space state and time. For example, these source-current perturbations can be generated by the drift change $\bF_t\to\bF_t+\sum_{i=1}^n\epsilon_i\bg_t^i$. The response matrix is defined by
\begin{equation}
	\Rm_{\alpha i} \equiv \left.\frac{\partial\E_{\vb*{\epsilon}}[J^\alpha_\tau]}{\partial\epsilon_i}\right|_{\vb*{\epsilon}=0}.
\label{eq:Rdef}
\end{equation}
Applying Proposition~\ref{prop:und.perturb} to each perturbation field gives
\begin{equation}
	\Rm_{\alpha i}=\int_0^\tau\ev{\bu_t^\alpha\cdot\bg^i_t}_t\dd{t}=\int_0^\tau \inp{\bu_t^\alpha}{\bh_t^i}_{\D,p_t} \dd{t},
\label{eq:R}
\end{equation}
where $\bh_t^i\equiv\D^{-1}\bg_t^i$. Define the matrix of quadratic perturbation costs by
\begin{equation}
	\Am_{ij} \equiv \int_0^\tau\ev{(\bg^i_t)^{\top}\D^{-1}\bg^j_t}_t\dd{t}=\int_0^\tau\inp{\bh_t^i}{\bh_t^j}_{\D,p_t}\dd{t}.
\label{eq:A}
\end{equation}
Under the time-integrated inner product $\int_0^\tau\inp{\cdot}{\cdot}_{\D,p_t}\dd{t}$, the collection $(\bu^1,\dots,\bu^m,\bh^1,\dots,\bh^n)$ has the positive semidefinite Gram matrix
\begin{equation}
	\Gm=
\begin{pmatrix}
	[\Cm-\Cm_0]/2 & \Rm\\[3pt]
	\Rm^{\top} & \Am
\end{pmatrix} \succeq\ 0,
\label{eq:gram}
\end{equation}
where $\Cm$, $\Rm$, and $\Am$ are given by Eqs.~\eqref{eq:C}, \eqref{eq:R}, and \eqref{eq:A}, respectively.
If $\Am\succ0$, the positive semidefiniteness of the Schur complement of $\Gm$ yields the matrix fluctuation-response inequality
\begin{equation}
\tcboxmath{
	\Cm-\Cm_0\succeq 2\Rm\Am^{-1}\Rm^{\top}.
}
\end{equation}
Similarly, if $\Cm-\Cm_0\succ0$, the complementary Schur complement gives
\begin{equation}
\tcboxmath{
	\Am\succeq 2\Rm^{\top}(\Cm-\Cm_0)^{-1}\Rm.
}
\end{equation}

\subsection{Frequency-resolved fluctuation-response relations}
We next show that the finite-time fluctuation-response framework admits a direct frequency-domain representation. No new stochastic construction is required: a Fourier-filtered observable is itself a member of the general additive-observable class considered in the main text, with explicitly time-dependent weights.  The stationary infinite-time relations then follow by taking the long-time limit of the finite-time identities derived below.

\subsubsection{Finite-time Fourier-filtered observable and exact variance identity}
For a real frequency $\omega$, define
\begin{equation}
    J_\tau[\omega]\equiv\int_0^\tau e^{-i\omega t}\dd{J_t}
    =\int_0^\tau e^{-i\omega t}\phi_t(\bz_t)\dd{t}+\int_0^\tau e^{-i\omega t}\bw_t(\bz_t)\circ\dd{\bv_t}.
\label{eq:freq.observable}
\end{equation}
Thus $J_\tau[0]=J_\tau$, while for $\omega\neq 0$ the observable is obtained from the original class by the replacements
\begin{equation}
    \phi_t\longrightarrow e^{-i\omega t}\phi_t,
    \qquad
    \bw_t\longrightarrow e^{-i\omega t}\bw_t.
\end{equation}
Equivalently,
\begin{equation}
    J_\tau[\omega]=J_{\tau,c}[\omega]-iJ_{\tau,s}[\omega],
\end{equation}
where
\begin{align}
    J_{\tau,c}[\omega]&\equiv\int_0^\tau \cos(\omega t)\dd{J_t},\\
    J_{\tau,s}[\omega]&\equiv\int_0^\tau \sin(\omega t)\dd{J_t}.
\end{align}
Hence, the complex notation is merely a compact representation of the two real quadratures. For complex random variables, we use
\begin{equation}
    \Var(X)\equiv\E\qty[\abs{X-\E[X]}^2],
    \qquad
    \Cov(X,Y)\equiv\E\qty[(X-\E[X])(Y-\E[Y])^*].
\end{equation}
Applying the finite-time martingale construction to Eq.~\eqref{eq:freq.observable} gives
\begin{equation}
\tcboxmath{
    \Var(J_\tau[\omega])-\Var_0(J_\tau[\omega])=2\int_0^\tau\ev{\bu_t[\omega]^\dagger\D\bu_t[\omega]}_t\dd{t},
}
\label{eq:freq.finite.identity}
\end{equation}
where
\begin{equation}
    \bu_t[\omega]\equiv e^{-i\omega t}\bw_t+\nabla_{\bv}G_t[\omega],
\label{eq:freq.finite.kernel}
\end{equation}
and $G_t[\omega]$ solves the backward equation
\begin{equation}
    \partial_tG_t[\omega]+\Lg_t^\dagger G_t[\omega]=-e^{-i\omega t}\Phi_t,
    \qquad
    G_\tau[\omega]=0.
\label{eq:freq.backward.eq}
\end{equation}
Equation~\eqref{eq:freq.finite.identity} is exact for arbitrary initial ensembles and arbitrary time-dependent driving.

\subsubsection{Finite-time response inequality}
Consider an infinitesimal injected velocity-space current
\begin{equation}
    \delta\bj_{t,\mathrm{src}}^{\bv}=\epsilon \bg_t p_t,
\label{eq:freq.source}
\end{equation}
where the physical perturbation field $\bg_t$ is real; complex fields may alternatively be understood as a shorthand for two real quadratures.  The response of the Fourier-filtered observable is
\begin{equation}
    \Rc_\omega[\bg]\equiv\left.\partial_\epsilon\E_\epsilon[J_\tau[\omega]]\right|_{\epsilon=0}=\int_0^\tau r_t^\omega[\bg]\dd{t},
    \qquad
    r_t^\omega[\bg]\equiv\ev{\bu_t[\omega]\cdot\bg_t}_t.
\label{eq:freq.response}
\end{equation}
The quantity $r_t^\omega[\bg]$ is the response density resolved at the perturbation time; equivalently, if the perturbation amplitude is allowed to vary as $\epsilon\to\epsilon_t$, then
\begin{equation}
    r_t^\omega[\bg]=\left.\frac{\delta\E[J_\tau[\omega]]}{\delta\epsilon_t}\right|_{\epsilon=0}.
\label{eq:freq.functional.response}
\end{equation}

For complex vector fields, define the Hermitian inner product
\begin{equation}
    \inp{\vb*{a}}{\vb*{b}}_{\D,p}\equiv\int\vb*{a}(\bz)^\dagger\D\vb*{b}(\bz)p(\bz)\dd{\bz},
    \qquad
    \lVert\vb*{a}\rVert_{\D,p}^2\equiv\inp{\vb*{a}}{\vb*{a}}_{\D,p}.
\label{eq:freq.inner.product}
\end{equation}
Because the primitive fields $\Phi_t$ and $\bw_t$ are real,
\begin{equation}
    \bu_t[\omega]^*=\bu_t[-\omega].
\end{equation}
Writing $\bh_t=\D^{-1}\bg_t$, one has
\begin{equation}
    r_t^\omega[\bg]=\inp{\bu_t[-\omega]}{\bh_t}_{\D,p_t}.
\end{equation}
The Cauchy-Schwarz inequality therefore gives
\begin{equation}
    \abs{r_t^\omega[\bg]}^2\le\lVert\bu_t[-\omega]\rVert_{\D,p_t}^2\lVert\bh_t\rVert_{\D,p_t}^2=\ev{\bu_t[\omega]^\dagger\D\bu_t[\omega]}_ta_t[\bg],
    \qquad
    a_t[\bg]\equiv\ev{\bg_t^\top\D^{-1}\bg_t}_t.
\label{eq:freq.pointwise.CS}
\end{equation}
Combining this with Eq.~\eqref{eq:freq.finite.identity}, and then applying the Cauchy-Schwarz inequality in time, yields the hierarchy
\begin{equation}
\tcboxmath{
\begin{aligned}
    \Var(J_\tau[\omega])-\Var_0(J_\tau[\omega])&\ge 2\int_0^\tau\frac{\abs{r_t^\omega[\bg]}^2}{a_t[\bg]}\dd{t}\\
    &\ge\frac{2\abs{\Rc_\omega[\bg]}^2}{\Ac[\bg]},
\end{aligned}
}
\label{eq:freq.strong.FRI}
\end{equation}
where $\Ac[\bg]=\int_0^\tau a_t[\bg]\dd{t}$.
The first inequality in Eq.~\eqref{eq:freq.strong.FRI} is generally stronger than the scalar response bound on the second line because it retains the magnitude of the response before integrating over time. In particular, an oscillatory response can suffer strong phase cancellation in $\Rc_\omega[\bg]=\int r_t^\omega[\bg]\dd{t}$ even when $\abs{r_t^\omega[\bg]}$ remains finite. The time-resolved bound therefore retains the response magnitude at each perturbation time and avoids cancellation between contributions at different times.

\subsubsection{Power spectrum in a stationary state}
We now specialize to autonomous dynamics in a stationary state and take the observable fields to be time independent. Define the finite-window spectral density by
\begin{equation}
    S_\tau(\omega)\equiv\frac{1}{\tau}\Var(J_\tau[\omega]).
\label{eq:freq.finite.spectrum}
\end{equation}
The stationary covariance of the observable rate is $C_{\dot J}(t)\equiv\Cov(\dot J_t,\dot J_0)$. We define its Fourier transform by
\begin{equation}
    S_{\dot J}(\omega)\equiv\int_{-\infty}^{\infty}e^{-i\omega t}C_{\dot J}(t)\dd{t}.
\label{eq:freq.PSD}
\end{equation}
Stationarity gives the exact finite-window representation
\begin{align}
    S_\tau(\omega)&=\int_{-\tau}^{\tau}\left(1-\frac{\abs{t}}{\tau}\right)e^{-i\omega t}C_{\dot J}(t)\dd{t}\label{eq:freq.triangular}\\
    &=\int_{-\infty}^{\infty}\frac{\dd{\nu}}{2\pi}F_\tau(\omega-\nu)S_{\dot J}(\nu),
    \qquad
    F_\tau(\Omega)\equiv\frac{4\sin^2(\Omega\tau/2)}{\tau\Omega^2}.
\label{eq:freq.Fejer}
\end{align}
Here, $F_\tau(0)=\tau$ is understood by continuity. To verify the second equality, insert the inverse Fourier transform
\begin{equation}
    C_{\dot J}(t)=\int_{-\infty}^{\infty}\frac{\dd{\nu}}{2\pi}e^{i\nu t}S_{\dot J}(\nu).
\end{equation}
After interchanging the order of integration, we find
\begin{equation}
    S_\tau(\omega)=\int_{-\infty}^{\infty}\frac{\dd{\nu}}{2\pi}S_{\dot J}(\nu)\int_{-\tau}^{\tau}\left(1-\frac{\abs{t}}{\tau}\right)e^{-i(\omega-\nu)t}\dd{t}.
\end{equation}
The Fourier transform of the triangular window is
\begin{align}
    \int_{-\tau}^{\tau}\left(1-\frac{\abs{t}}{\tau}\right)e^{-i\Omega t}\dd{t}
    &=2\int_0^\tau\left(1-\frac{t}{\tau}\right)\cos(\Omega t)\dd{t}\notag\\
    &=\frac{2[1-\cos(\Omega\tau)]}{\tau\Omega^2}
    =\frac{4\sin^2(\Omega\tau/2)}{\tau\Omega^2}
    =F_\tau(\Omega),
\end{align}
which proves Eq.~\eqref{eq:freq.Fejer}.
Thus, $S_\tau(\omega)$ is the Fej\'er-smoothed stationary spectrum, with a frequency resolution of order $2\pi/\tau$. Under standard mixing conditions ensuring that the regular part of $C_{\dot J}$ is integrable, the Fej\'er kernels form an approximate identity. Consequently, at every continuity point of the power spectral density,
\begin{equation}
    \lim_{\tau\to\infty}S_\tau(\omega)=S_{\dot J}(\omega).
\end{equation}
When the regular part of the correlation function is absolutely integrable, the same result follows directly by applying dominated convergence to Eq.~\eqref{eq:freq.triangular}.

We next derive an exact representation of the stationary power spectrum.
Let $\Psi_\omega$ denote the centered solution of the resolvent equation
\begin{equation}
    (\Lg^\dagger-i\omega)\Psi_\omega=-(\Phi-\jr_{\st}),
    \qquad
    \ev{\Psi_\omega}_{\st}=0,
\label{eq:freq.resolvent.eq}
\end{equation}
or, equivalently,
\begin{equation}
    \Psi_\omega=\int_0^\infty e^{-i\omega t}e^{t\Lg^\dagger}(\Phi-\jr_{\st})\dd{t}=(i\omega-\Lg^\dagger)^{-1}(\Phi-\jr_{\st}).
\label{eq:freq.resolvent.eq2}
\end{equation}
Define the stationary frequency-domain response kernel by
\begin{equation}
    \bu_\omega\equiv\bw+\nabla_{\bv}\Psi_\omega.
\label{eq:freq.stationary.kernel}
\end{equation}
To derive the spectral identity, introduce the centered process
\begin{equation}
    Y_t\equiv J_t[\omega]-\jr_{\st}\int_0^t e^{-i\omega s}\dd{s}+e^{-i\omega t}\Psi_\omega(\bz_t).
\end{equation}
It\^o's formula and Eq.~\eqref{eq:freq.resolvent.eq} give
\begin{align}
    \dd{Y_t}&=e^{-i\omega t}\qty[(\Lg^\dagger-i\omega)\Psi_\omega+\Phi-\jr_{\st}](\bz_t)\dd{t}+e^{-i\omega t}\sqrt{2\D}\bu_\omega(\bz_t)\cdot\dd{\bW_t}\notag\\
    &=e^{-i\omega t}\sqrt{2\D}\bu_\omega(\bz_t)\cdot\dd{\bW_t}.
\end{align}
Integration over $[0,\tau]$ yields
\begin{equation}
    J_\tau[\omega]-\jr_{\st}\int_0^\tau e^{-i\omega t}\dd{t}=M_\tau[\omega]+B_\tau[\omega],
\label{eq:freq.path.decomp}
\end{equation}
where
\begin{align}
    M_\tau[\omega]&\equiv\int_0^\tau e^{-i\omega t}\sqrt{2\D}\bu_\omega(\bz_t)\cdot\dd{\bW_t},\\
    B_\tau[\omega]&\equiv\Psi_\omega(\bz_0)-e^{-i\omega\tau}\Psi_\omega(\bz_\tau).
\end{align}
Because $\ev{\Psi_\omega}_{\st}=0$, both $M_\tau[\omega]$ and $B_\tau[\omega]$ have zero mean. Hence,
\begin{equation}
	\Var(J_\tau[\omega])=\E\qty[\abs{M_\tau[\omega]}^2]+\E\qty[\abs{B_\tau[\omega]}^2]+2\operatorname{Re}\E\qty[M_\tau[\omega]^*B_\tau[\omega]].
\label{eq:freq.variance.decomp}
\end{equation}
The It\^o isometry and stationarity yield
\begin{equation}
    \E\qty[\abs{M_\tau[\omega]}^2]=2\tau\lVert\bu_\omega\rVert_{\D,p_{\st}}^2.
\label{eq:freq.Mvariance}
\end{equation}
The boundary term remains bounded as $\tau\to\infty$:
\begin{equation}
    \E\qty[\abs{B_\tau[\omega]}^2]\le 4\ev{\abs{\Psi_\omega}^2}_{\st}=O(1),
\label{eq:Bbound}
\end{equation}
and the Cauchy-Schwarz inequality implies
\begin{equation}
    \abs{\E\qty[M_\tau[\omega]^*B_\tau[\omega]]}=O(\sqrt{\tau}).
\end{equation}
After dividing Eq.~\eqref{eq:freq.variance.decomp} by $\tau$, the pure boundary term vanishes as $O(\tau^{-1})$, while the cross term vanishes as $O(\tau^{-1/2})$. Using Eq.~\eqref{eq:freq.finite.spectrum}, we therefore obtain the stationary power spectral density
\begin{equation}
\tcboxmath{
    S_{\dot J}(\omega)=2\lVert\bu_\omega\rVert_{\D,p_{\st}}^2=2\ev{(\bw+\nabla_{\bv}\Psi_\omega)^\dagger\D(\bw+\nabla_{\bv}\Psi_\omega)}_{\st}.
}
\label{eq:freq.spectral.identity}
\end{equation}
At zero frequency, $\Psi_0$ reduces to the stationary Poisson solution used in the main text, and hence
\begin{equation}
    S_{\dot J}(0)=2D_J.
\label{eq:freq.zero.spectrum}
\end{equation}
Under the resolvent regularity assumptions, $\Psi_\omega=O(\abs{\omega}^{-1})$ as $\abs{\omega}\to\infty$. The direct velocity-current term therefore determines the high-frequency white-noise floor:
\begin{equation}
    S_{\dot J}(\omega)\xrightarrow[\abs{\omega}\to\infty]{}2\ev{\bw^\top\D\bw}_{\st}.
\label{eq:freq.white.floor}
\end{equation}

\subsubsection{Recovering the stationary finite-frequency fluctuation-response inequality}
We now show explicitly that the stationary finite-frequency fluctuation-response inequality follows from the stronger finite-time bound in Eq.~\eqref{eq:freq.strong.FRI}:
\begin{equation}
    \Var(J_\tau[\omega])-\Var_0(J_\tau[\omega])\ge 2\int_0^\tau\frac{\abs{r_t^\omega[\bg]}^2}{a_t[\bg]}\dd{t}.
\label{eq:freq.finite.hierarchy.longtime}
\end{equation}
For autonomous dynamics, the finite-time backward solution can be written as
\begin{equation}
	G_s[\omega]=\int_s^\tau e^{-i\omega t}e^{(t-s)\Lg^\dagger}\Phi\dd{t}=e^{-i\omega s}\int_0^{\tau-s}e^{-i\omega t}e^{t\Lg^\dagger}\Phi\dd{t}.
\end{equation}
Introduce the truncated resolvent
\begin{equation}
    \Psi_\omega^{(s)}\equiv\int_0^s e^{-i\omega t}e^{t\Lg^\dagger}(\Phi-\jr_{\st})\dd{t}=e^{i\omega(\tau-s)}G_{\tau-s}[\omega]-\jr_{\st}\int_0^s e^{-i\omega t}\dd{t},
    \qquad 0\le s\le \tau,
\end{equation}
and the associated kernel
\begin{equation}
    \bu_\omega(s)\equiv\bw+\nabla_{\bv}\Psi_\omega^{(s)}.
\end{equation}
Because
\begin{equation}
	\nabla_{\bv}\Psi_\omega^{(s)}=e^{i\omega(\tau-s)}\nabla_{\bv}G_{\tau-s}[\omega],
\end{equation}
the finite-time martingale kernel factorizes as
\begin{equation}
	\bu_t[\omega]=e^{-i\omega t}\bu_\omega(\tau-t).
\end{equation}
Define the truncated susceptibility along the perturbation direction $\bg$ by
\begin{equation}
    \chi_{\bg}^{(s)}(\omega)\equiv\ev{\bu_\omega(s)\cdot\bg}_{\st}.
\end{equation}
The finite-time response density then satisfies the exact relation
\begin{equation}
	r_t^\omega[\bg]=e^{-i\omega t}\chi_{\bg}^{(\tau-t)}(\omega),
	\qquad
	\abs{r_t^\omega[\bg]}^2=\abs{\chi_{\bg}^{(\tau-t)}(\omega)}^2.
\end{equation}
Using the integral representation \eqref{eq:freq.resolvent.eq2},
\begin{equation}
	\Psi_\omega=\int_0^\infty e^{-i\omega t}e^{t\Lg^\dagger}(\Phi-\jr_{\st})\dd{t},
\end{equation}
we find
\begin{align}
    \Psi_\omega-\Psi_\omega^{(s)}&=\int_s^\infty e^{-i\omega t}e^{t\Lg^\dagger}(\Phi-\jr_{\st})\dd{t}\notag\\
    &=e^{-i\omega s}e^{s\Lg^\dagger}\int_0^\infty e^{-i\omega t}e^{t\Lg^\dagger}(\Phi-\jr_{\st})\dd{t}\notag\\
    &=e^{-i\omega s}e^{s\Lg^\dagger}\Psi_\omega.
\end{align}
Hence,
\begin{equation}
    \chi_{\bg}^{(s)}(\omega)=\chi_{\bg}(\omega)-e^{-i\omega s}\ev{\bg\cdot\nabla_{\bv}e^{s\Lg^\dagger}\Psi_\omega}_{\st},
\end{equation}
where $\chi_{\bg}(\omega)\equiv\ev{\bg\cdot\bu_\omega}_{\st}$ is the frequency-dependent susceptibility.
Under the same mixing and regularity assumptions used above, the second term vanishes as $s\to\infty$. Therefore,
\begin{equation}
    \lim_{s\to\infty}\chi_{\bg}^{(s)}(\omega)=\chi_{\bg}(\omega).
\label{eq:truncated.chi.limit}
\end{equation}
For time-independent $\bg$ and $\D$ in the stationary state,
\begin{equation}
	a_t[\bg]=a_{\st}[\bg]\equiv\ev{\bg^\top\D^{-1}\bg}_{\st}.
\end{equation}
For a nonzero perturbation direction with $a_{\st}[\bg]>0$, a change of variables from $t$ to $s=\tau-t$, followed by the Ces\`aro limit implied by Eq.~\eqref{eq:truncated.chi.limit}, gives
\begin{equation}
    \lim_{\tau\to\infty}\frac{1}{\tau}\int_0^\tau\frac{\abs{r_t^\omega[\bg]}^2}{a_t[\bg]}\dd{t} = \frac{1}{a_{\st}[\bg]}\lim_{\tau\to\infty}\frac{1}{\tau}\int_0^\tau\abs{\chi_{\bg}^{(s)}(\omega)}^2\dd{s} = \frac{\abs{\chi_{\bg}(\omega)}^2}{a_{\st}[\bg]}.
\label{eq:strong.term.exact.relation}
\end{equation}
Moreover, the initial-state contribution $\Var_0(J_\tau[\omega])$ remains $O(1)$ under these assumptions. Hence, Eq.~\eqref{eq:freq.finite.spectrum} implies
\begin{equation}
    \lim_{\tau\to\infty}\frac{\Var(J_\tau[\omega])-\Var_0(J_\tau[\omega])}{\tau} = S_{\dot J}(\omega),
\label{eq:dyn.variance.spectrum.limit}
\end{equation}
Combining Eqs.~\eqref{eq:freq.finite.hierarchy.longtime}, \eqref{eq:strong.term.exact.relation}, and \eqref{eq:dyn.variance.spectrum.limit}, we recover the stationary finite-frequency response inequality
\begin{equation}
\tcboxmath{
    \abs{\chi_{\bg}(\omega)}^2\le\frac{1}{2}S_{\dot J}(\omega)a_{\st}[\bg].
}
\label{eq:stationary.finite.frequency.FRI}
\end{equation}
Thus, the stationary finite-frequency response inequality reported in Ref.~\cite{Dechant.2026.PRL} emerges as the long-time projection of the exact finite-time fluctuation-response relation. By retaining the response locally in the perturbation time, Eq.~\eqref{eq:freq.strong.FRI} provides the stronger finite-time statement.

We next provide a physical interpretation of $\chi_{\bg}(\omega)$.
Consider the impulsive velocity-space source current
\begin{equation}
    \delta\bj_{t,\mathrm{src}}^{\bv}(\bz) = \epsilon\delta_+(t)\bg(\bz)p_{\st}(\bz),
\end{equation}
where $\delta_+(t)$ denotes a causal Dirac impulse normalized such that
$\int_0^\infty\delta_+(t)\dd{t}=1$.
The perturbed Fokker-Planck equation reads
\begin{equation}
    \partial_t p_t^\epsilon = \Lg p_t^\epsilon-\nabla_{\bv}\cdot\delta\bj_{t,\mathrm{src}}^{\bv}.
\label{eq:freq.perturbed.FP}
\end{equation}
Writing $p_t^\epsilon=p_{\st}+\delta p_t+O(\epsilon^2)$ and using $\Lg p_{\st}=0$, the linearized equation becomes
\begin{equation}
    \partial_t\delta p_t = \Lg\delta p_t-\epsilon\delta_+(t)\nabla_{\bv}\cdot(\bg p_{\st}).
\label{eq:freq.linearized.FP}
\end{equation}
Before the impulse, $\delta p_{0^-}=0$. Integrating Eq.~\eqref{eq:freq.linearized.FP} across $t=0$ gives the jump condition
\begin{equation}
	\delta p_{0^+}=-\epsilon\nabla_{\bv}\cdot(\bg p_{\st}).
\end{equation}
For $t>0$, the source vanishes and the perturbation evolves under the unperturbed dynamics:
\begin{equation}
    \delta p_t = -\epsilon e^{t\Lg}\nabla_{\bv}\cdot(\bg p_{\st}),
    \qquad t>0.
\label{eq:freq.density.response}
\end{equation}
Equivalently,
\begin{equation}
    \delta p_t=-\epsilon\Theta(t)e^{t\Lg}\nabla_{\bv}\cdot(\bg p_{\st}),
\label{eq:freq.density.response.theta}
\end{equation}
where $\Theta(t)$ denotes the Heaviside step function, $\Theta(t)=0$ for $t<0$ and $\Theta(t)=1$ for $t>0$.
We now compute the corresponding response of the observable rate. For an arbitrary density $p_t$, the mean instantaneous rate associated with $\dd{J_t}=\phi(\bz_t)\dd{t}+\bw(\bz_t)\circ\dd{\bv_t}$ can be expressed in terms of the velocity-space probability current as
\begin{equation}
    \E[\dot J_t] = \int\qty[\phi(\bz)p_t(\bz)+\bw(\bz)\cdot\Jc^{\bv}(p_t)(\bz)]\dd{\bz},
\label{eq:freq.mean.rate.current}
\end{equation}
where $\Jc^{\bv}(p)\equiv(\bF-\gamma\bv)p-\D\nabla_{\bv}p$ is the velocity-space probability current generated by the unperturbed dynamics. In the presence of the source, the first-order variation of the total current is
\begin{equation}
	\delta\bj_t^{\bv}=\Jc^{\bv}(\delta p_t)+\delta\bj_{t,\mathrm{src}}^{\bv}.
\end{equation}
Retaining terms linear in $\epsilon$ therefore gives
\begin{align}
    \delta\E[\dot J_t]&=\int\phi\delta p_t\dd{\bz}+\int\bw\cdot\Jc^{\bv}(\delta p_t)\dd{\bz}+\int\bw\cdot\delta\bj_{t,\mathrm{src}}^{\bv}\dd{\bz}\notag\\
    &=\int\Phi\delta p_t\dd{\bz}+\int\bw\cdot\delta\bj_{t,\mathrm{src}}^{\bv}\dd{\bz}\notag\\
    &=-\epsilon\Theta(t)\int\Phi e^{t\Lg}\nabla_{\bv}\cdot(\bg p_{\st})\dd{\bz}+\epsilon\delta_+(t)\ev{\bw\cdot\bg}_{\st}\notag\\
    &=\epsilon\delta_+(t)\ev{\bw\cdot\bg}_{\st}+\epsilon\Theta(t)\ev{\bg\cdot\nabla_{\bv}e^{t\Lg^\dagger}\Phi}_{\st}.
\label{eq:freq.rate.response.Phi}
\end{align}
In the last equality, integration by parts and adjointness were used. Since the gradient annihilates the constant stationary mean $\jr_{\st}$, defining the causal response kernel by $\delta\E[\dot J_t]=\epsilon r_t[\bg]$ gives
\begin{equation}
    r_t[\bg]=\delta_+(t)\ev{\bw\cdot\bg}_{\st}+\Theta(t)\ev{\bg\cdot\nabla_{\bv}e^{t\Lg^\dagger}(\Phi-\jr_{\st})}_{\st}.
\label{eq:freq.causal.kernel}
\end{equation}
We now evaluate the one-sided Fourier transform of this causal response. By the normalization of $\delta_+$, the impulsive term in Eq.~\eqref{eq:freq.causal.kernel} contributes $\ev{\bw\cdot\bg}_{\st}$. The regular term gives
\begin{align}
    \int_0^\infty e^{-i\omega t}\ev{\bg\cdot\nabla_{\bv}e^{t\Lg^\dagger}(\Phi-\jr_{\st})}_{\st}\dd{t}&=\ev{\bg\cdot\nabla_{\bv}\left[\int_0^\infty e^{-i\omega t}e^{t\Lg^\dagger}\dd{t}\right](\Phi-\jr_{\st})}_{\st}\notag\\
    &=\ev{\bg\cdot\nabla_{\bv}\Psi_\omega}_{\st},
\label{eq:freq.propagated.Fourier}
\end{align}
where Eq.~\eqref{eq:freq.resolvent.eq2} was used. Combining the impulsive and regular contributions, we obtain
\begin{equation}
    \int_0^\infty e^{-i\omega t}r_t[\bg]\dd{t}=\ev{\bg\cdot(\bw+\nabla_{\bv}\Psi_\omega)}_{\st}=\ev{\bg\cdot\bu_\omega}_{\st}=\chi_{\bg}(\omega).
\label{eq:freq.susceptibility.overlap}
\end{equation}
Thus, the frequency-dependent susceptibility is precisely the one-sided Fourier transform of the causal response kernel $r_t[\bg]$.

\subsection{Dissipation-coherence bound for underdamped limit cycles}

We apply the friction-response thermodynamic uncertainty relation (TUR) to the phase of an underdamped stochastic limit cycle. The resulting weak-noise bound relates the entropy produced per oscillation to the product of the number of coherent oscillations and the squared friction sensitivity of the deterministic frequency. This response factor distinguishes the bound from the conventional dissipation-coherence bound based on the mean phase current, derived in Ref.~\cite{Nagayama.2025.arxiv} for overdamped limit cycles.

Below, we define the phase, derive its fluctuations and the friction sensitivity of the frequency, and then evaluate each term in the friction-response TUR over one reference period.

\subsubsection{Dynamics and asymptotic phase}
\label{sm:coh.dynamics}

We consider the autonomous dynamics with weak thermal noise,
\begin{align}
	\dd{\bx_t}&=\bv_t\dd{t},\\
	\dd{\bv_t}&=[\bF(\bx_t)-\gamma\bv_t]\dd{t}+\sqrt{2\D_\gamma^\epsilon}\dd{\bW_t},
\label{eq:coh.dynamics}
\end{align}
where $\D_\gamma^\epsilon=\epsilon\gamma\T$ and $0<\epsilon\ll1$ controls the noise strength.
We fix a reference friction coefficient $\gamma$ and introduce
\begin{equation}
	\bb_\gamma(\bz)\equiv
	\begin{pmatrix}\bv\\ \bF(\bx)-\gamma\bv\end{pmatrix},
	\qquad\D_\gamma\equiv\gamma\T.
\label{eq:coh.reference}
\end{equation}
Assume that the deterministic dynamics $\dot{\bz}=\bb_\gamma(\bz)$ admits a family of stable limit cycles that depends smoothly on the friction coefficient near its reference value, with period $\tau_p(\gamma)$ and angular frequency $\omega_\gamma\equiv 2\pi/\tau_p(\gamma)>0$. Such cycles require a nonconservative force. Here, stability means that small transverse displacements relax back to the cycle. Let $\bz_*(t)$ denote the periodic orbit at the reference friction coefficient, so that
\begin{equation}
	\dot{\bz}_*(t)=\bb_\gamma(\bz_*(t)),\qquad \bz_*(t+\tau_p)=\bz_*(t),\quad t\ge0.
\end{equation}

For $\epsilon>0$, assume that a smooth stationary density $p_\st^\epsilon$ exists at the reference friction coefficient and concentrates on this cycle as $\epsilon\to0$. More precisely, for the functions used below, we assume convergence of the stationary averages to the corresponding time averages along the cycle:
\begin{equation}
	\ev{g}_\epsilon\equiv\int g(\bz)p_\st^\epsilon(\bz)\dd{\bz}\xrightarrow{\epsilon\to0}\overline{g}\equiv\frac{1}{\tau_p}\int_0^{\tau_p}g(\bz_*(t))\dd{t}.
  \label{eq:coh.stationary.limit}
\end{equation}

Let $\varphi_t^\gamma(\bz)$ denote the deterministic flow. The asymptotic phase $\Theta_\gamma(\bz)$, defined modulo $2\pi$, specifies the phase shift of the trajectory on the cycle that the trajectory starting at $\bz$ approaches asymptotically \cite{Nakao.2015.CP}. Choosing $\Theta_\gamma(\bz_*(0))=0$ fixes the phase origin, and the phase is characterized by
\begin{equation}
  \lim_{s\to\infty}\left\|\varphi_s^\gamma(\bz)-\bz_*\qty(s+\frac{\Theta_\gamma(\bz)}{\omega_\gamma})\right\|=0,
\label{eq:coh.phase.definition}
\end{equation}
where $\lVert\cdot\rVert$ denotes the standard Euclidean norm.
The flow property gives $\varphi_s^\gamma(\varphi_t^\gamma(\bz))=\varphi_{s+t}^\gamma(\bz)$.
Together with Eq.~\eqref{eq:coh.phase.definition}, this implies
\begin{equation}
	\Theta_\gamma(\varphi_t^\gamma(\bz))=\Theta_\gamma(\bz)+\omega_\gamma t\pmod{2\pi}.
\label{eq:coh.phase.flow}
\end{equation}
Differentiating this identity with respect to $t$ at $t=0$ gives the phase equation
\begin{equation}
	\bb_\gamma(\bz)\cdot\nabla_{\bz}\Theta_\gamma(\bz)=\bv\cdot\nabla_{\bx}\Theta_\gamma+[\bF(\bx)-\gamma\bv]\cdot\nabla_{\bv}\Theta_\gamma=\omega_\gamma.
\label{eq:coh.phase.pde}
\end{equation}
This equation relates the phase gradient to the full deterministic drift and holds wherever the asymptotic phase is smooth, not only on the cycle. We define the phase sensitivity function along the cycle as
\begin{equation}
	\bZ_t\equiv\nabla_{\bz}\Theta_\gamma(\bz_*(t))
	=\begin{pmatrix}\bZ_{\bx}(t)\\ \bZ_{\bv}(t)\end{pmatrix},
\label{eq:coh.Z}
\end{equation}
which satisfies $\bZ_t^\top\dot{\bz}_*(t)=\omega_\gamma$ and $\bZ_{t+\tau_p}=\bZ_t$.
Let $\M_t\equiv\nabla_{\bz}\bb_{\gamma}(\bz_*(t))$ denote the Jacobian of the deterministic drift along the cycle. Taking the gradient of Eq.~\eqref{eq:coh.phase.pde} and evaluating it at $\bz_*(t)$, with $\dot{\bZ}_t=(\nabla_{\bz}^2\Theta_\gamma)(\bz_*(t))\dot{\bz}_*(t)$, yields the adjoint equation
\begin{equation}
	\dot{\bZ}_t=-\M_t^\top\bZ_t.
\label{eq:coh.adjoint}
\end{equation}
In components, $\dot{\bZ}_{\bx}=-(\nabla_{\bx}\bF)^\top\bZ_{\bv}$ and $\dot{\bZ}_{\bv}=-\bZ_{\bx}+\gamma\bZ_{\bv}$, where the force Jacobian is evaluated along the cycle.

\subsubsection{The number of coherent oscillations}
\label{sm:coh.coherence}

We consider the unwrapped phase process $\Theta_t\equiv\operatorname{unwrap}[\Theta_\gamma(\bz_t)]$, which tracks the accumulated phase continuously along a noisy trajectory. Unwrapping retains the number of completed turns rather than reducing the phase modulo $2\pi$. Applying It\^o's formula at the reference parameters gives
\begin{align}
	\dd{\Theta_t}&=[\omega_\gamma+\epsilon c(\bz_t)]\dd{t}+\nabla_{\bv}\Theta_\gamma(\bz_t)^\top\sqrt{2\epsilon\D_\gamma}\dd{\bW_t},
\label{eq:coh.phase.ito}\\
	c(\bz)&\equiv \D_\gamma:\nabla_{\bv}^2\Theta_\gamma(\bz).
\label{eq:coh.c}
\end{align}
By Eq.~\eqref{eq:coh.phase.pde}, the leading-order drift is exactly $\omega_\gamma$, while $\epsilon c$ is the It\^o correction. Both this correction and the noise amplitude depend on the transverse coordinates. Under the weak-noise assumptions, transverse deviations remain small after relaxation. We approximate these coefficients by evaluating them at the point on the deterministic cycle with the same phase and denote the resulting reduced phase process by $\theta_t$. Define the $2\pi$-periodic functions
\begin{equation}
	c_*(\theta)\equiv c\qty(\bz_*(\theta/\omega_\gamma)),\qquad
	q_*(\theta)\equiv \bZ_{\bv}(\theta/\omega_\gamma)^\top\D_\gamma\bZ_{\bv}(\theta/\omega_\gamma).
\label{eq:coh.phase.coefficients}
\end{equation}
The reduced phase process is then defined by the closed It\^o equation
\begin{equation}
	\dd{\theta_t}=[\omega_\gamma+\epsilon c_*(\theta_t)]\dd{t}+\sqrt{2\epsilon q_*(\theta_t)}\dd{B_t},
\end{equation}
where $B_t$ is a scalar Wiener process.
The reduced backward generator through first order in $\epsilon$ is
\begin{equation}
	\mathcal L_\epsilon^\dagger\equiv\omega_\gamma\partial_\theta+\epsilon\qty[c_*(\theta)\partial_\theta+q_*(\theta)\partial_\theta^2].
\label{eq:coh.phase.generator}
\end{equation}
Introduce the cycle-integrated phase diffusion coefficient and the cycle averages
\begin{equation}
	\mathcal{V}\equiv\int_0^{\tau_p}\bZ_{\bv}(t)^\top\D_\gamma\bZ_{\bv}(t)\dd{t},\qquad
	\bar q\equiv\frac{\mathcal{V}}{\tau_p},\qquad
	\bar c\equiv\frac{1}{\tau_p}\int_0^{\tau_p}c(\bz_*(t))\dd{t},
\label{eq:coh.B}
\end{equation}
and assume $\mathcal{V}>0$.

To determine the decay of oscillatory correlations, we analyze the eigenvalues of the reduced backward generator using the Fourier modes $e^{in\theta}$, with $n\in\mathbb Z$. At zero noise, $\mathcal L_0^\dagger e^{in\theta}=in\omega_\gamma e^{in\theta}$. For each fixed $n$, we substitute an eigenfunction $e^{in\theta}+\epsilon u_n(\theta)+o(\epsilon)$ and an eigenvalue $\lambda_n^\epsilon=in\omega_\gamma+\epsilon\lambda_n^{(1)}+o(\epsilon)$ into the eigenvalue equation. We then multiply the terms of order $\epsilon$ by $e^{-in\theta}$ and integrate over the circle. The eigenfunction correction drops out because $\int_0^{2\pi}e^{-in\theta}(\partial_\theta-in)u_n(\theta)\dd{\theta}=0$ by periodicity and integration by parts. Thus,
\begin{align}
	\lambda_n^\epsilon&=in\omega_\gamma+
    \frac{\epsilon}{2\pi}\int_0^{2\pi}e^{-in\theta}[c_*\partial_\theta+q_*\partial_\theta^2]e^{in\theta}\dd{\theta}+o(\epsilon)\notag\\
	&=in\omega_\gamma+\epsilon(in\bar c-n^2\bar q)+o(\epsilon).
\label{eq:coh.eigenvalues}
\end{align}
To leading order in the weak-noise expansion, the slowest-decaying nonzero phase modes correspond to the fundamental pair $n=\pm1$, whose eigenvalues take the form
\begin{equation}
	\lambda_{\pm1}^\epsilon=-r_\epsilon\pm i\Omega_\epsilon,\qquad
	r_\epsilon=\epsilon\bar q +o(\epsilon),\qquad
	\Omega_\epsilon=\omega_\gamma+\epsilon\bar c+o(\epsilon).
\label{eq:coh.rates}
\end{equation}
For a real observable $A$ with a nonzero contribution from the fundamental phase modes, the stationary connected correlation $C_A(t)=\langle A(\bz_t)A(\bz_0)\rangle_\epsilon-\langle A\rangle_\epsilon^2$ can be approximated, when this pair dominates, by
\begin{equation}
	C_A(t)\simeq\operatorname{Re}\left[C_1e^{(-r_\epsilon+i\Omega_\epsilon)t}\right].
\label{eq:coh.correlation}
\end{equation}
Here, $C_1$ is an observable-dependent complex amplitude. This approximation describes phase coherence at times when transverse transients and higher phase harmonics are negligible; exceeding the transverse relaxation time alone does not ensure that the fundamental pair dominates. The correlation envelope decays to $1/e$ of its initial value over the coherence time $t_{\rm coh}=1/r_\epsilon$, while the oscillation period is $\tau_{\rm osc}=2\pi/\Omega_\epsilon$. With this convention, the number of coherent oscillations is
\begin{equation}
	\mathcal N_\epsilon\equiv\frac{t_{\rm coh}}{\tau_{\rm osc}}=\frac{\Omega_\epsilon}{2\pi r_\epsilon}=\frac{1}{\epsilon\mathcal{V}}+o(\epsilon^{-1}).
\label{eq:coh.N}
\end{equation}

\subsubsection{Frequency sensitivity}
\label{sm:coh.sensitivity}

Both the limit cycle and its frequency depend on the friction coefficient $\gamma$. Choosing the phase origin smoothly as a function of $\gamma$, we define
\begin{equation}
	h(\bz)\equiv\partial_\gamma\Theta_\gamma(\bz),\qquad
	\omega_\gamma'\equiv\dv{\omega_\gamma}{\gamma}.
\label{eq:coh.h}
\end{equation}
Here, the partial derivative defining $h$ is taken at fixed $\bz$, with the force field $\bF$ held fixed. Since $\partial_\gamma\bb_\gamma=(\vb*{0},-\bv)$, differentiating Eq.~\eqref{eq:coh.phase.pde} with respect to $\gamma$ gives the exact deterministic identity
\begin{equation}
	\omega_\gamma'=\bb_{\gamma}(\bz)\cdot\nabla_{\bz}h(\bz)-\bv\cdot\nabla_{\bv}\Theta_\gamma(\bz).
\label{eq:coh.sensitivity.local}
\end{equation}
Along the reference limit cycle, $\bb_\gamma\cdot\nabla_{\bz}h$ is the total time derivative of $h(\bz_*(t))$. Hence,
\begin{equation}
	\omega_\gamma'=\dv{t}h(\bz_*(t))-\bZ_{\bv}(t)^\top\bv_*(t).
\label{eq:coh.sensitivity.cycle}
\end{equation}
Because $h(\bz_*(t))$ is periodic with period $\tau_p$, its total derivative integrates to zero over one cycle. Averaging Eq.~\eqref{eq:coh.sensitivity.cycle} over the reference period and multiplying by $\gamma$, we obtain
\begin{align}
	\gamma\omega_\gamma'&=\frac{1}{\tau_p}\int_0^{\tau_p}\bZ_{\bv}(t)^\top[-\gamma\bv_*(t)]\dd{t}.
\label{eq:coh.frequency.response}
\end{align}
We define the dimensionless friction sensitivity of the frequency as
\begin{equation}
	\chi_\gamma\equiv\gamma\partial_\gamma\ln\omega_\gamma=\frac{\gamma\omega_\gamma'}{\omega_\gamma}=\frac{1}{2\pi}\int_0^{\tau_p}\bZ_{\bv}(t)^\top[-\gamma\bv_*(t)]\dd{t},
\label{eq:coh.chi}
\end{equation}
where the last equality follows from $\omega_\gamma\tau_p=2\pi$.

\subsubsection{Coherence-dissipation bound from the friction-response TUR}
\label{sm:coh.tur.evaluation}

Recall that for a fixed time-integrated observable
\begin{equation}
	J_\tau=\int_0^\tau[\phi(\bz_t)\dd{t}+\bw(\bz_t)\circ\dd{\bv_t}],
\label{eq:coh.additive}
\end{equation}
the exact friction-response TUR reads
\begin{equation}
	[\Var(J_\tau)-\Var_0(J_\tau)]\Sigma_\tau\geq 2\Rc_\epsilon(\tau)^2,\qquad
	\Rc_\epsilon(\tau)\equiv\left.\partial_\delta\E_\delta[J_\tau]\right|_{\delta=0},
\label{eq:coh.TUR}
\end{equation}
where the perturbation rescales the friction as $\gamma\to(1+\delta)\gamma$. The entropy production is evaluated for the reference stationary process. The response in Eq.~\eqref{eq:coh.TUR} is $\gamma\partial_\gamma\E_\gamma[J_\tau]$, with the temperature, noise-strength parameter $\epsilon$, initial density, observable fields, and observation time held fixed. We evaluate the terms in this bound using the underlying exact variance and response representations:
\begin{align}
	\Var(J_\tau)-\Var_0(J_\tau)&=2\epsilon\int_0^\tau\ev{\bu_t^\top\D_\gamma\bu_t}_t\dd{t},
\label{eq:coh.variance.identity}\\
	\Rc_\epsilon(\tau)&=\int_0^\tau\ev{\bu_t^\top\bnu_t^{\rm irr}}_t\dd{t}.
\label{eq:coh.response.identity}
\end{align}

We choose the increment of the reference phase as the observable over the fixed observation interval $\tau=\tau_p$:
\begin{equation}
	J_{\tau_p}=\Theta_{\tau_p}-\Theta_0=\int_0^{\tau_p}\qty[\nabla_{\bx}\Theta_\gamma(\bz_t)\cdot\bv_t\dd{t}+\nabla_{\bv}\Theta_\gamma(\bz_t)\circ\dd{\bv_t}].
  \label{sm:coh:observable}
\end{equation}
Thus, $\phi=\nabla_{\bx}\Theta_\gamma\cdot\bv$ and $\bw=\nabla_{\bv}\Theta_\gamma$. These fields, as well as $\tau_p$, remain fixed at their reference values when the friction is perturbed. All averages below are evaluated for the reference process initialized in $p_\st^\epsilon$.

First, we evaluate the dynamical variance. Integrating Eq.~\eqref{eq:coh.phase.ito} from $t$ to $\tau_p$ and conditioning on $\bz_t=\bz$ gives
\begin{align}
	G_t(\bz)&\equiv\E[\Theta_{\tau_p}-\Theta_t\mid\bz_t=\bz]=\omega_\gamma(\tau_p-t)+\epsilon K_t(\bz),
\label{eq:coh.G}\\
	K_t(\bz)&\equiv\E\qty[\int_t^{\tau_p}c(\bz_s)\dd{s}\big\mid\bz_t=\bz],\qquad \bu_t=\nabla_{\bv}\Theta_\gamma+\epsilon\nabla_{\bv}K_t.
\label{eq:coh.kernel}
\end{align}
Although not indicated explicitly, $K_t$ depends on $\epsilon$ through the reference dynamics. For the following expansions, we assume that the stationary moments involving $\nabla_{\bv}K_t$ and $\nabla_{\bv}^2K_t$ that appear below remain bounded uniformly for $t\in[0,\tau_p]$ as $\epsilon\to0$, and that all boundary terms in the integrations by parts vanish. Substituting Eq.~\eqref{eq:coh.kernel} into the exact variance identity, Eq.~\eqref{eq:coh.variance.identity}, yields
\begin{align}
	\Var(J_{\tau_p})-\Var_0(J_{\tau_p})&=2\epsilon\int_0^{\tau_p}\ev{(\nabla_{\bv}\Theta_\gamma+\epsilon\nabla_{\bv}K_t)^\top\D_\gamma(\nabla_{\bv}\Theta_\gamma+\epsilon\nabla_{\bv}K_t)}_t\dd{t}\notag\\
	&=2\epsilon\tau_p\langle\nabla_{\bv}\Theta_\gamma^\top \D_\gamma\nabla_{\bv}\Theta_\gamma\rangle_\epsilon+O(\epsilon^2)\notag\\
	&=2\epsilon\mathcal{V}+o(\epsilon),
\label{eq:coh.dynamic.variance}
\end{align}
where stationarity is used in the second equality and Eq.~\eqref{eq:coh.stationary.limit} in the last.

Next, we evaluate the friction response. Define $a_\Theta(\bz)\equiv-\gamma\bv\cdot\nabla_{\bv}\Theta_\gamma(\bz)$. Substituting Eq.~\eqref{eq:coh.kernel} into Eq.~\eqref{eq:coh.response.identity} and integrating the terms containing $\nabla_{\bv}\ln p_\st^\epsilon$ by parts yields
\begin{align}
	\Rc_\epsilon(\tau_p)&=\int_0^{\tau_p}\ev{(\nabla_{\bv}\Theta_\gamma+\epsilon\nabla_{\bv}K_t)^\top(-\gamma\bv-\D_\gamma^\epsilon\nabla_{\bv}\ln p_\st^\epsilon)}_\epsilon\dd{t}\notag\\
	&=\tau_p\ev{a_\Theta}_\epsilon+\epsilon\tau_p\ev{c}_\epsilon-\epsilon\gamma\int_0^{\tau_p}\ev{\bv\cdot\nabla_{\bv}K_t}_\epsilon\dd{t}
	+\epsilon^2\int_0^{\tau_p}\ev{\D_\gamma:\nabla_{\bv}^2K_t}_\epsilon\dd{t}.
\label{eq:coh.perturbed.phase}
\end{align}
Under the stated boundedness assumptions, the last three terms vanish as $\epsilon\to0$. Using Eqs.~\eqref{eq:coh.stationary.limit} and \eqref{eq:coh.chi}, we therefore obtain
\begin{equation}
	\Rc_\epsilon(\tau_p)=\int_0^{\tau_p}a_\Theta(\bz_*(t))\dd{t}+o(1)=\int_0^{\tau_p}\bZ_{\bv}(t)^\top[-\gamma\bv_*(t)]\dd{t}+o(1)=2\pi\chi_\gamma+o(1).
\label{eq:coh.response.evaluated}
\end{equation}

Finally, we evaluate the entropy production. Let $\bs\equiv\nabla_{\bv}\ln p_\st^\epsilon$. Expanding the stationary entropy production rate and using $\ev{\bv\cdot\bs}_\epsilon=-d$, obtained by integration by parts, gives
\begin{align}
	\sigma_\epsilon&=\frac{\gamma^2}{\epsilon}
       \ev{\bv^\top\D_\gamma^{-1}\bv}_\epsilon+2\gamma\ev{\bv\cdot\bs}_\epsilon+\epsilon\ev{\bs^\top\D_\gamma\bs}_\epsilon\notag\\
	&=\frac{\gamma^2}{\epsilon}\ev{\bv^\top\D_\gamma^{-1}\bv}_\epsilon-2\gamma d+\epsilon I_{\bv}[p_\st^\epsilon],
\label{eq:coh.EP.expanded}\\
	I_{\bv}[p]&\equiv\int(\nabla_{\bv}\ln p)^\top\D_\gamma(\nabla_{\bv}\ln p)p\dd{\bz}.
\label{eq:coh.Fisher}
\end{align}
The weighted velocity-space Fisher information can be evaluated exactly in the stationary state. For the Shannon entropy $S[p_t]\equiv-\int p_t\ln p_t\dd{\bz}$, the Fokker-Planck equation gives
\begin{equation}
	\dv{S[p_t]}{t}=\ev{\nabla_{\bz}\cdot\bb_{\gamma}}_t+\epsilon I_{\bv}[p_t]=-\gamma d+\epsilon I_{\bv}[p_t].
\label{eq:coh.entropy.balance}
\end{equation}
Stationarity therefore implies $\epsilon I_{\bv}[p_\st^\epsilon]=\gamma d$. Substitution into Eq.~\eqref{eq:coh.EP.expanded} yields
\begin{equation}
	\sigma_\epsilon=\frac{\gamma^2}{\epsilon}\ev{\bv^\top\D_\gamma^{-1}\bv}_\epsilon-\gamma d.
\label{eq:coh.EP.exact}
\end{equation}
Define the dissipative cycle integral
\begin{equation}
	\bA_t\equiv -\gamma\bv_*(t),\qquad
	\mathcal{E}\equiv\int_0^{\tau_p}\bA_t^\top\D_\gamma^{-1}\bA_t\dd{t}.
\label{eq:coh.E}
\end{equation}
Equations~\eqref{eq:coh.stationary.limit} and \eqref{eq:coh.EP.exact} give the entropy produced over one reference period:
\begin{equation}
	\Sigma_{\tau_p}=\tau_p\sigma_\epsilon=\frac{\mathcal{E}}{\epsilon}+o(\epsilon^{-1}).
  \label{sm:coh:EP-asymptotic}
\end{equation}
The entropy produced per observed oscillation is
\begin{equation}
	\Sigma_{\rm cyc}^\epsilon\equiv\frac{2\pi}{\Omega_\epsilon}\sigma_\epsilon=\frac{\mathcal{E}}{\epsilon}+o(\epsilon^{-1}),
\label{eq:coh.EP.per.cycle}
\end{equation}
since $2\pi/\Omega_\epsilon=\tau_p+O(\epsilon)$.
Substituting the asymptotic expressions for the response, dynamical variance, and entropy production into the friction-response TUR gives
\begin{equation}
	[2\epsilon\mathcal{V}+o(\epsilon)][\mathcal{E}/\epsilon+o(\epsilon^{-1})]\geq 2[2\pi\chi_\gamma+o(1)]^2.
\label{eq:coh.TUR.insertion}
\end{equation}
Taking $\epsilon\to0$ with the reference oscillator fixed, we obtain
\begin{equation}
	\mathcal{E}\mathcal{V}\geq 4\pi^2\chi_\gamma^2.
\label{eq:coh.VE.bound}
\end{equation}
Combining this inequality with Eqs.~\eqref{eq:coh.N} and \eqref{eq:coh.EP.per.cycle} establishes the weak-noise coherence-dissipation bound:
\begin{equation}
\tcboxmath{
	\frac{\Sigma_{\rm cyc}^\epsilon}{\mathcal N_\epsilon}\geq 4\pi^2\qty(\gamma\partial_\gamma\ln\omega_\gamma)^2.
}
\label{eq:coh.main.bound}
\end{equation}
This result generalizes the existing coherence-dissipation bound for overdamped limit cycles \cite{Santolin.2025.PRL,Nagayama.2025.arxiv,Kolchinsky.2026.PRE}, which is recovered in the overdamped limit, where $\gamma\partial_\gamma\ln\omega_\gamma\to-1$.

\subsubsection{An exactly saturating example}
\label{sm:coh.example}
We demonstrate exact saturation of the coherence-dissipation bound using a driven rotor. Let $\alpha$ be an angle defined modulo $2\pi$, and consider
\begin{equation}
	\dd{\alpha_t}=v_t\dd{t},\qquad
	\dd{v_t}=(F-\gamma v_t)\dd{t}+\sqrt{2\epsilon\gamma T}\dd{W_t}.
\label{eq:coh.rotor}
\end{equation}
where $F>0$, $\gamma>0$, $T>0$, and $\epsilon>0$. The stable deterministic cycle has constant velocity $v_*=F/\gamma$ and angular frequency $\omega_\gamma=F/\gamma$. Up to an additive phase-origin constant, the asymptotic phase and its sensitivities are
\begin{equation}
	\Theta_\gamma(\alpha_t,v_t)=\alpha_t+\frac{v_t}{\gamma}\pmod{2\pi},\qquad
	\bb_\gamma\cdot\nabla_{\bz}\Theta_\gamma=\frac F\gamma,\qquad
	Z_v=\frac{1}{\gamma},\qquad \chi_\gamma=-1.
\label{eq:coh.rotor.phase}
\end{equation}
The reference stationary density is uniform in $\alpha$ and Gaussian in $v$, with mean $F/\gamma$ and variance $\epsilon T$. The corresponding irreversible local mean velocity is $\nu^{\rm irr}=-F$, and the It\^o correction vanishes, $c=0$. Under the reference dynamics, the unwrapped phase increment over any fixed observation time $\tau$ is exactly
\begin{equation}
	J_\tau=\frac{F\tau}{\gamma}+\sqrt{\frac{2\epsilon T}{\gamma}}(W_\tau-W_0).
\label{eq:coh.rotor.J}
\end{equation}
Since the conditional mean of this increment is independent of the initial state, $\Var_0(J_\tau)=0$. In evaluating the friction response, we hold the initial density, the reference phase field, and $\tau$ fixed. Because $c=0$, we have $K_t=0$, and the response follows from $a_\Theta=-v$ and $\ev{v}_\epsilon=F/\gamma$. Thus, the terms in the friction-response TUR are
\begin{equation}
	\Var(J_\tau)-\Var_0(J_\tau)=\frac{2\epsilon T\tau}{\gamma},\qquad
	\Rc_\epsilon(\tau)=-\frac{F\tau}{\gamma},\qquad
	\Sigma_\tau=\frac{F^2\tau}{\epsilon\gamma T}.
\label{eq:coh.rotor.terms}
\end{equation}
These expressions satisfy $[\Var(J_\tau)-\Var_0(J_\tau)]\Sigma_\tau=2\Rc_\epsilon(\tau)^2$, so the finite-time friction-response TUR is saturated. The decay rate and oscillation frequency are $r_\epsilon=\epsilon T/\gamma$ and $\Omega_\epsilon=F/\gamma$, respectively. It follows that
\begin{equation}
	\mathcal N_\epsilon=\frac{F}{2\pi\epsilon T},\qquad
	\Sigma_{\rm cyc}^\epsilon=\frac{2\pi F}{\epsilon T}=4\pi^2\mathcal N_\epsilon.
\label{eq:coh.rotor.saturation}
\end{equation}
For this example, $\chi_\gamma^2=1$, and both the friction-response TUR and the coherence-dissipation bound are saturated at every positive noise strength. The general bound in Eq.~\eqref{eq:coh.main.bound} is established only in the weak-noise limit.

\subsubsection{A limit cycle with enhanced friction sensitivity}
\label{sm:coh.enhanced}

The magnitude of the friction sensitivity $\chi_\gamma=\gamma\partial_\gamma\ln\omega_\gamma$ can exceed unity, yielding a coherence-dissipation prefactor larger than the overdamped value $4\pi^2$. We demonstrate this using an explicit two-dimensional oscillator whose deterministic frequency and friction sensitivity can be evaluated analytically.

In dimensionless units, consider
\begin{align}
    \dd{\bx_t}&=\bv_t\dd{t},\notag\\
    \dd{\bv_t}&=[\bF(\bx_t)-\gamma\bv_t]\dd{t}+\sqrt{2\epsilon\gamma T}\,\dd{\bW_t},
\label{eq:coh.enhanced.dynamics}
\end{align}
where $\bx,\bv\in\mathbb R^2$ and
\begin{equation}
	\bF(\bx)=(1-\|\bx\|^4)\bx+\|\bx\|^2\mathsf{J}\bx,\qquad
    \mathsf J=
    \begin{pmatrix}
        0&-1\\
        1&0
    \end{pmatrix}.
\label{eq:coh.enhanced.force}
\end{equation}
The first term is the conservative force generated by the radial potential $U(r)=r^6/6-r^2/2$, with $r=\|\bx\|$, whereas the second term provides a nonconservative rotational drive. The force field is independent of $\gamma$ and is held fixed when evaluating the friction response.

We first derive the deterministic equations in polar coordinates. Setting $\epsilon=0$ in Eq.~\eqref{eq:coh.enhanced.dynamics} gives
\begin{equation}
    \dot{\bx}=\bv,\qquad
    \dot{\bv}=(1-\|\bx\|^4)\bx+\|\bx\|^2\mathsf{J}\bx-\gamma\bv.
\end{equation}
For $r=\|\bx\|>0$, introduce the polar angle $\alpha$ and the orthonormal basis vectors
\begin{equation}
    \vb*{e}_r=
    \begin{pmatrix}
        \cos\alpha\\
        \sin\alpha
    \end{pmatrix},
    \qquad
    \vb*{e}_\alpha=
    \begin{pmatrix}
        -\sin\alpha\\
        \cos\alpha
    \end{pmatrix}
    =\mathsf J\vb*{e}_r.
\label{eq:coh.enhanced.polar.basis}
\end{equation}
Writing $\bx=r\vb*{e}_r$, we define the radial and angular velocities by
\begin{equation}
    u=\dot r,\qquad w=\dot\alpha.
\label{eq:coh.enhanced.polar.definitions}
\end{equation}
Because the basis vectors depend on $\alpha$, their time derivatives are
\begin{equation}
    \dot{\vb*{e}}_r=w\vb*{e}_\alpha,
    \qquad
    \dot{\vb*{e}}_\alpha=-w\vb*{e}_r.
\label{eq:coh.enhanced.polar.basis.derivatives}
\end{equation}
Differentiating $\bx=r\vb*{e}_r$ therefore gives
\begin{equation}
    \bv=\dot r \vb*{e}_r+r\dot{\vb*{e}}_r=u\vb*{e}_r+rw\vb*{e}_\alpha.
\label{eq:coh.enhanced.polar.velocity}
\end{equation}
Taking the time derivative once more yields
\begin{align}
    \dot{\bv}&=\dot u\vb*{e}_r+u\dot{\vb*{e}}_r+\frac{\dd(rw)}{\dd{t}}\vb*{e}_\alpha+rw\dot{\vb*{e}}_\alpha\notag\\
    &=\dot u\vb*{e}_r+uw\vb*{e}_\alpha+(\dot r w+r\dot w)\vb*{e}_\alpha-rw^2\vb*{e}_r\notag\\
    &=(\dot u-rw^2)\vb*{e}_r+(2uw+r\dot w)\vb*{e}_\alpha,
\label{eq:coh.enhanced.polar.acceleration}
\end{align}
where $\dot r=u$ was used in the last equality.
We next express the force in this basis. Since $\mathsf{J}\bx=r\vb*{e}_\alpha$, we have
\begin{equation}
    \bF(\bx)=r(1-r^4)\vb*{e}_r+r^3\vb*{e}_\alpha,
\label{eq:coh.enhanced.polar.force}
\end{equation}
while the friction force becomes
\begin{equation}
    -\gamma\bv=-\gamma u\vb*{e}_r-\gamma rw\vb*{e}_\alpha.
\end{equation}
Substituting these expressions into $\dot{\bv}=\bF-\gamma\bv$ gives
\begin{equation}
    (\dot u-rw^2)\vb*{e}_r+(2uw+r\dot w)\vb*{e}_\alpha=[r(1-r^4)-\gamma u]\vb*{e}_r+[r^3-\gamma rw]\vb*{e}_\alpha.
\end{equation}
Equating the radial and tangential components yields
\begin{align}
    \dot u-rw^2&=r(1-r^4)-\gamma u,\notag\\
    2uw+r\dot w&=r^3-\gamma rw.
\end{align}
Rearranging the first equation and dividing the second by $r>0$, we obtain
\begin{equation}
\label{eq:coh.enhanced.polar}
\begin{aligned}
    \dot r&=u,\\
    \dot u&=r(w^2+1-r^4)-\gamma u,\\
    \dot w&=r^2-\gamma w-\frac{2uw}{r},\\
    \dot\alpha&=w.
\end{aligned}
\end{equation}
For each fixed $\gamma>1$, we seek a circular orbit of nonzero radius and constant angular velocity $\omega_\gamma$:
\begin{equation}
    r_t=R_\gamma>0,
    \qquad
    \alpha_t=\omega_\gamma t+\alpha_0,
\end{equation}
where $\alpha_0$ is an arbitrary initial angle. The radius $R_\gamma$ and frequency $\omega_\gamma$ depend on $\gamma$ but not on time. Consequently,
\begin{equation}
    u=\dot r=0,\qquad
    \dot u=0,\qquad
    w=\dot\alpha=\omega_\gamma,\qquad
    \dot w=0.
\end{equation}
Substituting these conditions into Eq.~\eqref{eq:coh.enhanced.polar} gives
\begin{align}
    0&=R_\gamma(\omega_\gamma^2+1-R_\gamma^4),\\
    0&=R_\gamma^2-\gamma\omega_\gamma.
\end{align}
Because $R_\gamma>0$, these equations imply $R_\gamma^4=1+\omega_\gamma^2$ and $R_\gamma^2=\gamma\omega_\gamma$. For $\gamma>1$, their unique positive solution is
\begin{equation}
    \omega_\gamma=\frac{1}{\sqrt{\gamma^2-1}},
    \qquad
    R_\gamma=\qty(\frac{\gamma^2}{\gamma^2-1})^{1/4},
    \qquad
    R_\gamma^2=\frac{\gamma}{\sqrt{\gamma^2-1}}.
\label{eq:coh.enhanced.cycle}
\end{equation}
To establish stability in the three transverse variables, define $\vb*{\eta}\equiv(r-R_\gamma,u,w-\omega_\gamma)^\top$.
Linearizing Eq.~\eqref{eq:coh.enhanced.polar} gives
\begin{equation}
    \dot{\vb*{\eta}}=\mathsf{M}_\gamma\vb*{\eta}+O(\|\vb*{\eta}\|^2),
    \qquad
    \mathsf{M}_\gamma\equiv
    \begin{pmatrix}
        0&1&0\\
        -4R_\gamma^4&-\gamma&2R_\gamma\omega_\gamma\\
        2R_\gamma&-2\omega_\gamma/R_\gamma&-\gamma
    \end{pmatrix}.
\label{eq:coh.enhanced.stability.matrix}
\end{equation}
Using $R_\gamma^2=\gamma\omega_\gamma$ and $R_\gamma^4=1+\omega_\gamma^2$, the characteristic polynomial of $\mathsf{M}_\gamma$ reduces to
\begin{equation}
    \det(\lambda\mathds{I} -\mathsf M_\gamma)=\lambda^3+2\gamma\lambda^2+(\gamma^2+4+8\omega_\gamma^2)\lambda+4\gamma.
\label{eq:coh.enhanced.characteristic}
\end{equation}
All coefficients are positive, and
\begin{equation}
    2\gamma(\gamma^2+4+8\omega_\gamma^2)-4\gamma=2\gamma(\gamma^2+2+8\omega_\gamma^2)>0.
\end{equation}
For a cubic polynomial $\lambda^3+a_1\lambda^2+a_2\lambda+a_3$, the Routh-Hurwitz conditions are $a_1,a_2,a_3>0$ and $a_1a_2>a_3$. They are satisfied here, so all three transverse eigenvalues of $\mathsf M_\gamma$ have negative real parts. The circular orbit is therefore a stable limit cycle.

Differentiating the exact frequency in Eq.~\eqref{eq:coh.enhanced.cycle} yields
\begin{equation}
	\chi_\gamma=\gamma\partial_\gamma\ln\omega_\gamma=-\frac{\gamma^2}{\gamma^2-1}.
\label{eq:coh.enhanced.chi}
\end{equation}
Thus, $|\chi_\gamma|>1$ for every $\gamma>1$. Moreover, the sensitivity is unbounded within this family: $|\chi_\gamma|\to\infty$ as $\gamma\to1^+$.

\section{Strong violation of TUR for free diffusion}

Consider a one-dimensional underdamped Brownian particle of unit mass ($m=1$), moving in a potential $U(x)$ and coupled to a thermal reservoir at temperature $T$.
With $k_B=1$, the dynamics is governed by the Langevin equations
\begin{align}
	\dd{x_t}&=v_t\dd{t},\notag\\
	\dd{v_t}&=[-\gamma v_t+f(x_t)]\dd{t}+\sqrt{2\gamma T}\dd{W_t},
\end{align}
where $x_t$ and $v_t$ are the position and velocity at time $t$, respectively, and $\gamma$ is the friction coefficient.
The total force is
\begin{equation}
	f(x)=-\partial_xU(x)+f_{\mathrm{ext}}(x),
\end{equation}
where $f_{\mathrm{ext}}$ denotes an externally applied force.
We assume that the particle moves on a ring of circumference $L>0$ and that the potential is periodic, $U(x+L)=U(x)$.
In the following, we specialize to a flat potential, $U(x)=U_0$, and a constant external force, $f_{\mathrm{ext}}(x)=f$.

We assume that the system reaches a nonequilibrium steady state under these dynamics.
Consider the time-integrated current accumulated along a stochastic trajectory over the interval $[0,\tau]$:
\begin{equation}
	J_\tau=\int_0^\tau \psi(v_t)\circ\dd{x_t},
\end{equation}
where $\circ$ denotes the Stratonovich product and the weight satisfies $\psi(-v)=\psi(v)$.
Because $\dd{x_t}=v_t\dd{t}$, this time-antisymmetric current can equivalently be written as
\begin{equation}
	J_\tau=\int_0^\tau\Phi(v_t)\dd{t},
\end{equation}
where $\Phi(v) = v\psi(v)$.
The steady-state probability density is uniform in position and Gaussian in velocity:
\begin{equation}
	p_{\st}(x,v)=\frac{1}{L}p_{\st}(v),
\end{equation}
where
\begin{equation}
	p_{\st}(v)=\frac{1}{\sqrt{2\pi T}}\exp\qty[-\frac{(v-\vbar)^2}{2T}].
\end{equation}
Here, $\vbar\equiv\ev{v}_{\st}=f/\gamma$ denotes the mean velocity. For convenience, introduce the dimensionless driving strength
\begin{equation}
	\kappa\equiv\frac{\vbar^2}{T}\ge0.
\end{equation}

Let $u\equiv v-\vbar$ and $z\equiv u/\sqrt{T}$.
Using the probabilists' Hermite polynomials $\He_n$,
with $\He_0=1$, $\He_1=z$, $\He_2=z^2-1$, and $\He_3=z^3-3z$, define
\begin{equation}
	\varphi_n(z)\equiv\He_n(z).
\end{equation}
These functions have three properties that will be used repeatedly.
\begin{enumerate}
	\item \emph{Eigenfunction property}: The backward generator is
$\Lg_0^\dagger\equiv(f-\gamma v)\partial_v+\gamma T\partial_v^2$.
Because $f-\gamma v=-\gamma u$, it can be written as
$\Lg_0^\dagger=-\gamma u\partial_u+\gamma T\partial_u^2$.
Using the Hermite differential equation $\He_n''-z\He_n'=-n\He_n$, we obtain
\begin{equation}
	\Lg_0^\dagger \varphi_n=\gamma\big(\He_n''-z\He_n'\big)=-\gamma n\varphi_n.
\label{eq:eig}
\end{equation}

	\item \emph{Orthogonality}: With respect to the steady-state probability density, the eigenfunctions satisfy
\begin{equation}
	\ev{\varphi_n\varphi_m}_{\st}\equiv\int\varphi_n(z)\varphi_m(z)p_{\st}(v)\dd{v}=n!\delta_{nm}.
\label{eq:ortho}
\end{equation}

	\item \emph{Derivative ladder}: Since $\He_n'=n\He_{n-1}$, differentiation with respect to $v$ gives
\begin{equation}
	\partial_v\varphi_n=\frac{n}{\sqrt{T}}\varphi_{n-1}.
\label{eq:ladder}
\end{equation}
Equivalently, $\partial_z\varphi_n=n\varphi_{n-1}$.
\end{enumerate}

\subsection{Algebraic suppression}
We choose the weight function
\begin{equation}
	\psi(v)=1+c\qty(\frac{v^2}{T}-1),
\end{equation}
where $c$ is an arbitrary real parameter.
For $c=0$, the current reduces to the unwrapped displacement, $J_\tau=x_\tau-x_0$.

We next expand the current integrand $\Phi$ in the Hermite basis.
The chosen weight gives
\begin{equation}
	\Phi(v)=\psi(v)v=(1-c)v+\frac{c}{T}v^3.
\end{equation}
The required powers of $u=v-\vbar$ can be expressed as
\begin{align}
	1&=\varphi_0,
	&u&=\sqrt{T}\varphi_1,\notag\\
	u^2&=T(\varphi_2+\varphi_0),
	&u^3&=T^{3/2}(\varphi_3+3\varphi_1),
\end{align}
as follows directly from the definitions of $\He_1$, $\He_2$, and $\He_3$.
Using $v=u+\vbar$, we then obtain
\begin{align}
	v&=\vbar\varphi_0+\sqrt{T}\varphi_1,\\
	v^3
	&=\qty(\vbar^3+3\vbar T)\varphi_0
	+\qty(3\vbar^2\sqrt{T}+3T^{3/2})\varphi_1+3\vbar T\varphi_2+T^{3/2}\varphi_3.
\end{align}
Collecting terms mode by mode and using $\vbar^2=\kappa T$ gives
\begin{equation}
	\Phi=\underbrace{\vbar(1+2c+c\kappa)}_{\textstyle a_0}\varphi_0
	+\underbrace{\sqrt{T}(1+2c+3c\kappa)}_{\textstyle a_1}\varphi_1
	+\underbrace{3c\vbar}_{\textstyle a_2}\varphi_2
	+\underbrace{c\sqrt{T}}_{\textstyle a_3}\varphi_3
	=\sum_{n=0}^3a_n\varphi_n.
\label{eq:Phimodes}
\end{equation}

Because $\ev{\varphi_n}_{\st}=0$ for $n\ge1$, the steady-state current average is given by the $\varphi_0$ coefficient in Eq.~\eqref{eq:Phimodes}:
\begin{equation}
	\jr_{\st}=\ev{\Phi}_{\st}=\vbar(1+2c+c\kappa).
\label{eq:j}
\end{equation}
Solving the Poisson equation $\Lg_0^\dagger\Psi=-(\Phi-\jr_{\st})$ mode by mode and using $\Lg_0^\dagger\varphi_n=-\gamma n\varphi_n$ gives
\begin{equation}
	\Psi=\sum_{n=1}^{3}\frac{a_n}{\gamma n}\varphi_n.
\end{equation}
The derivative ladder in Eq.~\eqref{eq:ladder} then yields
\begin{equation}
	\partial_v\Psi=\frac{1}{\gamma\sqrt{T}}\sum_{n=1}^3a_n\varphi_{n-1},
\end{equation}
or, explicitly,
\begin{equation}
	\partial_v \Psi=\frac{1+2c+3c\kappa}{\gamma}\varphi_0
	+\frac{3c\vbar}{\gamma\sqrt{T}}\varphi_1
	+\frac{c}{\gamma}\varphi_2.
\label{eq:dvpsi}
\end{equation}
Using orthogonality, together with $\ev{\varphi_0^2}_{\st}=1$, $\ev{\varphi_1^2}_{\st}=1$, $\ev{\varphi_2^2}_{\st}=2$, and $\vbar^2/T=\kappa$, we obtain
\begin{equation}
	\ev{(\partial_v\Psi)^2}_{\st}=\frac{1}{\gamma^2}\qty[(1+2c+3c\kappa)^2+c^2(9\kappa+2)].
\end{equation}
Equation~\eqref{eq:long.var}, with the velocity-space diffusion coefficient $\gamma T$, therefore gives the long-time current diffusivity
\begin{equation}
	D_J=\frac{T}{\gamma}\qty[(1+2c+3c\kappa)^2+c^2(9\kappa+2)].
\label{eq:sigma2}
\end{equation}
The steady-state entropy production rate is
\begin{equation}
	\sigma_{\st}=\dot\Sigma=\frac{f\ev{v}_{\st}}{T}=\frac{\gamma\vbar^2}{T}=\gamma\kappa.
\label{eq:Sdot}
\end{equation}
We define the TUR uncertainty factor as
\begin{equation}
	\Qc(c,\kappa)\equiv\frac{D_J\sigma_{\st}}{\jr_{\st}^2},
\end{equation}
such that $\Qc<1$ indicates a violation of the conventional TUR.
Combining Eqs.~\eqref{eq:j}, \eqref{eq:sigma2}, and \eqref{eq:Sdot}, we find
\begin{equation}
\tcboxmath{
	\Qc(c,\kappa)=\frac{(1+2c+3c\kappa)^2+c^2(9\kappa+2)}{(1+2c+c\kappa)^2}.
}
\label{eq:Qf}
\end{equation}

For fixed $\kappa>0$, we minimize the uncertainty factor in Eq.~\eqref{eq:Qf} over $c$, excluding $c=-1/(\kappa+2)$, for which the current average vanishes. Differentiation gives
\begin{equation}
	\pdv{c}\Qc(c,\kappa)=\frac{2c(\kappa+2)(6\kappa+1)+4\kappa}{[c(\kappa+2)+1]^3}.
\end{equation}
Thus, the minimizing value is
\begin{equation}
	c^\star=-\frac{2\kappa}{(\kappa+2)(6\kappa+1)}.
\label{eq:cstar}
\end{equation}
Substitution into Eq.~\eqref{eq:Qf} yields
\begin{equation}
\tcboxmath{
	\Qc_{\mathrm{min}}(\kappa)\equiv\min_c\Qc(c,\kappa)=\frac{9\kappa+2}{4\kappa^2+9\kappa+2}\le 1.
}
\label{eq:Rmin}
\end{equation}
Its limiting behavior is
\begin{equation}
	\Qc_{\mathrm{min}}(\kappa)=1-2\kappa^2+O(\kappa^3)
	\quad(\kappa\to0),
	\qquad
	\Qc_{\mathrm{min}}(\kappa)\sim\frac{9}{4\kappa}
	\quad(\kappa\to\infty).
\end{equation}
Near equilibrium, the violation is weak and appears only at second order in $\kappa$. Far from equilibrium, $\Qc_{\mathrm{min}}\to0$, so the reciprocal violation factor $1/\Qc_{\mathrm{min}}$ grows without bound. The optimal coefficient is negative, causing the velocity-dependent weight to suppress the contribution of large-speed fluctuations to the current diffusivity more strongly than it suppresses the current average. This mechanism relies on resolving the velocity degree of freedom. By contrast, an overdamped position-weighted current, such as $\int_0^\tau g(x_t)\circ\dd{x_t}$, obeys the conventional long-time TUR $D_J\sigma_{\st}\ge \jr_{\st}^2$ and cannot exhibit this violation.

\begin{figure}[t]
\centering
\includegraphics[width=1.0\linewidth]{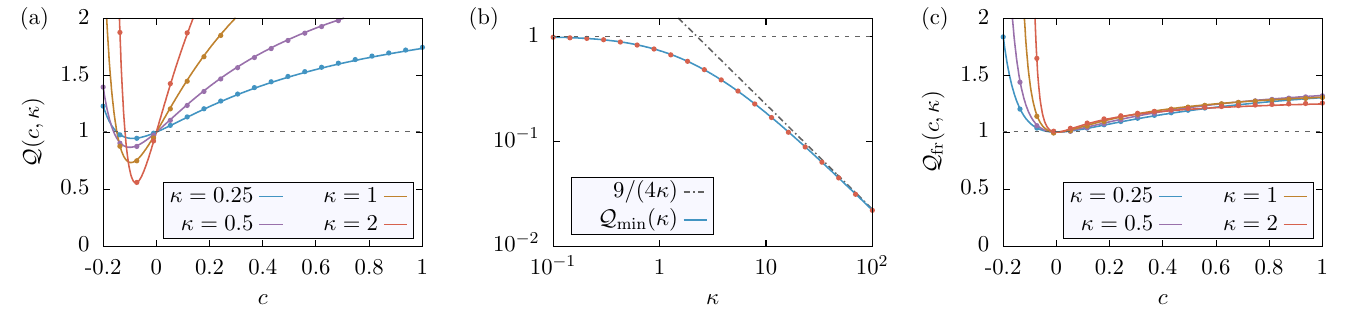}
\protect\caption{Comparison of conventional and friction-response TURs. (a) Conventional and (c) friction-response TUR uncertainty factors as functions of the weighting parameter $c$ for several values of $\kappa$. (b) Optimized conventional uncertainty factor $\mathcal{Q}_{\mathrm{min}}$ as a function of $\kappa$. Circles denote Langevin simulations at $\tau=400$, while solid lines show the analytical $\tau\to\infty$ results. Parameters are $\gamma=1$, $T=1$, and $f=\sqrt{\kappa}$.}\label{fig:1D.alg}
\end{figure}

By contrast, the friction-response uncertainty factor remains bounded below by unity. At fixed $f$, $T$, and $c$, we have $\gamma\partial_\gamma \jr_{\st}=-\vbar(1+2c+3c\kappa)$ and hence
\begin{equation}
\tcboxmath{
	\Qc_{\mathrm{fr}}\equiv\frac{D_J\sigma_{\st}}{(\gamma\partial_\gamma \jr_{\st})^2}=1+\frac{c^2(9\kappa+2)}{(1+2c+3c\kappa)^2}\ge 1.
}
\end{equation}
Numerical demonstration of these results is shown in Fig.~\ref{fig:1D.alg}.

\subsection{Exponential suppression}
We choose the weight function
\begin{equation}
	\psi(v)=\frac{1}{v}\erf\qty(\frac{\alpha v}{\sqrt{T}}),
\end{equation}
where $\alpha$ is a positive constant. Note that $\psi(v)$ is even, smooth, and bounded with $\psi(0)=2\alpha/\sqrt{\pi T}$; thus, $J_\tau=\int_0^\tau \psi(v_t)\circ\dd{x_t}$ is an admissible velocity-resolved antisymmetric current. The chosen weight gives 
\begin{equation}
	\Phi(v)=\erf\qty(\frac{\alpha v}{\sqrt{T}}).
\end{equation}
For later convenience, define
\begin{equation}
    \lambda\equiv\frac{2\alpha^2}{1+2\alpha^2}~(0<\lambda <1),
    \qquad
    \eta\equiv\sqrt{\frac{\lambda\kappa}{2}}.
\label{eq:poisson.c}
\end{equation}

The mean current is
\begin{equation}
    \jr_{\st}\equiv\ev{\Phi}_{\st}=\E_z\qty[\erf(\alpha(z+\sqrt\kappa))],
\label{eq:poisson.mean.def}
\end{equation}
where $z=(v-\vbar)/\sqrt{T}\sim\mca{N}(0,1)$ in the steady state. For a standard normal variable $z$, the Gaussian convolution identity holds:
\begin{equation}
    \E_z[\erf(a+bz)]=\erf\qty(\frac{a}{\sqrt{1+2b^2}}).
\label{eq:poisson.erf.convolution}
\end{equation}
Applying Eq.~\eqref{eq:poisson.erf.convolution} with $a=\alpha\sqrt\kappa$ and $b=\alpha$ gives
\begin{equation}
\tcboxmath{
    \jr_{\st}=\erf\qty(\frac{\alpha\sqrt\kappa}{\sqrt{1+2\alpha^2}})=\erf(\eta).
}
\label{eq:poisson.mean}
\end{equation}
In particular, $\jr_{\st}\to 1$ as $\kappa\to\infty$ for every fixed $0<\lambda<1$.

Let
\begin{equation}
    \Theta(z)\equiv\Phi(\vbar+\sqrt{T}z)-\jr_{\st}=\erf(\alpha(z+\sqrt\kappa))-\jr_{\st},
    \qquad
    \ev{\Theta}_{\st}=0,
\label{eq:poisson.centered.Theta}
\end{equation}
where we take $f>0$ without loss of generality, so that $\vbar/\sqrt{T}=\sqrt{\kappa}$.
The stationary Poisson equation is
\begin{equation}
    \Lg_0^\dagger \Psi(z)=-\Theta(z),
    \qquad
    \ev{\Psi}_{\st}=0.
\label{eq:poisson.equation}
\end{equation}
Expand the centered readout as
\begin{equation}
    \Theta=\sum_{n=1}^\infty a_n\varphi_n,
    \qquad
    a_n=\frac{1}{n!}\ev{\Theta\varphi_n}_{\st}=\frac{1}{n!}\ev{\Phi\varphi_n}_{\st}.
\label{eq:poisson.Hermite.expansion}
\end{equation}
The $n=0$ coefficient vanishes by construction. Since $\Lg_0^\dagger\varphi_n=-\gamma n\varphi_n$, inverting mode by mode yields
\begin{equation}
    \Psi=\frac{1}{\gamma}\sum_{n=1}^\infty\frac{a_n}{n}\varphi_n.
\label{eq:poisson.g.Hermite}
\end{equation}
Substituting Eq.~\eqref{eq:poisson.g.Hermite} into the long-time variance formula \eqref{eq:long.var} and using $\partial_z\varphi_n = n\varphi_{n-1}$ gives
\begin{align}
    D_J&=\gamma\ev{(\partial_z \Psi)^2}_{\st}\notag\\
    &=\frac{1}{\gamma}\sum_{n,m\ge 1}a_na_m\ev{\varphi_{n-1}\varphi_{m-1}}_{\st}\notag\\
    &=\frac{1}{\gamma}\sum_{n=1}^\infty a_n^2(n-1)!.
\label{eq:poisson.D.spectral}
\end{align}

We next evaluate $a_n$ exactly. For a standard normal variable $z$ and a sufficiently smooth function $q$, the Hermite-Stein identity
\begin{equation}
    \ev{q(z)\varphi_n(z)}_{\st}=\ev{q^{(n)}(z)}_{\st}
\label{eq:poisson.Stein}
\end{equation}
follows from $\varphi_n(z)e^{-z^2/2}=(-1)^n\partial_z^n e^{-z^2/2}$ and $n$ repeated integrations by parts. Here, $q^{(n)}$ is the $n$-th derivative with respect to $z$. For $q(z)=\erf[\alpha(z+\sqrt\kappa)]$, one has
\begin{equation}
    q^{(n)}(z)=\frac{2}{\sqrt\pi}(-1)^{n-1}\alpha^nH_{n-1}(\alpha(z+\sqrt\kappa))e^{-\alpha^2(z+\sqrt\kappa)^2},
\label{eq:poisson.erf.derivative}
\end{equation}
where $H_n$ denotes the physicists' Hermite polynomial. The remaining Gaussian average can be evaluated entirely by a generating function. Define $s=z+\sqrt\kappa\sim\mathcal N(\sqrt\kappa,1)$ and
\begin{equation}
    I_n\equiv\E_s\qty[H_n(\alpha s)e^{-\alpha^2s^2}].
\label{eq:poisson.Ik.def}
\end{equation}
Using
\begin{equation}
    e^{2xt-t^2}=\sum_{n=0}^\infty\frac{H_n(x)}{n!}t^n,
\label{eq:poisson.Hphys.gen}
\end{equation}
we obtain
\begin{align}
    \sum_{n=0}^\infty\frac{I_n}{n!}t^n&=e^{-t^2}\E_s\qty[e^{-\alpha^2s^2+2\alpha ts}]\notag\\
    &=\sqrt{1-\lambda}e^{-\lambda\kappa/2}\exp\qty[2\eta\sqrt{1-\lambda}t-(1-\lambda)t^2]\notag\\
    &=\sqrt{1-\lambda}e^{-\lambda\kappa/2}\sum_{n=0}^\infty\frac{H_n(\eta)}{n!}\qty(\sqrt{1-\lambda}t)^n.
\label{eq:poisson.Ik.gen}
\end{align}
Here, we used $1-\lambda =(1+2\alpha^2)^{-1}$ and $\eta=\sqrt{\lambda\kappa/2}$.  Comparing coefficients of $t^n$ gives
\begin{equation}
    I_n=e^{-\lambda\kappa/2}(1-\lambda)^{(n+1)/2}H_n(\eta).
\label{eq:poisson.Ik}
\end{equation}
Combining Eqs.~\eqref{eq:poisson.Stein}, \eqref{eq:poisson.erf.derivative}, and \eqref{eq:poisson.Ik}, and using $\alpha^2(1-\lambda)=\lambda/2$ yields
\begin{equation}
    a_n=\frac{(-1)^{n-1}}{n!}\frac{2}{\sqrt\pi}e^{-\lambda\kappa/2}\qty(\frac{\lambda}{2})^{n/2}H_{n-1}(\eta),\qquad \forall n\ge 1.
\label{eq:poisson.an}
\end{equation}
Substituting Eq.~\eqref{eq:poisson.an} into Eq.~\eqref{eq:poisson.D.spectral} gives
\begin{align}
    \gamma D_J&=\frac{4}{\pi}e^{-\lambda\kappa}\sum_{n=1}^\infty\frac{(n-1)!}{(n!)^2}\qty(\frac{\lambda}{2})^nH_{n-1}(\eta)^2\notag\\
    &=\frac{2\lambda}{\pi}e^{-\lambda\kappa}\sum_{n=0}^\infty\frac{\lambda^n}{2^n n!}\frac{H_n(\eta)^2}{(n+1)^2}.
\label{eq:poisson.D.series}
\end{align}
Equation~\eqref{eq:poisson.D.series} is already an exact Poisson-solution formula for the diffusivity. The exponential large-$\kappa$ behavior is not, however, visible from any fixed truncation of this series: as the drive grows, an increasing set of Hermite modes contributes.  A uniform asymptotic analysis is obtained by resumming the entire series.

The spectral denominator can be represented as
\begin{equation}
    \frac{1}{(n+1)^2}=\int_0^1 \varsigma^n\ln\qty(\frac{1}{\varsigma})\dd{\varsigma},
\label{eq:poisson.weight.identity}
\end{equation}
which follows directly from $\int_0^1\varsigma^n\dd{\varsigma}=(n+1)^{-1}$ by differentiating with respect to $n$. Since every term in Eq.~\eqref{eq:poisson.D.series} is nonnegative, the sum and the $\varsigma$ integral can be exchanged:
\begin{equation}
    \gamma D_J=\frac{2\lambda}{\pi}e^{-\lambda\kappa}\int_0^1\ln\qty(\frac{1}{\varsigma})\left[\sum_{n=0}^\infty\frac{H_n(\eta)^2}{2^n n!}(\lambda \varsigma)^n\right]\dd{\varsigma}.
\label{eq:poisson.before.Mehler}
\end{equation}
For $|t|<1$, Mehler's formula for the physicists' Hermite polynomials reads
\begin{equation}
    \sum_{n=0}^\infty\frac{H_n(x)H_n(y)}{2^nn!}t^n=\frac{1}{\sqrt{1-t^2}}\exp\qty[\frac{2xyt-(x^2+y^2)t^2}{1-t^2}].
\label{eq:poisson.Mehler}
\end{equation}
At coincident arguments $x=y=\eta$, this simplifies to
\begin{align}
    \sum_{n=0}^\infty\frac{H_n(\eta)^2}{2^nn!}t^n&=\frac{1}{\sqrt{1-t^2}}\exp\qty[\frac{2\eta^2t(1-t)}{1-t^2}]\notag\\
    &=\frac{1}{\sqrt{1-t^2}}\exp\qty(\frac{2\eta^2t}{1+t})\notag\\
    &=\frac{1}{\sqrt{1-t^2}}\exp\qty(\frac{\lambda\kappa t}{1+t}),
\label{eq:poisson.Mehler.coincident}
\end{align}
where $2\eta^2=\lambda\kappa$. Using Eq.~\eqref{eq:poisson.Mehler.coincident} and substituting $t=\lambda\varsigma\,(<1)$ into Eq.~\eqref{eq:poisson.before.Mehler} gives
\begin{equation}
    \gamma D_J=\frac{2\lambda}{\pi}e^{-\lambda\kappa}\int_0^1\frac{\ln(1/\varsigma)}{\sqrt{1-\lambda^2\varsigma^2}}\exp\qty(\frac{\lambda^2\kappa \varsigma}{1+\lambda\varsigma})\dd{\varsigma}.
\label{eq:poisson.w.integral}
\end{equation}
Finally, let $r=\lambda\varsigma$. Then, $\dd{\varsigma}=\dd{r}/\lambda$ and
\begin{equation}
    -\lambda\kappa+\frac{\lambda\kappa r}{1+r}=-\frac{\lambda\kappa}{1+r}.
\label{eq:poisson.exp.recombine}
\end{equation}
We arrive at the exact single-integral representation of the diffusion coefficient
\begin{equation}
\tcboxmath{
    D_J=\frac{2}{\pi\gamma}\int_0^\lambda\frac{\ln(\lambda/r)}{\sqrt{1-r^2}}\exp\qty(-\frac{\lambda\kappa}{1+r})\dd{r}.
}
\label{eq:poisson.D.exact.integral}
\end{equation}

We now determine the large-drive asymptotic form of the current diffusivity for fixed $0<\lambda<1$. Equation~\eqref{eq:poisson.D.exact.integral} can be written in the Laplace form
\begin{equation}
    D_J=\frac{2}{\pi\gamma}\int_0^\lambda F(r)e^{-\kappa\theta(r)}\dd{r},
    \qquad
    F(r)=\frac{\ln(\lambda/r)}{\sqrt{1-r^2}},
    \qquad
    \theta(r)=\frac{\lambda}{1+r}.
\label{eq:poisson.Laplace.form}
\end{equation}
Because $\theta(r)$ decreases monotonically, the dominant contribution comes from the upper endpoint $r=\lambda$. Setting $r=\lambda-\delta$, with $\delta\ll1$, we obtain
\begin{equation}
    F(\lambda-\delta)=\frac{\delta}{\lambda\sqrt{1-\lambda^2}}+O(\delta^2),
    \qquad
    \theta(\lambda-\delta)=a+b\delta+O(\delta^2),
\end{equation}
where
\begin{equation}
	a\equiv\frac{\lambda}{1+\lambda},
	\qquad
	b\equiv\frac{\lambda}{(1+\lambda)^2}.
\end{equation}
The relevant endpoint region has width $\delta=O(\kappa^{-1})$. We may therefore extend the upper limit of the $\delta$ integral to infinity at leading order, since contributions away from the endpoint are exponentially smaller. This gives
\begin{align}
    D_J&\sim\frac{2e^{-a\kappa}}{\pi\gamma\lambda\sqrt{1-\lambda^2}}\int_0^\infty\delta e^{-b\kappa\delta}\dd{\delta}\notag\\
    &=\frac{2e^{-a\kappa}}{\pi\gamma b^2\lambda\sqrt{1-\lambda^2}}\kappa^{-2}\notag\\
    &=\frac{2}{\pi\gamma}\frac{(1+\lambda)^4}{\lambda^3\sqrt{1-\lambda^2}}\kappa^{-2}e^{-a\kappa}.
\label{eq:poisson.D.leading}
\end{align}
Defining
\begin{equation}
    \ell(\lambda)\equiv\frac{2(1+\lambda)^4}{\pi\lambda^3\sqrt{1-\lambda^2}},
\end{equation}
we arrive at
\begin{equation}
\tcboxmath{
    D_J\sim\frac{\ell(\lambda)}{\gamma}\kappa^{-2}\exp\qty(-\frac{\lambda}{1+\lambda}\kappa).
}
\label{eq:poisson.D.asymptotic}
\end{equation}

The conventional long-time uncertainty factor is
\begin{equation}
    \Qc\equiv\frac{D_J\sigma_{\st}}{\jr_{\st}^2}=\frac{\gamma\kappa D_J}{\jr_{\st}^2}.
\label{eq:poisson.Q.def}
\end{equation}
To determine the corresponding asymptotics of $\Qc$, we use the large-$z$ expansion
\begin{equation}
    \operatorname{erfc}(z)=\frac{e^{-z^2}}{\sqrt\pi z}\qty[1-\frac{1}{2z^2}+O(z^{-4})],
\label{eq:poisson.erfc.asymptotic}
\end{equation}
of the complementary error function. Setting $z=\eta=\sqrt{\lambda\kappa/2}$, Eq.~\eqref{eq:poisson.mean} gives
\begin{equation}
	\jr_{\st}=1-\sqrt{\frac{2}{\pi\lambda\kappa}}e^{-\lambda\kappa/2}\qty[1-\frac{1}{\lambda\kappa}+O(\kappa^{-2})].
\label{eq:poisson.mean.asymptotic}
\end{equation}
Consequently,
\begin{equation}
    \frac{1}{\jr_{\st}^2}=1+O\qty(\kappa^{-1/2}e^{-\lambda\kappa/2}),
\label{eq:poisson.mean.inverse}
\end{equation}
so the mean approaches unity exponentially fast and does not affect the leading asymptotics of the uncertainty factor. Combining Eqs.~\eqref{eq:poisson.D.asymptotic}, \eqref{eq:poisson.Q.def}, and \eqref{eq:poisson.mean.inverse}, we obtain
\begin{equation}
\tcboxmath{
    \Qc\sim \ell(\lambda)\kappa^{-1}\exp(-a\kappa).
}
\label{eq:poisson.Q.asymptotic}
\end{equation}
Thus, even though the underlying dynamics is linear and Gaussian, the uncertainty factor of this nonlinear current decreases exponentially with the entropy production rate.

The friction-response TUR, however, remains valid. Using Eq.~\eqref{eq:poisson.mean}, together with $\dd{\erf(x)}/\dd{x}=2e^{-x^2}/\sqrt\pi$ and $\gamma\partial_\gamma\kappa=-2\kappa$, gives
\begin{equation}
	\gamma\partial_\gamma \jr_{\st}=-\sqrt{\frac{2\lambda\kappa}{\pi}}e^{-\lambda\kappa/2}.
\end{equation}
Combining this identity with Eq.~\eqref{eq:poisson.D.series}, we obtain the friction-response uncertainty factor
\begin{equation}
\tcboxmath{
	\Qc_{\mathrm{fr}}\equiv\frac{D_J\sigma_{\st}}{(\gamma\partial_\gamma \jr_{\st})^2}=1+\sum_{n=1}^\infty\frac{\lambda^n}{2^n n!}\frac{H_n(\eta)^2}{(n+1)^2}\ge 1.
}
\end{equation}

\section{Overdamped Langevin dynamics}
\label{sec:ovd.langevin}

\subsection{Setup}
Consider general overdamped Langevin dynamics in a $d$-dimensional space. The position $\bx_t$ evolves according to
\begin{equation}
	\dd{\bx_t}=\bF_t(\bx_t)\dd{t}+\sqrt{2\D}\dd{\bW_t},
\end{equation}
where $\D=[D_{mn}]\in\mathbb{R}^{d\times d}$ is a symmetric positive-definite diffusion matrix.
The corresponding Fokker-Planck equation is
\begin{equation}
	\dot p_t = \Lg_t(p_t) = -\nabla\cdot \bj_t,
\end{equation}
where $\bj_t\equiv\bF_tp_t-\D\nabla p_t$ is the probability current and $\vb*{\nu}_t\equiv\bj_t/p_t$ is the associated local mean velocity.
The backward generator is $\Lg_t^\dagger=\bF_t\cdot\nabla+\D:\nabla\nabla$.
The total entropy production is $\Sigma_\tau=\int_0^\tau\sigma_t\dd{t}$, with entropy production rate
\begin{equation}
	\sigma_t =\ev{\vb*{\nu}_t^{\top} \D^{-1}\vb*{\nu}_t}_t.
\end{equation}
Consider the time-integrated observable
\begin{equation}
	J_\tau=\int_0^\tau [\phi_t(\bx_t)\dd{t}+\bw_t(\bx_t)\circ\dd{\bx_t}],
\end{equation}
where $\circ$ denotes the Stratonovich product, $\bw_t$ is a vector-valued weight, and $\phi_t$ is a state-dependent rate. In It\^o form, its increment is
\begin{equation}
	\dd{J_t}=\Phi_t(\bx_t)\dd{t}+\qty[\sqrt{2\D}\bw_t(\bx_t)]\cdot\dd{\bW_t},
\end{equation}
where $\Phi_t=\phi_t+\bw_t\cdot\bF_t+\D:\nabla\bw_t$.
Assuming that the boundary terms vanish, the observable average can be written as
\begin{equation}
	\E[J_\tau]=\int_0^\tau\qty[\inprod{\phi_t}{p_t}+\inprod{\bw_t}{\bj_t}]\dd{t}.
\end{equation}

\subsection{Quadratic expression for the observable variance}
Proceeding as in the underdamped case, It\^o's formula gives
\begin{equation}
	\dd{G_t}=(\partial_t G_t+\Lg_t^\dagger G_t)\dd{t}+(\sqrt{2\D}\nabla G_t)\cdot\dd{\bW_t}.\label{eq:dG.ovd}
\end{equation}
Since $G_t$ satisfies $\partial_tG_t+\Lg_t^\dagger G_t=-\Phi_t$, the drift terms cancel in the martingale $M_t=J_t+G_t(\bx_t)$, yielding
\begin{equation}
	\dd{M_t}=\qty[\sqrt{2\D}\big(\bw_t(\bx_t)+\nabla G_t(\bx_t)\big)]\cdot\dd{\bW_t}=\qty[\sqrt{2\D}\vb*{u}_t(\bx_t)]\cdot\dd{\bW_t},
\end{equation}
where $\vb*{u}_t(\bx)\equiv\bw_t(\bx)+\nabla G_t(\bx)$.
The It\^o isometry then yields the exact quadratic representation
\begin{equation}
\tcboxmath{
	\Var(J_\tau) = \Var_0(J_\tau) + 2\int_0^\tau\ev{\vb*{u}_t^{\top} \D\vb*{u}_t}_t\dd{t},
}
\label{eq:varlangevin}
\end{equation}
where $\Var_0(J_\tau)\equiv \Var(\E[J_\tau\cond \bx_0])$.

\subsection{Fluctuation-response relations}
We formulate the response directly in terms of an arbitrary infinitesimal source $\delta\bj_{t,\mathrm{src}}$ injected into the probability current. Such a source may arise from a perturbation of the drift, the diffusion, or both. Proposition~\ref{prop:ovd.perturb} gives the corresponding first variation of the observable average as
\begin{equation}
    \delta\E[J_\tau]=\int_0^\tau\inprod{\bu_t}{\delta\bj_{t,\mathrm{src}}}\dd{t},
\end{equation}
where $\bu_t(\bx)=\bw_t(\bx)+\nabla G_t(\bx)$.
It follows from the definition of the functional derivative that
\begin{equation}
	\frac{\delta\E[J_\tau]}{\delta\bj_{t,\mathrm{src}}}=\vb*{u}_t.
\end{equation}
Combining this identity with Eq.~\eqref{eq:varlangevin} yields the fluctuation-response relation
\begin{equation}
\tcboxmath{
	\Var(J_\tau)-\Var_0(J_\tau) = 2\int_0^\tau\ev{\qty(\frac{\delta\E[J_\tau]}{\delta\bj_{t,\mathrm{src}}})^{\top}\D\qty(\frac{\delta\E[J_\tau]}{\delta\bj_{t,\mathrm{src}}})}_t\dd{t}.
}
\end{equation}

We next choose a specific perturbation direction in the space of injected probability currents, $\delta\bj_{t,\mathrm{src}}=\epsilon\bg_t p_t$, where $\epsilon$ is a small scalar perturbation parameter and $\bg_t(\bx)\in\mathbb{R}^d$ is an arbitrary field. For example, this source current is generated by the drift perturbation $\bF_t\to\bF_t+\epsilon\bg_t$. We define the corresponding linear response and quadratic cost by
\begin{equation}
    \Rc[\bg]\equiv\left.\partial_\epsilon\E_\epsilon[J_\tau]\right|_{\epsilon=0},
    \qquad
    \Ac[\bg]\equiv\int_0^\tau\ev{\bg_t^{\top}\D^{-1}\bg_t}_t\dd{t},
\end{equation}
where $\E_\epsilon$ denotes the expectation under the perturbed dynamics. Using the response relation above gives
\begin{equation}
    \Rc[\bg]=\int_0^\tau\inprod{\bu_t}{\bg_t p_t}\dd{t}=\int_0^\tau\ev{\bu_t\cdot\bg_t}_t\dd{t}.
\end{equation}
The Cauchy-Schwarz inequality then gives the fluctuation-response inequality
\begin{equation}
\tcboxmath{
	\Rc[\bg]^2\le \frac{\Ac[\bg]}{2}\qty[\Var(J_\tau) - \Var_0(J_\tau)].
}
\end{equation}

\begin{proposition}\label{prop:ovd.perturb}
Consider an arbitrary admissible infinitesimal injection $\delta\bj_{t,\mathrm{src}}(\bx)$ of probability current, while the initial density and the observable fields $\phi_t$ and $\bw_t$ are held fixed. Then the corresponding variation of the observable average is
\begin{equation}
    \delta\E[J_\tau]=\int_0^\tau\inprod{\bu_t}{\delta\bj_{t,\mathrm{src}}}\dd{t},
\label{eq:overresp}
\end{equation}
where $\bu_t=\bw_t+\nabla G_t$.
\end{proposition}

\begin{proof}
The induced variation $\delta p_t$ of the probability density satisfies
\begin{equation}
    \partial_t\delta p_t=\Lg_t(\delta p_t)-\nabla \cdot\delta\bj_{t,\mathrm{src}},
    \qquad
    \delta p_0=0.
\label{eq:ovd.forced}
\end{equation}
The mean of the additive observable can be expressed in terms of the probability current as
\begin{equation}
    \E[J_\tau]=\int_0^\tau\qty[\inprod{\phi_t}{p_t}+\inprod{\bw_t}{\bj_t}]\dd{t}.
\label{eq:ovd.mean.current.rep}
\end{equation}
Under an arbitrary infinitesimal current injection $\delta\bj_{t,\mathrm{src}}$, the total variation of the probability current is
\begin{equation}
    \delta\bj_{t}=\Jc_t(\delta p_t)+\delta\bj_{t,\mathrm{src}},
\label{eq:ovd.total.current.variation}
\end{equation}
where $\Jc_t(p)\equiv\bF_tp-\D\nabla p$ denotes the probability-current operator.
Taking the variation of Eq.~\eqref{eq:ovd.mean.current.rep} therefore gives
\begin{align}
    \delta\E[J_\tau]&=\int_0^\tau\qty[\inprod{\phi_t}{\delta p_t}+\inprod{\bw_t}{\Jc_t(\delta p_t)}+\inprod{\bw_t}{\delta\bj_{t,\mathrm{src}}}]\dd{t}\notag\\
    &=\int_0^\tau\qty[\inprod{\Phi_t}{\delta p_t}+\inprod{\bw_t}{\delta\bj_{t,\mathrm{src}}}]\dd{t},
\label{eq:ovd.response.current.rep}
\end{align}
where integration by parts and $\Phi_t=\phi_t+\bw_t\cdot\bF_t+\D:\nabla\bw_t$ have been used.

Let $\mca{U}(t,s)$ and $\mca{U}^\dagger(t,s)$ denote the forward propagator generated by $\Lg_t$ and its adjoint propagator, respectively, given by
\begin{equation}
	\mca{U}(t,s)=\mca{T}\exp(\int_s^t\Lg_r\dd{r}),\quad \mca{U}^\dagger(t,s)=\overline{\mca{T}}\exp(\int_s^t\Lg_r^\dagger\dd{r}).
\end{equation}
Here, $\mca{T}$ and $\overline{\mca{T}}$ denote chronological and anti-chronological time ordering, respectively. The backward equation $\partial_t G_t+\Lg_t^\dagger G_t=-\Phi_t$, together with the terminal condition $G_\tau=0$, gives 
\begin{equation}
	G_s=\int_s^\tau\mca{U}^\dagger(t,s)\Phi_t\dd{t}.
\label{eq:ovd.G.prop}
\end{equation}
The variation of the probability density can also be written directly as
\begin{equation}
    \delta p_t=-\int_0^t\mca{U}(t,s)\nabla\cdot\delta\bj_{s,\mathrm{src}}\dd{s}.
\label{eq:ovd.delta.p}
\end{equation}
Using Eqs.~\eqref{eq:ovd.G.prop} and \eqref{eq:ovd.delta.p}, the first term becomes
\begin{align}
    \int_0^\tau\inprod{\Phi_t}{\delta p_t}\dd{t}&=-\int_0^\tau\int_0^t\inprod{\Phi_t}{\mca{U}(t,s)\nabla\cdot\delta\bj_{s,\mathrm{src}}}\dd{s}\dd{t}\notag\\
    &=-\int_0^\tau\int_s^\tau\inprod{\mca{U}^\dagger(t,s)\Phi_t}{\nabla\cdot\delta\bj_{s,\mathrm{src}}}\dd{t}\dd{s}\notag\\
    &=-\int_0^\tau\inprod{G_s}{\nabla\cdot\delta\bj_{s,\mathrm{src}}}\dd{s}\notag\\
    &=\int_0^\tau\inprod{\nabla G_t}{\delta\bj_{t,\mathrm{src}}}\dd{t}.
\end{align}
Substitution into Eq.~\eqref{eq:ovd.response.current.rep} gives
\begin{align}
    \delta\E[J_\tau]&=\int_0^\tau\inprod{\bw_t+\nabla G_t}{\delta\bj_{t,\mathrm{src}}}\dd{t}\notag\\
    &=\int_0^\tau\inprod{\bu_t}{\delta\bj_{t,\mathrm{src}}}\dd{t},
\end{align}
which proves the result.
\end{proof}

\subsection{Improved finite-time TUR}
Let $\Lg_t=\Lg_{\lambda(\upsilon t)}$, where $\lambda(\upsilon t)$ is a time-dependent control protocol and $\upsilon$ sets the protocol speed. We restrict attention to time-antisymmetric currents, for which $\phi_t(\bx)=0$ and $J_\tau=\int_0^\tau\bw_t(\bx_t)\circ\dd{\bx_t}$. All quantities in the section are evaluated at the reference speed $\upsilon=1$. We hold the initial distribution fixed and assume that all explicit time dependence of the current weights is rescaled together with the protocol when $\upsilon$ is varied.

Applying the Cauchy-Schwarz inequality over both state space and time, together with Proposition~\ref{prop:avg.pd.form}, gives
\begin{align}
	&\qty[(\tau\partial_\tau-\upsilon\partial_\upsilon)\E[J_\tau]]^2=\qty[\int_0^\tau\ev{\vb*{u}_t\cdot\vb*{\nu}_t}_t\dd{t}]^2\notag\\
	&\qquad\le \underbrace{\int_0^\tau\ev{\vb*{u}_t^{\top}\D\vb*{u}_t}_t\dd{t}}_{\textstyle =[\Var(J_\tau)-\Var_0(J_\tau)]/2}\underbrace{\int_0^\tau\ev{\vb*{\nu}_t^{\top}\D^{-1}\vb*{\nu}_t}_t\dd{t}}_{\textstyle =\Sigma_\tau}.
\end{align}
Therefore, for an arbitrary time-dependent protocol, we obtain the improved finite-time TUR
\begin{equation}
\tcboxmath{
	\frac{\Var(J_\tau)-\Var_0(J_\tau)}{\qty[\qty(\tau\partial_\tau-\upsilon\partial_\upsilon)\E[J_\tau]]^2}\ge\frac{2}{\Sigma_\tau}.
}
\end{equation}

In the stationary state with time-independent dynamics and current weights, $\qty(\tau\partial_\tau-\upsilon\partial_\upsilon)\E[J_\tau]=\E[J_\tau]$, and the improved TUR reduces to
\begin{equation}
	\frac{\Var(J_\tau)-\Var_0(J_\tau)}{\E[J_\tau]^2}\ge \frac{2}{\Sigma_\tau}.
\end{equation}
The saturation condition follows directly from the exact quadratic expression for the current variance. With respect to the $\D$-weighted inner product, decompose $\vb*{u}_t$ as
\begin{equation}
	\vb*{u}_t=\frac{\jr_{\st}}{\sigma_{\st}}\D^{-1}\vb*{\nu}_{\st}+\vb*{r}_t,
	\qquad \ev{\vb*{r}_t^{\top}\vb*{\nu}_{\st}}_{\st}=0,
\end{equation}
where $\jr_{\st}$ and $\sigma_{\st}$ are the steady-state mean observable and entropy production rates, respectively. Substituting this decomposition into the variance formula gives
\begin{equation}
	[\Var(J_\tau)-\Var_0(J_\tau)] - \frac{2\E[J_\tau]^2}{\Sigma_\tau}=2\int_0^\tau\ev{\vb*{r}_t^{\top}\D\vb*{r}_t}_{\st}\dd{t}\ge 0.
\end{equation}
Equality holds if and only if $\vb*{r}_t=0$ almost everywhere, or equivalently,
\begin{equation}
	\vb*{u}_t=\bw+\nabla G_t=\frac{\jr_{\st}}{\sigma_{\st}}\D^{-1}\vb*{\nu}_{\st}
\end{equation}
for all $t\in[0,\tau]$, assuming the relevant fields are continuous. Because the right-hand side is time independent and the terminal condition $G_\tau(\bx)=0$ implies $\nabla G_\tau(\bx)=0$, finite-time saturation requires both (i) $\bw=(\jr_{\st}/\sigma_{\st})\D^{-1}\vb*{\nu}_{\st}$, so that the current is proportional to the entropy-production current, and (ii) $\nabla G_t=0$ throughout the interval.

We now show that $\nabla G_t=0$ if and only if $\Phi(\bx)=\jr_{\st}$. If $\Phi(\bx)=\jr_{\st}$, then $G_t=\jr_{\st}(\tau-t)$ solves the backward equation $\partial_tG_t+\Lg_t^\dagger G_t=-\Phi$ with the terminal condition $G_\tau=0$. By uniqueness of the solution to this terminal-value problem, $G_t$ is spatially uniform, and hence $\nabla G_t=0$. Conversely, if $\nabla G_t=0$, then $G_t$ is spatially uniform and can be written as $G_t(\bx)=c(t)$ for some function $c$. Substitution into the backward equation gives $\Phi(\bx)=-\dot c(t)$. Because the left-hand side depends only on $\bx$, whereas the right-hand side depends only on $t$, both must equal a constant, say $b_0$. Therefore, $\jr_{\st}=\ev{\Phi}_{\st}=b_0$, and thus $\Phi(\bx)=\jr_{\st}$. In what follows, we further characterize the dynamics that saturate the TUR.

Using $\bF=\vb*{\nu}_{\st}+\D\nabla\ln p_{\st}$, we obtain
\begin{equation}
	\bw\cdot\bF = \frac{\jr_{\st}}{\sigma_{\st}}\vb*{\nu}_{\st}^{\top}\D^{-1}\qty(\vb*{\nu}_{\st}+\D\nabla\ln p_{\st})
	=\frac{\jr_{\st}}{\sigma_{\st}}\vb*{\nu}_{\st}^{\top}\D^{-1}\vb*{\nu}_{\st}
	+\frac{\jr_{\st}}{\sigma_{\st}}\vb*{\nu}_{\st}^{\top}\nabla\ln p_{\st}.
\end{equation}
Moreover, the steady-state condition $\nabla\cdot\bj_{\st}=\nabla\cdot(\vb*{\nu}_{\st}p_{\st})=0$ implies $\nabla\cdot\vb*{\nu}_{\st}=-\vb*{\nu}_{\st}^{\top}\nabla\ln p_{\st}$. Because $\D$ is spatially uniform, $\D:\nabla(\D^{-1}\vb*{\nu}_{\st})=\nabla\cdot\vb*{\nu}_{\st}$.
It follows that
\begin{align}
	\Phi&=\bw\cdot\bF + \D:\nabla\bw\notag\\
	&=\frac{\jr_{\st}}{\sigma_{\st}}\vb*{\nu}_{\st}^{\top}\D^{-1}\vb*{\nu}_{\st}+\frac{\jr_{\st}}{\sigma_{\st}}\vb*{\nu}_{\st}^{\top}\nabla\ln p_{\st}+\frac{\jr_{\st}}{\sigma_{\st}}\nabla\cdot\vb*{\nu}_{\st}\notag\\
	&=\frac{\jr_{\st}}{\sigma_{\st}}\vb*{\nu}_{\st}^{\top}\D^{-1}\vb*{\nu}_{\st}.
\end{align}
Together with $\Phi(\bx)=\jr_{\st}$, this result implies that the local entropy production rate must be spatially uniform:
\begin{equation}
	\vb*{\nu}_{\st}^{\top}\D^{-1}\vb*{\nu}_{\st}=\sigma_{\st}.
\end{equation}
This condition agrees with that obtained using the Cram{\'e}r-Rao approach \cite{Hasegawa.2019.PRE}.

\begin{proposition}\label{prop:avg.pd.form}
For any time-dependent protocol $\lambda(\upsilon t)$ and any initial probability density $p_0$, the following relation holds:
\begin{equation}
	\int_0^\tau\ev{\vb*{u}_t\cdot\vb*{\nu}_t}_t\dd{t}=\qty(\tau\partial_\tau-\upsilon\partial_\upsilon)\E[J_\tau].
\end{equation}
\end{proposition}
\begin{proof}
Using $\vb*{u}_t=\bw_t+\nabla G_t$, we rewrite the left-hand side as
\begin{align}
	\int_0^\tau\ev{\vb*{u}_t\cdot\vb*{\nu}_t}_t\dd{t}&=\int_0^\tau\ev{\bw_t\cdot\vb*{\nu}_t}_t\dd{t}+\int_0^\tau\ev{\nabla G_t\cdot\vb*{\nu}_t}_t\dd{t}\notag\\
	&=\int_0^\tau\inprod{\bw_t}{\bj_t}\dd{t}+\int_0^\tau\inprod{\nabla G_t}{\bj_t}\dd{t}\notag\\
	&=\E[J_\tau] + \int_0^\tau\inprod{G_t}{\partial_tp_t}\dd{t}.
\end{align}
Here, we used $\partial_t p_t=-\nabla\cdot\bj_t$ and integration by parts in the last equality.
For any differentiable function $f(\tau,\upsilon)$,
\begin{equation}
	(\tau\partial_\tau-\upsilon\partial_\upsilon)f(\tau,\upsilon)=\frac{\dd}{\dd\alpha}f(\alpha\tau,\upsilon/\alpha)\Big|_{\alpha=1}.
\end{equation}
Under the transformation $(\tau,\upsilon)\to(\alpha\tau,\upsilon/\alpha)$, followed by the time change $s=\alpha t$, the protocol becomes $\lambda(\upsilon t)$ on $t\in[0,\tau]$, while the generator and the It\^o current rate are rescaled as $\Lg_t\to\alpha\Lg_t$ and $\Phi_t\to\alpha\Phi_t$, respectively.
Therefore,
\begin{equation}
	(\tau\partial_\tau-\upsilon\partial_\upsilon)\E[J_\tau]=\dv{\alpha}\E_\alpha[J_\tau]\Big|_{\alpha=1},
\end{equation}
where $\E_\alpha[J_\tau]$ denotes the current average under the dynamics generated by $\alpha\Lg_t$.
More explicitly,
\begin{equation}
	\E_\alpha[J_\tau]=\int_0^\tau\inprod{\alpha\Phi_t}{p_t^\alpha}\dd{t},
	\qquad 
	\partial_t p_t^\alpha=\alpha\Lg_t p_t^\alpha,
\end{equation}
with $p_0^\alpha=p_0$.
Differentiating with respect to $\alpha$ gives
\begin{equation}
	\frac{\dd}{\dd\alpha}\E_\alpha[J_\tau]\Big|_{\alpha=1}=\underbrace{\int_0^\tau\inprod{\Phi_t}{p_t}\dd{t}}_{\textstyle =\E[J_\tau]}+\int_0^\tau\inprod{\Phi_t}{\widetilde{p}_t}\dd{t},
\end{equation}
where $\widetilde{p}_t\equiv\partial_\alpha p_t^{\alpha}|_{\alpha=1}$ obeys
\begin{equation}
	\partial_t\widetilde{p}_t=\Lg_t\widetilde{p}_t+\Lg_t p_t,
	\qquad 
	\tilde p_0=0.
\end{equation}
The second term can be written as
\begin{equation}
	\int_0^\tau\inprod{\Phi_t}{\widetilde{p}_t} \dd{t}=\int_0^\tau\inprod{G_t}{\partial_tp_t}\dd{t}.
\label{eq:adjoint}
\end{equation}
To verify this identity, we differentiate $\inprod{G_t}{\widetilde{p}_t}$ and use the backward equation together with the boundary conditions:
\begin{align}
	0&=\inprod{G_\tau}{\widetilde{p}_\tau} - \inprod{G_0}{\widetilde{p}_0}\notag\\
	&=\int_0^\tau\dv{t}\inprod{G_t}{\widetilde{p}_t}\dd{t}\notag\\
	&=\int_0^\tau\qty[\inprod{\partial_tG_t}{\widetilde{p}_t}+\inprod{G_t}{\partial_t \widetilde{p}_t}]\dd{t}\notag\\
	&=\int_0^\tau\qty[\inprod{-\Phi_t-\Lg_t^\dagger G_t}{\widetilde{p}_t}+\inprod{G_t}{\Lg_t \widetilde{p}_t+\Lg_t p_t}]\dd{t}\notag\\
	&=\int_0^\tau\big[-\inprod{\Phi_t}{\widetilde{p}_t}+\underbrace{\inprod{-\Lg_t^\dagger G_t}{\widetilde{p}_t}+\inprod{G_t}{\Lg_t \widetilde{p}_t}}_{\textstyle =0}+\inprod{G_t}{\Lg_t p_t}\big]\dd{t}\notag\\
	&=-\int_0^\tau\inprod{\Phi_t}{\widetilde{p}_t}\dd{t} + \int_0^\tau\inprod{G_t}{\partial_t p_t}\dd{t}.
\end{align}
Combining these results proves the stated relation.
\end{proof}

\section{Markov jump processes}
\label{sec:jump}

\subsection{Setup}
Consider a Markov jump process governed by the transition-rate matrix $\msf{W}_t=[w_{mn}]$, where $w_{mn}\ge 0$ denotes the time-dependent transition rate from state $n$ to state $m\,(\neq n)$.
We assume irreducibility and microscopic reversibility, $w_{mn}>0\Leftrightarrow w_{nm}>0$, and set $w_{nn}=-\sum_{m(\neq n)}w_{mn}$.
Let $\vb*{p}_t$ denote the probability distribution at time $t$.
The forward generator is $\Lg_t=\msf{W}_t$, and the corresponding backward generator, $\Lg_t^\dagger=\msf{W}_t^\dagger$, acts on a state-dependent function $\vb*{q}$ as
\begin{equation}
	(\Lg_t^\dagger\vb*{q})_n=\sum_{m(\neq n)}w_{mn}(q_m-q_n).
\end{equation}
Although we consider general time-dependent dynamics, we suppress explicit time arguments below when no confusion can arise.
For each edge, define the probability current and traffic by
\begin{equation}
	j_{mn}=w_{mn}p_n-w_{nm}p_m=-j_{nm},
	\qquad
	a_{mn}=w_{mn}p_n+w_{nm}p_m=a_{nm},
\end{equation}
respectively.
The master equation then takes the form of Kirchhoff's law, $\sum_m j_{nm}=\dot p_n$, for each state $n$.
The entropy production rate and dynamical activity are given by
\begin{align}
	\sigma_t&=\sum_{m<n}j_{mn}\ln\frac{w_{mn}p_n}{w_{nm}p_m},\notag\\
	a_t&=\sum_{m<n}a_{mn},
\end{align}
and their time-integrated counterparts are $\Sigma_\tau=\int_0^\tau\sigma_t\dd{t}$ and $A_\tau=\int_0^\tau a_t\dd{t}$, respectively.

A generalized observable assigns an antisymmetric increment $d_{mn}(t)=-d_{nm}(t)$ to each transition $n\to m$ and a state-dependent contribution $\phi_n(t)$ to the occupation of state $n$:
\begin{equation}
	J_\tau=\sum_{m\neq n}\int_0^\tau d_{mn}(t)\dd{\hat N_{mn}(t)}+\sum_n\int_0^\tau \phi_n(t)\mathbbm{1}_{\{z_t=n\}}\dd{t},
\label{eq:mjp.general.observable}
\end{equation}
where $\dd{\hat N_{mn}(t)}$ counts the number of transitions from $n$ to $m$ during the interval $[t,t+\dd{t})$, and $\mathbbm{1}_{\{z_t=n\}}$ is the occupation indicator of state $n$.
The corresponding instantaneous mean-observable rate is
\begin{equation}
	\Phi_t(n)\equiv\sum_{m(\neq n)}w_{mn}(t)d_{mn}(t)+\phi_n(t),
	\qquad
	\jr_t=\vb*{\Phi}_t\cdot\vb*{p}_t=\sum_{m<n}d_{mn}(t)j_{mn}(t)+\sum_n \phi_n(t)p_n(t).
\end{equation}
Consequently, for an arbitrary initial probability distribution,
\begin{equation}
	\E[J_\tau]=\int_0^\tau \jr_t\dd{t}=\int_0^\tau\qty[\sum_{m<n}d_{mn}(t)j_{mn}(t)+\sum_n \phi_n(t)p_n(t)]\dd{t}.
\end{equation}

Closely related dynamical fluctuation-response relations for nonautonomous Markov jump processes were recently derived in Ref.~\cite{Aslyamov.2026.arxiv}. Here, we recover this structure from the martingale and source-current formulation, providing a discrete counterpart of the underdamped construction developed above.

\subsection{Quadratic expression for the observable variance}
Let
\begin{equation}
	G_t(z)\equiv\E[J_\tau-J_t\cond z_t=z]
\end{equation}
denote the expected remaining observable conditioned on the state at time $t$. As in the underdamped case, $M_t\equiv\E[J_\tau\cond\mca{F}_t]=J_t+G_t(z_t)$ is a martingale. We first derive the backward equation for $\vb*{G}_t$.

Fix a state $n$ and consider a small time interval $[t,t+h]$, with $h>0$. We decompose the future observable $J_\tau-J_t$ according to the events occurring within this interval, retaining terms up to first order in $h$. There are three mutually exclusive possibilities: (i) exactly one transition $n\to m$ occurs, with probability $w_{mn}h$; the increment $d_{mn}$ is accumulated, after which the expected remaining observable is $G_{t+h}(m)$; (ii) no transition occurs, with probability $1+w_{nn}h$, and the expected remaining observable is $G_{t+h}(n)$; or (iii) two or more transitions occur, with probability $O(h^2)$.
To first order, the occupation contribution accumulated during this interval is $\phi_nh$.
The tower property therefore gives
\begin{equation}
	G_t(n)=\phi_n(t)h+\sum_{m(\neq n)}w_{mn}(t)h\qty[d_{mn}(t)+G_{t+h}(m)]+\qty[1+w_{nn}(t)h]G_{t+h}(n)+O(h^2).
\label{eq:firststep}
\end{equation}
Expanding $G_{t+h}=G_t+h\partial_tG_t+O(h^2)$, we note that the time-derivative term in the transition contribution is already of order $h^2$ and may be omitted.
Equation~\eqref{eq:firststep} then becomes
\begin{align}
	G_t(n)&=G_t(n)+h\qty[\phi_n(t)+\sum_{m(\neq n)}w_{mn}(t)d_{mn}(t)+\sum_{m(\neq n)}w_{mn}(t)G_t(m)+w_{nn}(t)G_t(n)+\partial_tG_t(n)]+O(h^2).
\end{align}
Canceling $G_t(n)$, dividing by $h$, and taking $h\to0$ yields
\begin{equation}
	0=\underbrace{\phi_n(t)+\sum_{m(\neq n)}w_{mn}(t)d_{mn}(t)}_{\textstyle =\Phi_t(n)}+\underbrace{\sum_{m(\neq n)}w_{mn}(t)G_t(m)+w_{nn}(t)G_t(n)}_{\textstyle (\mathrm{I})}+\partial_tG_t(n).
\end{equation}
Using $w_{nn}=-\sum_{m(\neq n)}w_{mn}$, the term $(\mathrm{I})$ becomes
\begin{equation}
	\sum_{m(\neq n)}w_{mn}(t)\qty[G_t(m)-G_t(n)]=(\Lg_t^\dagger G_t)(n).
\end{equation}
Hence,
\begin{equation}
	\partial_t\vb*{G}_t+\Lg_t^\dagger\vb*{G}_t=-\vb*{\Phi}_t,
	\qquad 
	\vb*{G}_\tau=0.
\end{equation}

Next, we derive a quadratic expression for the observable variance.
The corresponding compensated counting-process increment is
\begin{equation}
	\dd\widetilde{N}_{mn}(t)\equiv\dd\hat{N}_{mn}(t)-w_{mn}(t)\delta_{z_{t^-},n}\dd{t},
\end{equation}
where $z_{t^-}$ denotes the state immediately before time $t$.
By construction, $\E[\dd\widetilde{N}_{mn}(t)\cond\Fc_{t^-}]=0$.
The stochastic increment of $G_t(z_t)$ is
\begin{align}
	\dd{G_t(z_t)}
	&=\partial_tG_t(z_{t^-})\dd{t}+\sum_{m\neq n}[G_t(m)-G_t(n)]\dd\hat{N}_{mn}(t)\notag\\
	&=\qty[\partial_tG_t(z_{t^-})+(\Lg_t^\dagger G_t)_{z_{t^-}}]\dd{t}+\sum_{m\neq n}[G_t(m)-G_t(n)]\dd\widetilde{N}_{mn}(t).
\end{align}
Similarly, the observable increment is
\begin{align}
	\dd{J_t}
	&=\phi_{z_{t^-}}(t)\dd{t}+\sum_{m\neq n}d_{mn}(t)\dd\hat{N}_{mn}(t)\notag\\
	&=\Phi_t(z_{t^-})\dd{t}+\sum_{m\neq n}d_{mn}(t)\dd\widetilde{N}_{mn}(t).
\end{align}
Consequently, the stochastic increment of the martingale $M_t=J_t+G_t(z_t)$ is
\begin{equation}
	\dd{M_t}=\qty[\partial_tG_t+\Lg_t^\dagger G_t+\Phi_t](z_{t^-})\dd{t}+\sum_{m\neq n}u_{mn}(t)\dd\widetilde{N}_{mn}(t)=\sum_{m\neq n}u_{mn}(t)\dd\widetilde{N}_{mn}(t),
\end{equation}
where $u_{mn}(t)\equiv d_{mn}(t)+G_t(m)-G_t(n)=-u_{nm}(t)$.
Integrating over $[0,\tau]$ gives
\begin{equation}
	J_\tau=\E[J_\tau\cond z_0]+\int_0^\tau\sum_{m\neq n}u_{mn}(t)\dd\widetilde{N}_{mn}(t).
\end{equation}
The compensated counting processes satisfy, to leading order in $\dd{t}$,
\begin{equation}
	\E\qty[\dd\widetilde{N}_{mn}(t)\dd\widetilde{N}_{m'n'}(t)\cond\Fc_{t^-}]=\delta_{m,m'}\delta_{n,n'}w_{mn}(t)\delta_{z_{t^-},n}\dd{t}.
\end{equation}
Using the martingale isometry, $\E[\delta_{z_{t^-},n}]=p_n(t)$, and $u_{mn}=-u_{nm}$, we obtain the quadratic expression
\begin{equation}
\tcboxmath{
	\Var(J_\tau)=\Var_0(J_\tau)+\int_0^\tau\sum_{m<n}u_{mn}(t)^2a_{mn}(t)\dd{t}.
}
\label{eq:varjump}
\end{equation}
Here, $\Var_0(J_\tau)\equiv\Var(\E[J_\tau\cond z_0])$.
Thus, the traffic $a_{mn}$ serves as a discrete diffusion metric, playing the same role as $2\D$ in the continuous case [cf.~Eq.~\eqref{eq:varlangevin}]. An analogous initial-variability structure appears in the dynamical formulation for Markov jump processes developed in Ref.~\cite{Aslyamov.2026.arxiv}.

\subsection{Fluctuation-response relations}
We formulate the response directly in terms of arbitrary infinitesimal source currents $\delta j_{mn}^{\mathrm{src}}(t)$ injected into the probability currents on the edges $m\leftrightarrow n$, with $\delta j_{mn}^{\mathrm{src}}=-\delta j_{nm}^{\mathrm{src}}$. Such source currents may arise from perturbations of the transition rates. According to Proposition~\ref{prop:mjp.perturb}, the corresponding first-order variation of the observable average is
\begin{equation}
    \delta\E[J_\tau]=\int_0^\tau\sum_{m<n}u_{mn}(t)\delta j_{mn}^{\mathrm{src}}(t)\dd{t},
\end{equation}
where $u_{mn}(t)=d_{mn}(t)+G_t(m)-G_t(n)$. Since the injected source currents are arbitrary, the definition of the functional derivative gives
\begin{equation}
	\frac{\delta\E[J_\tau]}{\delta j_{mn}^{\mathrm{src}}(t)}=u_{mn}(t).
\end{equation}
Combining this identity with Eq.~\eqref{eq:varjump} yields the fluctuation-response relation
\begin{equation}
\tcboxmath{
	\Var(J_\tau)-\Var_0(J_\tau) = \int_0^\tau\sum_{m<n}\qty[\frac{\delta\E[J_\tau]}{\delta j_{mn}^{\mathrm{src}}(t)}]^2a_{mn}(t)\dd{t}.
}
\end{equation}
For perturbations generated by changes of the transition rates, this relation is equivalent to the dynamical fluctuation-response relation of Ref.~\cite{Aslyamov.2026.arxiv}. The present source-current formulation removes the particular rate parametrization: the elementary response variable is directly the injected edge current, with $u_{mn}$ as its functional susceptibility.

We next consider a specific perturbation direction in the space of injected edge currents, $\delta j_{mn}^{\mathrm{src}}(t)=\epsilon g_{mn}(t)$, where $\epsilon$ is a small scalar perturbation parameter and $g_{mn}(t)=-g_{nm}(t)$ is an arbitrary antisymmetric edge field. For example, such a source-current perturbation can be generated by an infinitesimal change of the transition rates, $\msf{W}_t\to\msf{W}_t+\epsilon\msf{W}_t'$ and $g_{mn}(t)=w_{mn}'(t)p_n(t)-w_{nm}'(t)p_m(t)$. We define the corresponding linear response and quadratic cost by
\begin{align}
	\Rc[g]&\equiv\partial_\epsilon\E_\epsilon[J_\tau]\big|_{\epsilon=0},\\
	\Ac[g]&\equiv\int_0^\tau\sum_{m<n}\frac{g_{mn}(t)^2}{a_{mn}(t)}\dd{t},
\end{align}
where $\E_\epsilon$ denotes the expectation under the perturbed dynamics. Using the response relation above gives
\begin{equation}
	\Rc[g]=\int_0^\tau\sum_{m<n}u_{mn}(t)g_{mn}(t)\dd{t}.
\end{equation}
The Cauchy-Schwarz inequality then yields
\begin{equation}
\tcboxmath{
	\Rc[g]^2\le \Ac[g]\qty[\Var(J_\tau) - \Var_0(J_\tau)].
}
\end{equation}

\begin{proposition}\label{prop:mjp.perturb}
Consider an arbitrary infinitesimal source current $\delta j_{mn}^{\mathrm{src}}(t)=-\delta j_{nm}^{\mathrm{src}}(t)$ injected into the probability current on each edge, while the initial distribution and the observable fields are held fixed. Then the corresponding first variation of the observable average is
\begin{equation}
    \delta\E[J_\tau]=\int_0^\tau\sum_{m<n}u_{mn}(t)\delta j_{mn}^{\mathrm{src}}(t)\dd{t},
\label{eq:mjpresp}
\end{equation}
where $u_{mn}(t)=d_{mn}(t)+G_t(m)-G_t(n)$.
\end{proposition}

\begin{proof}
Let
\begin{equation}
    \Jc_{mn,t}(\vb*{p})\equiv w_{mn}(t)p_n-w_{nm}(t)p_m
\end{equation}
denote the probability-current operator on the edge $m\leftrightarrow n$. Under an arbitrary infinitesimal source-current injection, the total variation of the edge current is
\begin{equation}
    \delta j_{mn}(t)=\Jc_{mn,t}(\delta\vb*{p}_t)+\delta j_{mn}^{\mathrm{src}}(t).
\label{eq:mjp.total.current.variation}
\end{equation}
The source current induces the probability source
\begin{equation}
    \delta s_m(t)\equiv\sum_{n(\neq m)}\delta j_{mn}^{\mathrm{src}}(t),
\label{eq:mjp.source}
\end{equation}
so that the corresponding density variation is generated by the unperturbed dynamics.
Let
\begin{equation}
    \msf{U}(t,s)=\mca{T}\exp\qty(\int_s^t \msf{W}_r\dd{r})
\end{equation}
denote the forward propagator generated by $\msf{W}_t$. Since the initial distribution is held fixed, the induced density variation can be written directly as
\begin{equation}
    \delta\vb*{p}_t=\int_0^t\msf{U}(t,s)\delta\vb*{s}_s\dd{s},
\label{eq:mjp.delta.p}
\end{equation}
where $\delta\vb*{s}_t=(\delta s_1(t),\delta s_2(t),\ldots)^\top$.

Since $\E[J_\tau]=\int_0^\tau\qty[\sum_{m<n}d_{mn}(t)j_{mn}(t)+\sum_n \phi_n(t)p_n(t)]\dd{t}$, the first variation of the observable average consists of the density-induced contribution and the direct contribution from the injected edge currents:
\begin{align}
    \delta\E[J_\tau]&=\int_0^\tau\qty[\sum_{m<n}d_{mn}(t)\delta j_{mn}(t)+\sum_n \phi_n(t)\delta p_n(t)]\dd{t}\notag\\
    &=\int_0^\tau\vb*{\Phi}_t\cdot\delta\vb*{p}_t\dd{t}+\int_0^\tau\sum_{m<n}d_{mn}(t)\delta j_{mn}^{\mathrm{src}}(t)\dd{t}.
\label{eq:mjp.response.current.rep}
\end{align}
Let $\msf{U}^\dagger(t,s)$ denote the adjoint propagator. The backward equation $\partial_t\vb*{G}_t+\msf{W}_t^\dagger\vb*{G}_t=-\vb*{\Phi}_t$ together with its terminal condition $\vb*{G}_\tau=0$ gives
\begin{equation}
    \vb*{G}_s=\int_s^\tau\msf{U}^\dagger(t,s)\vb*{\Phi}_t\dd{t}.
\label{eq:mjp.G.prop}
\end{equation}
Using Eqs.~\eqref{eq:mjp.delta.p} and \eqref{eq:mjp.G.prop}, we obtain
\begin{align}
    \int_0^\tau\vb*{\Phi}_t\cdot\delta\vb*{p}_t\dd{t}&=\int_0^\tau\int_0^t\vb*{\Phi}_t\cdot\msf{U}(t,s)\delta\vb*{s}_s\dd{s}\dd{t}\notag\\
    &=\int_0^\tau\left[\int_s^\tau\msf{U}^\dagger(t,s)\vb*{\Phi}_t\dd{t}\right]\cdot\delta\vb*{s}_s\dd{s}\notag\\
    &=\int_0^\tau\vb*{G}_s\cdot\delta\vb*{s}_s\dd{s}.
\end{align}
Using Eq.~\eqref{eq:mjp.source} and the antisymmetry $\delta j_{mn}^{\mathrm{src}}=-\delta j_{nm}^{\mathrm{src}}$, this becomes
\begin{align}
    \vb*{G}_t\cdot\delta\vb*{s}_t&=\sum_mG_t(m)\sum_{n(\neq m)}\delta j_{mn}^{\mathrm{src}}(t)\notag\\
    &=\sum_{m<n}[G_t(m)-G_t(n)]\delta j_{mn}^{\mathrm{src}}(t).
\end{align}
Hence,
\begin{equation}
    \int_0^\tau\vb*{\Phi}_t\cdot\delta\vb*{p}_t\dd{t}=\int_0^\tau\sum_{m<n}[G_t(m)-G_t(n)]\delta j_{mn}^{\mathrm{src}}(t)\dd{t}.
\end{equation}
Substituting this result into
Eq.~\eqref{eq:mjp.response.current.rep} yields
\begin{align}
    \delta\E[J_\tau]&=\int_0^\tau\sum_{m<n}\qty[d_{mn}(t)+G_t(m)-G_t(n)]\delta j_{mn}^{\mathrm{src}}(t)\dd{t}\notag\\
    &=\int_0^\tau\sum_{m<n}u_{mn}(t)\delta j_{mn}^{\mathrm{src}}(t)\dd{t},
\end{align}
which proves the result.
\end{proof}

\subsection{Improved finite-time TUR}
Let $\Lg_t=\Lg_{\lambda(\upsilon t)}$, where $\lambda(\upsilon t)$ is a time-dependent control protocol and $\upsilon$ is a speed parameter.
We restrict attention to time-antisymmetric currents, for which $\phi_n=0$ and
\begin{equation}
	J_\tau=\sum_{m\neq n}\int_0^\tau d_{mn}(t)\dd{\hat N_{mn}(t)}.
\end{equation}
All quantities in the section are evaluated at the reference speed $\upsilon=1$.
We hold the initial distribution fixed and assume that all explicit time dependence of the current weights is rescaled together with the protocol when $\upsilon$ is varied.

Applying the Cauchy-Schwarz inequality and using Proposition~\ref{prop:mjp.avg.pd.form}, we obtain
\begin{align}
	\qty[(\tau\partial_\tau-\upsilon\partial_\upsilon)\E[J_\tau]]^2
	&=\qty[\int_0^\tau\sum_{m<n}u_{mn}(t)j_{mn}(t)\dd{t}]^2\notag\\
	&\le\underbrace{\int_0^\tau\sum_{m<n}u_{mn}(t)^2a_{mn}(t)\dd{t}}_{\textstyle =\Var(J_\tau)-\Var_0(J_\tau)}\underbrace{\int_0^\tau\sum_{m<n}\frac{j_{mn}(t)^2}{a_{mn}(t)}\dd{t}}_{\textstyle =\Sigma_\tau^{\mathrm{ps}}}.
\end{align}
This immediately yields the improved finite-time TUR for an arbitrary time-dependent protocol:
\begin{equation}
\tcboxmath{
	\frac{\Var(J_\tau)-\Var_0(J_\tau)}{\qty[(\tau\partial_\tau-\upsilon\partial_\upsilon)\E[J_\tau]]^2}\ge\frac{1}{\Sigma_\tau^{\mathrm{ps}}}\ge\frac{2}{\Sigma_\tau}.
}
\end{equation}
Here,
\begin{equation}
	\Sigma_\tau^{\mathrm{ps}}
	\equiv\int_0^\tau\sum_{m<n}\frac{j_{mn}(t)^2}{a_{mn}(t)}\dd{t}
\end{equation}
is the pseudo-entropy production, which satisfies $\Sigma_\tau^{\mathrm{ps}}\le\min\qty(\Sigma_\tau/2,A_\tau)$.
An equivalent strengthened TUR was obtained in Ref.~\cite{Aslyamov.2026.arxiv}; the present derivation follows directly from the quadratic martingale representation.

In the stationary state with time-independent dynamics and current weights, $(\tau\partial_\tau-\upsilon\partial_\upsilon)\E[J_\tau]=\E[J_\tau]$, and the improved TUR becomes
\begin{equation}
	\frac{\Var(J_\tau)-\Var_0(J_\tau)}{\E[J_\tau]^2}\ge \frac{2}{\Sigma_\tau}.
\end{equation}
We now characterize the saturation condition using the exact variance formula.
Define the steady-state pseudo-entropy production rate by
\begin{equation}
	\sigma^{\mathrm{ps}}\equiv\sum_{m<n}\frac{j_{mn}^2}{a_{mn}}.
\end{equation}
Thus, $\Sigma_\tau^{\mathrm{ps}}=\tau\sigma^{\mathrm{ps}}$.
For a nontrivial observable with $\jr_{\st}\neq 0$ and $\sigma^{\mathrm{ps}}>0$, decompose
\begin{equation}
	u_{mn}(t)=\frac{\jr_{\st}}{\sigma^{\mathrm{ps}}}\frac{j_{mn}}{a_{mn}}+r_{mn}(t),
	\qquad
	\sum_{m<n}j_{mn}r_{mn}(t)=0.
\end{equation}
Substitution into the variance formula gives the exact decomposition
\begin{equation}
	\qty[\Var(J_\tau)-\Var_0(J_\tau)]-\frac{\E[J_\tau]^2}{\Sigma_\tau^{\mathrm{ps}}}=\int_0^\tau\sum_{m<n}a_{mn}r_{mn}(t)^2\dd{t}.
\end{equation}
Therefore, the stronger pseudo-entropy-production TUR is saturated if and only if $r_{mn}(t)=0$ for every edge and all $t\in[0,\tau]$, or equivalently,
\begin{equation}
	u_{mn}(t)=d_{mn}+G_t(m)-G_t(n)=\frac{\jr_{\st}}{\sigma^{\mathrm{ps}}}\frac{j_{mn}}{a_{mn}}.
\end{equation}
Because the right-hand side is time independent and $G_\tau(n)=0$, finite-time saturation requires both
\begin{equation}
	d_{mn}=\frac{\jr_{\st}}{\sigma^{\mathrm{ps}}}\frac{j_{mn}}{a_{mn}}
\end{equation}
and $G_t(m)-G_t(n)=0$ for every edge and all $t$.
Saturation of the conventional TUR additionally requires $\Sigma_\tau^{\mathrm{ps}}=\Sigma_\tau/2$.

We next show that $G_t(m)-G_t(n)=0$ on every edge if and only if $\Phi(n)=\jr_{\st}$ for every state $n$.
If $\Phi(n)=\jr_{\st}$, then $\vb*{G}_t=\jr_{\st}(\tau-t)\vb*{1}$ solves the backward equation $\partial_t\vb*{G}_t+\Lg_t^\dagger\vb*{G}_t=-\vb*{\Phi}$ with $\vb*{G}_\tau=0$. Uniqueness of the solution therefore implies $G_t(m)-G_t(n)=0$. Conversely, if $G_t(m)-G_t(n)=0$ on every edge, then $G_t(n)=c(t)$ for some function $c$ on each connected component.
Substitution into the backward equation gives $\Phi(n)=-\dot c(t)$.
Because the left-hand side depends only on $n$, whereas the right-hand side depends only on $t$, both are equal to a constant, say $b_0$.
It follows that $\jr_{\st}=\vb*{\Phi}\cdot\vb*{p}_{\st}=b_0$, and hence $\Phi(n)=\jr_{\st}$.

Under the saturation condition on $d_{mn}$, $\Phi(n)$ can also be evaluated explicitly:
\begin{align}
	\Phi(n)
	&=\frac{\jr_{\st}}{\sigma^{\mathrm{ps}}}\sum_{m(\neq n)}\frac{j_{mn}}{a_{mn}}w_{mn}\notag\\
	&=\frac{\jr_{\st}}{\sigma^{\mathrm{ps}}}\frac{1}{2p_n}\sum_{m(\neq n)}\qty(\frac{j_{mn}^2}{a_{mn}}+j_{mn})\notag\\
	&=\frac{\jr_{\st}}{\sigma^{\mathrm{ps}}}\frac{1}{2p_n}\sum_{m(\neq n)}\frac{j_{mn}^2}{a_{mn}},
\end{align}
where stationarity, $\sum_{m(\neq n)}j_{mn}=0$, was used in the last equality.
Defining the state-resolved pseudo-entropy production rate as
\begin{equation}
	\sigma_n^{\mathrm{ps}}\equiv\sum_{m(\neq n)}\frac{j_{mn}^2}{2a_{mn}},
\end{equation}
we conclude that $\sigma_n^{\mathrm{ps}}/p_n$ must be independent of $n$; in fact,
$\sigma_n^{\mathrm{ps}}/p_n=\sigma^{\mathrm{ps}}$.

\begin{proposition}\label{prop:mjp.avg.pd.form}
For any time-dependent protocol $\lambda(\upsilon t)$ and any initial probability distribution $\vb*{p}_0$, the following relation holds:
\begin{equation}
	\int_0^\tau\sum_{m<n}u_{mn}(t)j_{mn}(t)\dd{t}=(\tau\partial_\tau-\upsilon\partial_\upsilon)\E[J_\tau].
\end{equation}
\end{proposition}
\begin{proof}
Using $u_{mn}(t)=d_{mn}(t)+G_t(m)-G_t(n)$, we rewrite the left-hand side as
\begin{align}
	\int_0^\tau\sum_{m<n}u_{mn}(t)j_{mn}(t)\dd{t}
	&=\int_0^\tau\sum_{m<n}d_{mn}(t)j_{mn}(t)\dd{t}+\int_0^\tau\sum_{m<n}[G_t(m)-G_t(n)]j_{mn}(t)\dd{t}\notag\\
	&=\E[J_\tau]+\int_0^\tau\sum_nG_t(n)\dot p_n(t)\dd{t}\notag\\
	&=\E[J_\tau]+\int_0^\tau\vb*{G}_t\cdot\dot{\vb*{p}}_t\dd{t},
\end{align}
where we used $\phi_n=0$ and Kirchhoff's law in the second equality.
For any differentiable function $f(\tau,\upsilon)$,
\begin{equation}
	(\tau\partial_\tau-\upsilon\partial_\upsilon)f(\tau,\upsilon)=\frac{\dd}{\dd\alpha}f(\alpha\tau,\upsilon/\alpha)\Big|_{\alpha=1}.
\end{equation}
Under the transformation $(\tau,\upsilon)\to(\alpha\tau,\upsilon/\alpha)$, followed by the time change $s=\alpha t$, the protocol becomes $\lambda(\upsilon t)$ on $t\in[0,\tau]$, while the generator and the instantaneous current rate are rescaled as $\Lg_t\to\alpha\Lg_t$ and $\vb*{\Phi}_t\to\alpha\vb*{\Phi}_t$, respectively.
Consequently,
\begin{equation}
	(\tau\partial_\tau-\upsilon\partial_\upsilon)\E[J_\tau]=\dv{\alpha}\E_\alpha[J_\tau]\Big|_{\alpha=1},
\end{equation}
where $\E_\alpha[J_\tau]$ denotes the current average under the dynamics generated by $\alpha\Lg_t$.
More explicitly,
\begin{equation}
	\E_\alpha[J_\tau]=\int_0^\tau\alpha\vb*{\Phi}_t\cdot\vb*{p}_t^\alpha\dd{t},
	\qquad
	\partial_t\vb*{p}_t^\alpha=\alpha\Lg_t\vb*{p}_t^\alpha,
\end{equation}
with $\vb*{p}_0^\alpha=\vb*{p}_0$.
Differentiating with respect to $\alpha$ gives
\begin{equation}
	\frac{\dd}{\dd\alpha}\E_\alpha[J_\tau]\Big|_{\alpha=1}=\underbrace{\int_0^\tau\vb*{\Phi}_t\cdot\vb*{p}_t\dd{t}}_{\textstyle =\E[J_\tau]}+\int_0^\tau\vb*{\Phi}_t\cdot\widetilde{\vb*{p}}_t\dd{t},
\end{equation}
where $\widetilde{\vb*{p}}_t\equiv\partial_\alpha\vb*{p}_t^\alpha|_{\alpha=1}$ obeys
\begin{equation}
	\partial_t\widetilde{\vb*{p}}_t=\Lg_t\widetilde{\vb*{p}}_t+\Lg_t\vb*{p}_t,
	\qquad
	\widetilde{\vb*{p}}_0=0.
\end{equation}
The second term can be written as
\begin{equation}
	\int_0^\tau\vb*{\Phi}_t\cdot\widetilde{\vb*{p}}_t\dd{t}=\int_0^\tau\vb*{G}_t\cdot\dot{\vb*{p}}_t\dd{t}.
\label{eq:mjp.adjoint}
\end{equation}
To verify this identity, we differentiate $\vb*{G}_t\cdot\widetilde{\vb*{p}}_t$ and use the backward equation together with the boundary conditions:
\begin{align}
	0
	&=\vb*{G}_\tau\cdot\widetilde{\vb*{p}}_\tau-\vb*{G}_0\cdot\widetilde{\vb*{p}}_0\notag\\
	&=\int_0^\tau\dv{t}\qty(\vb*{G}_t\cdot\widetilde{\vb*{p}}_t)\dd{t}\notag\\
	&=\int_0^\tau\qty[(-\vb*{\Phi}_t-\Lg_t^\dagger\vb*{G}_t)\cdot\widetilde{\vb*{p}}_t+\vb*{G}_t\cdot\qty(\Lg_t\widetilde{\vb*{p}}_t+\Lg_t\vb*{p}_t)]\dd{t}\notag\\
	&=-\int_0^\tau\vb*{\Phi}_t\cdot\widetilde{\vb*{p}}_t\dd{t}+\int_0^\tau\vb*{G}_t\cdot\dot{\vb*{p}}_t\dd{t},
\end{align}
where the terms containing $\Lg_t$ cancel by adjointness.
Combining the above identities proves the result.
\end{proof}


%